\documentclass[12pt]{article}
\usepackage{amsmath}
\usepackage{multirow}
\usepackage{amsfonts}
\usepackage{amssymb,graphics,psfrag}
\usepackage{array,epsfig,multirow,graphicx}
\usepackage{comment}
\usepackage{slashed}
\usepackage{hyperref}
\usepackage{tensor}
\usepackage{booktabs}
\usepackage{colortbl}
\usepackage{hhline}
\usepackage{mathtools}
\usepackage{enumitem}
\usepackage{mathrsfs}  
\usepackage{xfrac}
\usepackage{tikz}
\usepackage{amssymb}
\usepackage[nosort]{cite}
\usetikzlibrary{decorations.markings}
\usetikzlibrary{shapes,arrows}
\usetikzlibrary{decorations.pathreplacing}
\usetikzlibrary{arrows,positioning}

\def\hybrid{\topmargin -20pt    \oddsidemargin 0pt
        \headheight 0pt \headsep 0pt 
        \textwidth 6.25in      
        \textheight 9 in      
        \marginparwidth .875in
        \parskip 5pt plus 1pt
          \jot = 1.5ex
  }
\hybrid
\numberwithin{equation}{section}
\numberwithin{table}{section}

\newcommand{\beq}{\begin{equation}}
\newcommand{\eeq}{\end{equation}}
\newcommand{\be}{\begin{equation}}
\newcommand{\ee}{\end{equation}}
\newcommand{\bea}{\begin{eqnarray}}
\newcommand{\eea}{\end{eqnarray}}
\newcommand{\ben}{\begin{eqnarray*}}
\newcommand{\een}{\end{eqnarray*}}               
\newcommand{\ba}{\begin{align}}
\newcommand{\ea}{\end{align}}
\newcommand{\bt}{\begin{tabular}}
\newcommand{\et}{\end{tabular}}
\newcommand{\bc}{\begin{center}}
\newcommand{\ec}{\end{center}}
\newcommand{\ax}{\alpha}
\newcommand{\bx}{\beta}

\newcommand{\dx}{\delta}
\newcommand{\zx}{\zeta}
\newcommand{\ox}{\omega}

\newcommand{\cO}{\mathcal{O}}

\newcommand{\cL}{\mathcal{L}}

\newcommand{\cK}{\mathcal{K}}
\newcommand{\cN}{\mathcal{N}}
\newcommand{\cX}{\mathcal{X}}

\newcommand{\cG}{\mathcal{G}}

\newcommand{\cJ}{\mathcal{J}}
\newcommand{\cR}{\mathcal{R}}

\newcommand{\cV}{\mathcal{V}}

\newcommand{\cM}{\mathcal M}

\newcommand{\ib}{{\bar\imath }}
\newcommand{\jb}{{\bar\jmath }}

\newcommand{\ah}{{\hat{a}}}

\def\cZ{\mathcal{Z}}

\newcommand{\bbM}{\mathbb{M}}

\newcommand{\cref}{{\bf [check ref]}}

\newcommand{\M}{M}

\newcommand{\G}{\mathcal{I}}

\definecolor{mppgreen}{RGB}{17,102,86}
\definecolor{mppgray}{RGB}{221,222,214}

\newcommand{\wh}[1]{{\hat{#1}}}
\newcommand{\wt}[1]{{\tilde{#1}}}

\def\e{\text{e}}
\newcommand{\inds}[1]{\indices{#1}}
\def\tbz{{\scriptscriptstyle{(0)}}}
\def\tbo{{\scriptscriptstyle{(1)}}}
\def\la{\langle}
\def\ra{\rangle}
\def\pd{\partial}
\def\jb{{\bar{\jmath}}}
\def\pdb{{\bar{\partial}}}
\def\Kahler{{K\"ahler }}
\def\alh{\hat{\alpha}}

\def\upd{\mathrm{d}}

\def\Ccal{{\mathcal{Z}}^{\scriptscriptstyle{(0)}}}
\def\Cbar{{\mathcal{Z}}}
\def\lam{a}
\def\sig{b}
\def\tbfive{{\scriptscriptstyle{(5)}}}
\def\tbfour{{\scriptscriptstyle{(4)}}}
\def\Vm{\mathcal{V}_{\text{ex}}}
\def\Va{\mathcal{V}_{\text{IIA}}}
\def\Rcal{\mathscr{R}}
\def\Tcal{\mathscr{T}}
\def\om{\omega}

\def\del{\delta}
\def\ah{\hat{\alpha}}
\def\tba{{\scriptscriptstyle{(\alpha)}}}
\def\tbb{{\scriptscriptstyle{(\beta)}}}
\def\ze{\zeta}
\def\X{\mathcal{X}}
\def\M{\mathcal{M}}
\def\tbzero{{\scriptscriptstyle{(0)}}}
\def\nab{\nabla}
\def\V{\mathcal{V}}
\def\K{\mathcal{K}}

\def\tbf{{\scriptscriptstyle{(4)}}}
\def\nab{\nabla}
\def\am{a^{\scriptscriptstyle{(\text{\tiny{M}})}}}
\def\ah{\hat{\alpha}}
\def\aa{a^{\scriptscriptstyle{(\tiny{\text{A}})}}}

\def\Ycal{\mathscr{Y}}
\def\Tr{\text{Tr }}
\def\pa{\partial}
\def\G{\Gamma}
\def\fr{\frac}
\def\we{\wedge}
\def\lz{\ell_0}
\def\lo{\ell_1}
\def\Z0{Z^\tbz}

\def\blfootnote{\xdef\@thefnmark{}\@footnotetext}
\long\def\symbolfootnote[#1]#2{\begingroup%
\def\thefootnote{\fnsymbol{footnote}}\footnote[#1]{#2}\endgroup}

\begin{document}

\baselineskip=15pt

\begin{titlepage}
\begin{flushright}
\parbox[t]{1.8in}{\begin{flushright}  MPP-2017-2\\
IPMU17-0037 \end{flushright}}
\end{flushright}

\begin{center}

\vspace*{ 1.2cm}

{\Large \bf Higher derivatives in Type II and M-theory\\[.2cm]
                  on Calabi-Yau threefolds}

\vskip 1.2cm

\renewcommand{\thefootnote}{}
\begin{center}
 {Thomas W.~Grimm$^{\,1,2}$, Kilian Mayer$^{\, 1}$ and Matthias Weissenbacher$^{\, 3}$\ \footnote{t.w.grimm@uu.nl \  k.mayer@uu.nl \  matthias.weissenbacher@ipmu.jp}}
\end{center}
\vskip .2cm
\renewcommand{\thefootnote}{\arabic{footnote}}

{
$^1$ Institute for Theoretical Physics and \\
Center for Extreme Matter and Emergent Phenomena,\\
Utrecht University, Princetonplein 5, 3584 CE Utrecht, The Netherlands

$^2$ Max-Planck-Institut f\"ur Physik, \\
F\"ohringer Ring 6, 80805 Munich, Germany

$^3$ Kavli Institute for the Physics and Mathematics of the Universe, University of Tokyo, \\Kashiwa-no-ha 5-1-5, 277-8583, Japan}

 \vspace*{.2cm}

\end{center}

 \renewcommand{\thefootnote}{\arabic{footnote}}
 
\begin{center} {\bf Abstract } \end{center}

The four- and five-dimensional effective actions of Calabi-Yau threefold compactifications are derived with a focus on terms involving up to four space-time derivatives. The starting points for these 
reductions are the ten- and 
eleven-dimensional supergravity actions supplemented with the known eight-derivative corrections that have been inferred 
from Type II string amplitudes. The corrected background solutions are determined and the 
fluctuations of the K\"ahler structure of the compact space and the form-field background are discussed. 
It is concluded that the two-derivative effective actions for these fluctuations only 
takes the expected supergravity form if certain additional ten- and eleven-dimensional higher-derivative terms 
for the form-fields are included.
The main results on the four-derivative terms include a detailed treatment of higher-derivative gravity coupled to
K\"ahler structure deformations. This is supplemented by a derivation of the vector sector in 
reductions to five dimensions.  While the general result is only given as an expansion in the fluctuations, a complete 
treatment of the one-K\"ahler modulus case is presented for both Type II theories and M-theory.

\hfill {February, 2017}
\end{titlepage}

\tableofcontents

\newpage



\section{Introduction and Summary}

Compactifications of Type II string theories and M-theory on Calabi-Yau manifolds are interesting from various interrelated perspectives 
reaching from phenomenologically motivated model building to purely mathematical studies of quantum geometry. In particular, reductions on 
Calabi-Yau threefolds have been studied over several decades leading to intriguing discoveries such as mirror 
symmetry. Compactifications of Type IIA and Type IIB string theory on Calabi-Yau threefolds result in four-dimensional effective theories 
that admit $\cN = 2$ supersymmetry \cite{Blumenhagen:2013fgp}, or sixteen preserved
four-dimensional supercharges. Similarly the compactifications of M-theory on Calabi-Yau threefolds lead to
five-dimensional supersymmetric theories that preserve sixteen supercharges
or five-dimensional $\cN=2$. Four-dimensional $\cN=1$ supersymmetric effective theories are obtained 
from Calabi-Yau threefold compactifications including D-branes and orientifold planes \cite{Ibanez:2012zz,Blumenhagen:2006ci}. 
So far, however, the effective actions arising in Calabi-Yau threefold compactifications are only fully understood at the 
two-derivative level. In this work we aim to go beyond this and systematically include terms up to four derivatives into the 
effective action.

Our starting point for the compactifications are the ten- or eleven-dimensional low energy effective action 
of of Type II string theory or M-theory.
At the two-derivative level these are the well-known ten-dimensional Type IIA and Type IIB supergravity actions with 
$\cN=2$ supersymmetry, and the unique eleven-dimensional supergravity theory with $\cN=1$ supersymmetry. 
The Type II supergravity actions are modified by two types of stringy. First, there are $\ax'$-corrections 
imprinted in higher-derivative terms. Second, the worldsheet genus expansion in $g_s$ 
leading to higher dimensional operators in the effective theory depending on higher orders of the dilaton.
Using perturbative string theory these corrections can be computed by explicitly 
evaluation string scattering amplitudes. While this can be notoriously difficult, various corrections are known in the literature. 
For example, it is well established that the $\mathcal{R}^4$-terms in Type II theories at order $\ax'^3$ are the 
complete set of purely gravitational 
eight-derivative terms \cite{Gross:1986iv,Gross:1986mw}. 
Additionally, there are several known higher-derivative terms 
in ten dimensions involving the NS-NS two-form $B_2$ and dilaton \cite{Duff:1995wd,Green:1997di, Green:1997as,Kiritsis:1997em,Kehagias:1997cq,Kehagias:1997jg,Policastro:2006vt,Policastro:2008hg,Liu:2013dna,Minasian:2015bxa}. 
It was conjectured in \cite{Liu:2013dna} that the $B_2$-completion of the $\cR^4$-terms can almost be completely captured 
by introducing a connection with torsion, where $H_3=\upd B_2$ acquires the role of the torsion. A strategy to extract corrections 
to the eleven-dimensional supergravity action is to up-lift the Type IIA corrections computed by string amplitudes. From the known 
Type IIA terms one can thus infer $\mathcal{R}^4$-terms \cite{Russo:1997mk,Tseytlin:2000sf} and terms involving the 
M-theory three-form $C_3$ \cite{Peeters:2005tb,Liu:2013dna}. The supersymmetry completions of such terms have been studied in \cite{Hyakutake:2005rb,Hyakutake:2006aq,Hyakutake:2007sm}.
It is eminent that the only expansion parameter in eleven dimensions is the eleven-dimensional Planck length $\ell_M$.

 Obtaining effective actions in four and five dimensions taking into account higher-derivative corrections in ten 
 or eleven dimensions is an important yet challenging task.\footnote{We extensively use the packages 
 xAct, xTensor \cite{Martin-Garcia:2007bqa,Nutma:2013zea,Martin-Garcia} for tensor computer algebra in Mathematica
 to perform our computations.}
 Since the first corrections at order $\ax'^3$ involve terms quartic in the Riemann tensor, the derivation of the effective couplings 
 of the geometric moduli constitutes a computationally challenging task. Furthermore, although the $\cR^4$-couplings 
 in ten and eleven dimensions are well established, less is known about the completion of the action 
 at eight-derivative level for the full NS-NS and R-R sector.  
 Important partial results on the R-R sector can be found, for example, in \cite{Paulos:2008tn},
 but it is desirable to obtain a complete action via, for example, supersymmetric completion or 
 comparison with scattering amplitudes.  
 Another important challenge is the construction of general higher-derivative supergravity 
 theories in various dimensions independent of their string theory origins. 
While many standard formulations of supergravity at two derivatives are well established, there exist only 
partial results beyond the two-derivative truncation. In the considered string compactifications
a general higher-derivative supergravity action would not only allow us to check consistency with 
the expected preserved supersymmetry, but also suggest a clear systematics to determine 
physical couplings in terms of geometric data of the compactification space.

Higher-derivative corrections in string theory and M-theory have led to interesting modifications of the 
low energy dynamics and phenomenological models obtained in compactifications. In the following we will only 
give a very incomplete list of past applications for which higher-derivative corrections play a prominent role. 
It is well known that higher-derivative corrections give rise to corrections of the \Kahler potential of $\cN=2$ compactifications involving characteristic classes of the compactification Calabi-Yau threefold \cite{Candelas:1990rm,Antoniadis:1993ze,Antoniadis:1997eg}. It was later argued 
in \cite{Becker:2002nn,Bonetti:2016dqh} that some of these corrections survive the truncation to $\cN=1$ supergravity in compactifications 
including fluxes and an orientifold projection. 
Although obtaining a general higher-derivative theory in five dimensions is not yet available, an off-shell completion 
of $R^2$-terms in five-dimensional $\cN=2$ supergravity and its on shell version for pure supergravity 
is known \cite{Hanaki:2006pj,Ozkan:2013nwa}. These results were used to study the effect of higher-derivative 
corrections on five-dimensional black holes as reviewed, for example, in \cite{Castro:2008ne}. 
Four-derivative $\cN=1$ and $\cN=2$ supergravity theories in four spacetime dimensions are not yet 
completely understood.
Recently, progress towards classifying four-derivative superspace operators in the context of $\cN=1$ supergravity 
was made in \cite{Ciupke:2016agp}. This interest arose from the observation that higher-derivative terms in Type IIB orientifolds 
might stabilze K\"ahler moduli \cite{Ciupke:2015msa}.
Let us emphazise out that a first complete treatment of the kinetic terms for the \Kahler deformations originating from 
$\cR^4$-terms in eleven dimensions was presented in \cite{Grimm:2014xva, Grimm:2014efa,Grimm:2015mua}, 
where M-theory was compactified to three dimensions on a Calabi-Yau fourfold.

Our goal in this work is to study various compactifications of M-theory and Type II string theories on Calabi-Yau threefolds
taking into account the known eight-derivative terms. We review the considered known and conjectured terms in the 
ten and eleven-dimensional action contributing at order $\ax'^3$. This will allow us to solve the equations of motion 
determining background solutions possessing a compact, six-dimensional manifold which 
is topologically a Calabi-Yau threefold. 
We find that the modified solution in Type IIA involves a non-trivial background of the dilaton given by the Euler 
density in six dimensions on the lowest order Calabi-Yau geometry. Furthermore, the metric receives a correction 
leading to a deviation from Ricci flatness, which, however, does not alter the cohomology class of the curvature two-form. 
We then briefly comment on the relation between M-theory and Type IIA at order $\ax'^3$ and explain 
why adding a certain term in eleven dimensions is necessary.

The main part of this work is devoted to a detailed discussion of the dimensional reduction of M-theory and Type II 
on $Y_3$ including infinitesimal deformations of the \Kahler structure. In the M-theory reduction we also 
present results on the five-dimensional two- and four-derivative effective action also including vectors 
arsing from the M-theory three-form, while in Type II theories we comment on including the dilaton and
modes from the NS-NS two-from. 
We first derive the five-dimensional two-derivative effective action and show that 
eleven-dimensional eight-derivative terms only lead to a modification of the 
scalar field corresponding to the overall classical volume of $Y_3$ in accordance with \cite{Antoniadis:1997eg}. 
Furthermore, we then derive the five-dimensional four-derivative 
terms originating from the eleven-dimensional eight-derivative terms. Increasing 
computational complexity forces us to restrict to terms quadratic in the fluctuation.  
The main couplings at four-derivative level are found to be divisor integrals of the second Chern class of $Y_3$ and two new 
tensorial structures $Z_{i \bar{\jmath} k \bar{l}}$, $\cX^\tbz_{a i \bar{\jmath}}$, see \eqref{ZetaTens} and \eqref{XTens}, involving one and two 
Riemann tensors on $Y_3$. However, the physical significance of the latter two 
remains unclear, since there might exist a field redefinition ambiguity in the description of higher-derivative theories. 
We also preform the dimensional reduction of Type II ten-dimensional effective actions with a focus  
the deformations of the \Kahler structure and scalars $b^a$ arising from the zero mode expansion of the NS-NS two-form. The two-derivative 
four-dimensional effective action for the K\"ahler structure deformations contains the well known shift with the Euler characteristic 
of $Y_3$ \cite{Antoniadis:1997eg}. In Type IIA reductions we then use the fact that the K\"ahler structure deformations and the scalars 
from the NS-NS two-form combine into the complexified \Kahler moduli in $\cN=2$ vector multiplets. In order to obtain their kinetic terms 
from a $\cN=2$ prepotential \cite{Candelas:1990rm}, we find that a string tree-level structure for the kinetic terms of the 
scalars $b^a$ is missing. This forces us to add a minimal novel set of $H_3$-terms in ten dimensions. In other words, 
we use four-dimensional supersymmetry constraints to suggest missing higher-derivative terms in the ten-dimensional action. 

In the last part of this work we discuss the dimensional reduction when considering 
only a single K\"ahler modulus $u=\log \cV $, where $\cV$ is the volume of $Y_3$. 
In this case all computations simplify significantly and can be preformed exactly without 
restricting to quadratic order in the fluctuations. 
We derive all four-derivative couplings stemming from the $\cR^4$-terms in M-theory and both Type II theories. 
Furthermore, we discuss the possibility of higher-derivative field redefinitions. This allows us to show that there exists a 
field basis in which the four-derivative effective actions can be chosen 
to only consist of the Gauss-Bonnet term and a four-derivative interaction $(\pd u)^4$. We also comment on 
the orientifold truncation of this $\cN=2$ result to a minimal supersymmetric setting in four dimensions. 
After truncation we aim to make contact to the proposal of \cite{Ciupke:2015msa} to stabilize moduli 
using higher-derivative terms. Our reduction allows us to determine the exact coefficient of the $(\pd u)^4$-term. 
However, our findings do not indicate which supersymmetric completion of this term has to be chosen.
To clarify the phenomenological relevance of higher-derivative terms for moduli stabilization
a direct derivation of the scalar potential appears to be crucial.

\section{Higher derivatives and circle reduction}

The purpose of this section is to introduce the bosonic low-energy effective actions of Type IIA string theory and 
M-theory including known and conjectured eight-derivative terms. Furthermore, we comment on the duality 
of the Type IIA and M-theory low-energy effective actions once eight-derivative terms are taken into account. 
As already mentioned in the introduction many authors contributed significantly 
to the determination of the higher-derivative corrections 
to these low-energy effective actions in the past. The most recent results on these corrections
were obtained in \cite{Liu:2013dna}. We will therefore follow this reference closely, but 
also suggest new terms that seem to be required by supersymmetry arguments. 

\subsection{Ten-dimensional Type IIA  supergravity action at eight derivatives}\label{IIAderivatives}

In the following we will use a tilde to denote a field or operator in ten dimensions and capital letters at the beginning of the alphabet for tensor indices in ten dimensions. The bosonic field content of ten-dimensional Type IIA supergravity is given by the ten-dimensional metric $\wt{g}_{A B}$, the dilaton in ten dimensions $\wt{\phi}$ and the two-form $\wt{B}_2$ all of which descend as massless modes from the NS-NS sector of the RNS superstring. These terms are augmented by further fields from the R-R sector which however do not play a role in the following. We will therefore restrict ourselves completely to the NS-NS fields. 

The part of the $\cN=2$ low energy effective action of the Type IIA superstring we are considering takes the schematic form
\be\label{IIAaction}
S_{\text{IIA}}=S_{\text{IIA}}^{\text{class}}+ \alpha S^{\text{tree}}_{\wt{R}^4}+ \alpha S^{\text{loop}}_{\wt{R}^4}+ \alpha S_{\wt{H}^2}\, ,
\ee 
where we already introduced the expansion parameter
\be\label{IIAalpha}
\alpha=\frac{\alpha'^3}{3 \cdot 2^{11}}.
\ee
The classical action at lowest order in the parameter $\ax$ has for the NS-NS fields the form
\be\label{IIAactionclass}
2 \kappa_{10}^2 S_{\text{IIA}}^{\text{class}}=\int_{\cM_{10}}\! \!\e^{-2 \wt{\phi}}\Big( \wt{R}\, \wt{\ast} \,1 +4 \upd \wt{\phi} \wedge \wt{\ast}\, \upd \wt{\phi}-\frac{1}{2} \wt{H}_3 \wedge \wt{\ast} \, \wt{H}_3\Big)\,,
\ee
where $\cM_{10}$ is the spacetime manifold, $\wt{R}$ is the Ricci scalar in ten spacetime dimensions and $\wt{H}_3=\upd \wt{B}_2$ is the field strength of the NS-NS two-form. The gravitational coupling in ten dimensions, denoted by $\kappa_{10}$ is related to the Regge slope  $\ax'$ by the relation
\be 
2 \kappa_{10}^2=(2\pi)^7 \ax'^4.
\ee
At eight derivatives the action gets supplemented by additional terms quartic in the Riemann tensor at both tree-level and one-loop in the string coupling $g_s$. These pieces of the action take the schematic form 
\begin{align}
2 \kappa^2_{10} S^{\text{tree}}_{\wt{R}^4}&=\zeta(3)\int_{\cM_{10}} \e^{-2 \wt{\phi}}\Big(\wt{t}_{8} \wt{t}_8+\frac{1}{8}\epsilon_{10}\epsilon_{10} \Big)\wt{R}^4 \, \wt{\ast}\, 1\, ,\label{IIAtree}\\
2 \kappa^2_{10} S^{\text{loop}}_{\wt{R}^4}&=\frac{\pi^2}{3}\int_{\cM_{10}}\Big(\wt{t}_{8} \wt{t}_8-\frac{1}{8}\epsilon_{10}\epsilon_{10} \Big)\wt{R}^4 \, \wt{\ast}\, 1\label{IIA1loop}.
\end{align}
For the detailed form of these terms and the definition of the tensor $t_8$ in ten and eleven dimensions we refer the reader to appendix \ref{higherappendix}. Note that the relative sign flip of the one-loop contribution compared to the tree-level piece is characteristic for Type IIA and does not appear in Type IIB.  Additionally, there is an an eight-derivative coupling involving the NS-NS two-form $\wt{B}_2$ and four Riemann tensors, which will, however, not play a role in the Type IIA discussion. Its M-theory counterpart will be taken into account when the four-derivative couplings in five dimensions are computed.

It is furthermore conjectured in \cite{Liu:2013dna} that the completion of the eight-derivative terms with respect to the NS-NS two-form is almost completely captured by introducing a connection with torsion. It is then claimed that higher-derivative terms involving $\wt{B}_2$ can be obtained from the $\wt{R}^4$ terms by computing the latter with respect to the aforementioned torsionful connection. We will not collect all the structures emerging from this procedure but outline the strategy how to get the eight-derivative terms we need for our discussion.

Both the tree-level and one-loop contributions to the $\wt{R}^4$ action can be expressed in terms of the two `superinvariants'
\begin{align}
\cJ_0&=\Big(\wt{t}_8 \wt{t}_8+\frac{1}{8} \epsilon_{10}\epsilon_{10} \Big)\wt{R}^4\, ,\label{J0}\\
\cJ_1&=\wt{t}_8 \wt{t}_8 \wt{R}^4-\frac{1}{4}\wt{t}_8 \epsilon_{10} \wt{B}_2 \wt{R}^4\, \, \label{J1}
\end{align}
such that the $\wt{R}^4$-terms read
\begin{align}
2 \kappa^2_{10} S^{\text{tree}}_{\wt{R}^4}&=\zeta(3)\int_{\cM_{10}} \e^{-2 \wt{\phi}} \, \cJ_0 \wt{\ast}1\\
2 \kappa^2_{10} S^{\text{loop}}_{\wt{R}^4}&=\frac{\pi^2}{3}\int_{\cM_{10}}(2 \cJ_1-\cJ_0) \, \wt{\ast}\, 1.\label{loopinJ}
\end{align}
The last term in (\ref{J1}) corresponds to the eight-derivative coupling involving $\wt{B}_2$ that we do not consider in the Type IIA context and is therefore also ignore in \eqref{loopinJ}. The $\wt{B}_2$ field completion at eight derivatives is according to \cite{Liu:2013dna} then given by the replacements
\begin{align}
\mathcal{J}_0 \rightarrow &\,\, \mathcal{J}_0(\Omega_+)+\Delta \mathcal{J}_0(\Omega_+,\wt{H}_3)=\nonumber\\
&=\Big( \wt{t}_8 \wt{t}_8+\frac{1}{8} \epsilon_{10} \epsilon_{10}\Big){\wt{R}^{4}}(\Omega_+)+\frac{1}{3}\epsilon_{10}\epsilon_{10}{\wt{H}_3}^2{\wt{R}^{ 3}}(\Omega_+)\label{J0rep}\\
\mathcal{J}_1 \rightarrow &\,\, \mathcal{J}_1(\Omega_+)=\wt{t}_8 \wt{t}_8 \wt{R}^{4}(\Omega_+)-\frac{1}{8} \epsilon_{10}\,\wt{t}_8 \, \wt{B}_2 \left(\wt{R}^{4}(\Omega_+)+\wt{R}^{4}(\Omega_-) \right) \, ,\label{J1rep}
\end{align}
where the Riemann tensor with respect to the connection with torsion $\Omega_{\pm}$ is given in components by
\begin{equation}\label{torsion}
{\wt{R}(\Omega_{\pm})}\inds{_{A_1 A_2}^{B_1 B_2}}={\wt{R}}\inds{_{A_1 A_2}^{B_1 B_2}}\pm\wt{\nab}_{[A_1} {\wt{H}_3}{}_{A_2]}{}^{B_1 B_2}+\frac{1}{2}\wt{H}_3{}_{[A_1}{}^{B_1 B_3} \wt{H}_3{}_{A_2] B_3}{}^{B_2}\, .
\end{equation}
For the detailed structure of the coupling $\epsilon_{10}\epsilon_{10} \wt{H}^2_3 \wt{R}^3$ we again refer the reader to appendix \ref{higherappendix}. The terms generated by this replacement up to quadratic order in the NS-NS three-form field strength are denoted by $S_{\wt{H}^2}$ in (\ref{IIAaction}).

However, as we will explain in section \ref{IIAsection}, the replacements \eqref{J0rep} and \eqref{J1rep} appear
not consistent with four-dimensional supersymmetry. We therefore propose a corrected replacement  given by
\begin{align}
\mathcal{J}_0& \rightarrow \,\, \mathcal{J}_0(\Omega_+)+\Delta \mathcal{J}_0(\Omega_+,\wt{H}_3)+\dx \cJ\, \nonumber~\\
\mathcal{J}_1 &\rightarrow \,\, \mathcal{J}_1(\Omega_+)+\frac{1}{2} \dx \cJ\,  ,~~~~~\text{where}\nonumber\\
\dx \cJ&=-2 \int_{\M_{10}}\!\!\! \wt{t}_8\, \wt{t}_8\, \wt{H}_3^2 \, \wt{R}^3 \,\wt{\ast} \, 1\label{newrep}
\end{align}
and the explicit index expression of (\ref{newrep}) is given by
\begin{equation}\label{ttHR}
\wt{t}_8\, \wt{t}_8\,\wt{H}^2_3 \, \wt{R}^3=\wt{t}_8^{A_1 \dotsb A_2}\, \wt{t}_{8 \, B_1 \dotsb B_8}\, \wt{H}_{3\, A_1 A_2 C}\, \wt{H}_3^{B_1 B_2 C}\, \wt{R}^{B_3 B_4}_{~~~~~A_3 A_4}\dotsb\wt{R}^{B_7 B_8}_{~~~~~A_7 A_8}\, .
\end{equation}
Notice, that the structure of the modified replacement \eqref{newrep} is such that the tree-level terms get modified, whereas the one-loop terms remain untouched. Furthermore, note that the index contraction of \eqref{ttHR}  is such that it cannot be obtained via the metric replacement \eqref{torsion} applied to the $\cR^4$-terms.

\subsection{Corrected $Y_3$ background solution at order $\ax'^3$ 
in Type IIA}\label{IIAsolution}

At lowest order in $\alpha'$ a solution of the equations of motion with a six-dimensional compact internal space 
preserving four-dimensional $\cN=2$ supersymmetry is simply given by a product manifold $\cM_{10}=\bbM_{1,3} \times Y_3$, 
a constant dilaton, and no background 
fluxes. Here $\bbM_{1,3}$ is four-dimensional Minkowski space and $Y_3$ is a Calabi-Yau manifold. Due to the 
eight-derivative couplings at order $\ax'^3$ we next look for a 
corrected solution to the modified equations of motion. In doing that we demand that 
the corrected solution reduces to the classical $Y_3$ solution at lowest order in the expansion parameter $\ax$ introduced in (\ref{IIAalpha}). 
Therefore, we make the ansatz 
\begin{align}\label{IIAansatz}
\la \upd \wt{s}^2\ra &=\eta_{\mu \nu}\upd x^\mu \upd x^\nu +\big(g^\tbz_{m n}+\ax g^\tbo_{m n} \big)\upd y^m \upd y^n \, ,\\
\la \wt{\phi} \ra&= \phi_0+\ax \la \phi^\tbo(y) \ra \, ,\nonumber\\
\la \wt{H}_3 \ra &=0\, .\nonumber
\end{align}
In this ansatz $\eta_{\mu \nu}$ denotes the four-dimensional Minkowski metric, $g^\tbz_{mn}$ is the lowest order Calabi-Yau metric and $\phi_0$ is a constant. The goal is now to fix the correction to the dilaton $\la \phi^\tbo \ra$ as well as the correction to the metric $g^\tbo_{m n}$. Note that in this section we work entirely with real indices $m, n=1,...,6$.

Varying the corrected action (\ref{IIAaction}) with respect to the fields and evaluating the resulting equations of motion on the ansatz (\ref{IIAansatz}) yields the following equations:
\begin{itemize}
\item {\bf external Einstein equation.}
\begin{equation}\label{extEinstIIA}
0={g^\tbz}\inds{^{mn}}{R^\tbo}\inds{_{mn}} +4 \nabla^\tbz_n \nabla^{\tbz\,n} \la \phi^\tbo(y) \ra 
\end{equation}
\item {\bf internal Einstein equations.}
\begin{align}\label{intEinstIIA}
0=R^\tbo_{mn}-\frac{1}{2} g^\tbz_{mn}{g^\tbz}\inds{^{kl}}{R^\tbo}\inds{_{kl}}
&-2 g^\tbz_{mn}\nabla^\tbz_k \nabla^{\tbz\, k} \la \phi^\tbo(y) \ra+2 \nabla^\tbz_{m} \nabla^{\tbz}_n \la \phi^\tbo(y) \ra\\
&+768\, (2\pi)^3\Big(\zeta (3)+\frac{\pi^2}{3}\e^{2 \phi_0}\Big) {J^\tbz}\inds{_{\!\!\!m}^r}{J^\tbz}\inds{_{\!\!\!n}^{s}} \nabla^\tbz_{r} \nabla^\tbz_{s} (\ast^\tbz_6\, c_3^\tbz)\, .\nonumber
\end{align}
\end{itemize}
In (\ref{extEinstIIA}) and (\ref{intEinstIIA}) we have employed the definition $R_{m n}(g^\tbz+\ax g^\tbo)\equiv R^\tbz_{m n}+\ax R^\tbo_{m n} $ and furthermore used the complex structure ${J^\tbz}\inds{_{\!\!\!m}^{n}}$, the Hodge star $\ast_6^\tbz$ and the third Chern class $c_3^\tbz$ all of which are evaluated on the zeroth order Calabi-Yau background. The equation of motion for the NS-NS two-form is trivially satisfied and the equation of motion of the dilaton coincides with the external Einstein equation in the string frame. Compatibility of the external and internal Einstein equation is achieved by taking the trace of the internal Einstein equation and comparing it to the external Einstein equation, which in turn fixes the correction of the dilaton to
\begin{align}\label{dilatoncorr}
\la \phi^\tbo(y)\ra&=384\,(2\pi)^3 \Big(\zeta(3)+\frac{\pi^2}{3}\e^{2 \phi_0} \Big)\ast_6^\tbz  c_3^\tbz.
\end{align}
Plugging (\ref{dilatoncorr}) again into the internal Einstein equation results in
\begin{equation}\label{Riccicorrection}
{R^{\tbo}}\inds{_{\!\!\!\!\!\!mn}}=-768\,(2\pi)^3  \Big(\zeta(3)+\frac{\pi^2}{3}\e^{2 \phi_0} \Big)\left(\nabla^\tbz_m \nabla^\tbz_n+{J^\tbz}\inds{_{\!\!\!m}^r}{J^\tbz}\inds{_{\!\!\!n}^s} \nabla^\tbz_r \nabla^\tbz_s \right)\ast^\tbz_6 c_3^\tbz.
\end{equation}
Going to complex indices in (\ref{Riccicorrection}) shows that $R^\tbo_{i \bar{\jmath}}  \sim \pd_i \pdb_\jb ( \ast_6^\tbz c_3^\tbz)$, which in turn has a solution $g^\tbo_{i \jb} \sim \pd_i \pdb_\jb f(y)$ for some specific function $f$ which depends on the compact manifold and serves of as a \Kahler  potential. The holomorphic and antiholomorphic indices can take the values $i=1,2,3$ and $\jb=\bar{1}, \bar{2}, \bar{3}$.  The precise form of $f$ can be computed explicitly, which we will only do for the case of M-theory in section \ref{Mthsolution} since the procedure is the same and the precise form of $f$ is of no physical importance, as the metric correction turns out to completely decouple from low energy dynamics.

\subsection{Eleven-dimensional  supergravity action at eight derivatives}\label{Mthderivatives}

The higher-derivative corrections obtained for the low-energy limit of Type IIA superstring can be lifted to an 
eleven dimensions and are believed to comprise the low-energy effective action of M-theory. Regarding notation we will use hats to 
indicate that a certain object is defined in eleven dimensions and eleven-dimensional indices $M, N,...$ from the middle of the alphabet. 
At two-derivative level the effective action of M-theory is eleven-dimensional $\cN=1$ supergravity \cite{Cremmer:1978km}. Its bosonic part is given by
\begin{equation}\label{Mthclass}
S^{\text{class}}_{\text{M}}=\frac{1}{2 \kappa_{11}^2}\int_{\cM_{11}}\wh{R}\, \wh{\ast}\, 1-\frac{1}{2}\wh{G}_4 \wedge \wh{\ast}\, \wh{G}_4-\frac{1}{6} \wh{C}_3 \wedge \wh{G}_4 \wedge \wh{G}_4\,.
\end{equation}
The dynamical degrees of freedom are the eleven-dimensional metric $\wh{g}_{ M N}$ and the M-theory three-form $\wh{C}_3$ with its field strength $\wh{G}_4=\upd \wh{C}_3$. As for the Type IIA action we introduce an expansion parameter
\begin{equation}\label{Mthalpha}
\alh=\frac{(4\pi \kappa_{11}^2)^{2/3}}{(2\pi)^4 3^2 2^{13}}\, ,
\end{equation}
where $\kappa_{11}$ is related to the eleven-dimensional Planck length $\ell_M$ 
as $\kappa_{11}^2=\frac{1}{2} (2\pi)^8 \ell_M^9$ such that $\alh \propto \ell_M^6$.
The eight-derivative action of M-theory up to terms quadratic in $\wh{G}_4$ takes the following schematic form 
\be\label{Mthaction}
S_{\text{M}}=S^{\text{class}}_{\text{M}}+\alh S_{\wh{R}^4}+\alh S_{\wh{C} \wh{X}_8}+\alh S_{\wh{G}^2 \wh{R}^3}+\alh S_{(\wh{\nabla} \wh{G})^2 \wh{R}^2}\,.
\ee
Let us proceed by introducing the various pieces contributing to the eight-derivative action (\ref{Mthaction}). The most prominent part is the well known $\wh{R}^4$ combination, which is given by 
\be\label{MthR4}
S_{\wh{R}^4}=\frac{1}{2 \kappa^2_{11}} \int_{\mathcal{M}_{11}} \Big(\wh{t}_8 \wh{t}_8-\frac{1}{24} \epsilon_{11} \epsilon_{11}\Big){\wh{R}}^4\, \wh{\ast}\, 1
\ee
and gets supplemented by an $\wh{R}^4$ coupling to the M-theory three-form $\wh{C}_3$ via an eight-form curvature polynomial $\wh{X}_8$, namely
\be\label{MthCX8}
S_{\wh{C}\wh{X}_8}=-\frac{3^2 2^{13}}{2 \kappa_{11}^2}\int_{\cM_{11}} \wh{C}_3 \wedge\wh{X}_8\, .
\ee
The sector involving the four-form field strength $\wh{G}_4$ is obtained by lifting the conjectured terms, which we recalled in section \ref{IIAderivatives}, to eleven dimensions \cite{Liu:2013dna}. The terms we are considering are the ones quadratic in $\wh{G}_4$ and their corresponding actions in (\ref{Mthaction}) are
\begin{align}
S^{}_{\wh{G}^2 \wh{R}^3}&=-\frac{1}{2 \kappa_{11}^2}\int_{\mathcal{M}_{11}} \Big( \wh{t}_8 \wh{t}_8+\frac{1}{96}\epsilon_{11} \epsilon_{11}  \Big)\wh{G}^{2 }_4 \wh{R}^{3}\, \wh{\ast}\, 1 ,\label{GHCTerms1}\\
S^{}_{(\wh{\nabla} \wh{G})^2\wh{R}^2}&=\frac{1}{2 \kappa_{11}^2} \int_{\mathcal{M}_{11}} 
\wh{s}_{18} (\wh{\nabla} \wh{G}_4)^2 \wh{R}^{2}\, \wh{\ast}\, 1\,.\label{GHCTerms2}
\end{align}
For the detailed structure of all the terms in (\ref{Mthaction}) we once again refer the reader to appendix \ref{higherappendix}. Note however that the new tensorial structure $\wh{s}_{18}$ is not fully known but contains six unfixed coefficients $a_i$ \cite{Peeters:2005tb,Gross:1986mw}. The reason for this ambiguity is that the analysis carried out in \cite{Peeters:2005tb} is sensitive only to terms which do not vanish at the level of the four point function. However, there are six independent combinations of contractions which require a five point function analysis to fix their corresponding coefficient in the effective action.

Let us stress that we will argue in section \ref{MthIIAduality} that we need to add another term quadratic in $\hat G_4$ 
to the eleven-dimensional effective action to ensure compatibility of the M-theory and Type IIA reductions 
for the considered backgrounds.

\subsection{Corrected $Y_3$ background solution at order $\ell_M^6$ in M-theory}\label{Mthsolution}

We now apply a similar strategy as in section \ref{IIAsolution} for the case of Type IIA and determine a background solution of M-theory at eight derivatives. This solution should again have the property that it reduces to the classical direct product solution of five-dimensional Minkowski spacetime and a compact Calabi-Yau threefold $\cM^{\text{class}}_{11}=\bbM_{1, 4} \times Y_3$ as $\wh{\ax} \to 0$ considered in \cite{Cadavid:1995bk}. This problem was already solved in the case of a three-dimensional Minkowski spacetime and a complex four-dimensional compact internal space, which is at lowest order a Calabi-Yau fourfold \cite{Becker:1996gj,Becker:2001pm,Grimm:2014xva}, a Spin$(7)$ holonomy manifold \cite{Lu:2004ng} or an internal manifold with $G_2$ holonomy \cite{Lu:2003ze}.

The ansatz for the fourfold solution in \cite{Becker:2001pm,Grimm:2014xva} involves a warp factor, fluxes and an overall Weyl rescaling. The necessity for warping and fluxes can be traced back to the fact, that the eight-form curvature polynomial $\wh{X}_8$ does not vanish on the internal Calabi-Yau manifold and one therefore has to take into account fluxes in order to ensure, that the $\wh{C}_3$ equation of motion is satisfied. In our case the situation is different, since the eight-form $\wh{X}_8$ trivially vanishes on the Calabi-Yau threefold $Y_3$. Thus the modified ansatz for the background solution can be taken to have the following form 
\begin{align}\label{Mthbackground}
\langle \mathrm{d}\wh{s}^2  \rangle&=\text{e}^{\alh \langle\Phi^{\tbo}(y^m)\rangle} \big( \eta_{\mu \nu}\mathrm{d}x^{\mu} \mathrm{d}x^{\nu}+\left(g^{\tbz}_{m n}+\alh g^{\tbo}_{m n} \right)\mathrm{d}y^{m} \mathrm{d}y^{n} \big)\,,\\
\langle \wh{G}_4 \rangle&=0.\nonumber
\end{align}
The variation of the quantum corrected action (\ref{Mthaction}) then gives rise to the following conditions on the corrections $\la \Phi^\tbo (y^m) \ra$ and $g^\tbo_{m n}(y^m)$:
\begin{itemize}
\item{\bf external Einstein equation.}
\begin{equation}
 g^{{\scriptscriptstyle (0)} m n}R^{\tbo}_{m n}-9 g^{{\scriptscriptstyle (0)}m n} \nabla^{\scriptscriptstyle (0)}_m \nabla^{\scriptscriptstyle (0)}_n \langle \Phi^{\tbo}\rangle=0
 \end{equation}
\item {\bf internal Einstein equation.}
\begin{align}
0=R^{\tbo}_{m n}-\frac{1}{2} g^{\scriptscriptstyle (0)}_{m n}g^{{\scriptscriptstyle (0)} k l}R^{\tbo}_{k l}&-\frac{9}{2} \nabla_m  \nabla_n \langle \Phi^{\tbo}\rangle+\frac{9}{2}g^{\scriptscriptstyle (0)}_{m n}\nabla_k \nabla^k \langle \Phi^{\tbo}\rangle\\
&+768(2\pi)^3 J^{{\scriptscriptstyle (0)} l}_{m} J^{{\scriptscriptstyle (0)} k}_{n} \nabla_l \nabla_k (\ast^{\scriptscriptstyle (0)}_{\scriptscriptstyle 6}c^{\scriptscriptstyle (0)}_3 ) \, . \nonumber
\end{align}
\end{itemize}
 In order to fix the global Weyl factor $\langle \Phi^{\tbo}\rangle$ one again takes the trace over the internal Einstein equation and eliminates all expressions involving $R^{\tbo)}_{m n}$ by making use of the external Einstein equation resulting in 
\begin{align}
\langle \Phi^{\tbo}\rangle&=-\frac{512}{3}(2\pi)^3\, \ast^{\scriptscriptstyle (0)}_{\scriptscriptstyle 6} c^{\scriptscriptstyle (0)}_3,\label{MthPhi}\\
R^{\tbo}_{m n}&=-768 (2\pi)^3\, \left(\nabla^{\scriptscriptstyle (0)}_m \nabla^{\scriptscriptstyle (0)}_n+J_m^{\scriptscriptstyle (0) l} J_n^{\scriptscriptstyle (0) k} \nabla^{\scriptscriptstyle (0)}_l \nabla^{\scriptscriptstyle (0)}_k \right)\ast^{\scriptscriptstyle (0)}_{\scriptscriptstyle 6} c^{\scriptscriptstyle (0)}_3.
\end{align}
It is important to notice that the correction to the internal Ricci tensor is governed by an expression which is twice a covariant derivative of the Hodge dual of the third Chern form. The strategy to solve this equation is to split $\ast^{\scriptscriptstyle (0)}_{\scriptscriptstyle 6} c^{\scriptscriptstyle (0)}_3$ into a part which is constant on the internal space and therefore drops out of the equation of motion and a part which varies non trivially over $Y_3$. To make this separation one uses the fact that the third Chern form satisfies $\mathrm{d}c^{\scriptscriptstyle (0)}_3=0$ but $\mathrm{d}^{\dagger}c^{\scriptscriptstyle (0)}_3\neq 0$ and can therefore be expanded as
\begin{equation}\label{c3decomp}
c^{\scriptscriptstyle (0)}_3=\Pi_H c^{\scriptscriptstyle (0)}_3+i \partial \bar{\partial}   \xi
\end{equation}
by applying the $\pd \pdb$-Lemma, where $\Pi_H c^{\scriptscriptstyle (0)}$ is the harmonic part of $c^{\scriptscriptstyle (0)}_3$ which is unique by virtue of the Hodge decomposition theorem and $\xi$ is a (2,2)-form satisfying $\partial^{\dagger} \xi=\bar{\partial}^{\dagger} \xi=0$. The unspecified (2,2)- form clearly parametrizes the non-harmonicity of the third Chern form. Since $\Pi_H c^{\scriptscriptstyle (0)}_3$ is harmonic by definition and $\lbrack  \Delta_{\mathrm{d}},*_6 \rbrack=0$ it follows immediately that the scalar function $h(y^m) \equiv \ast^\tbz_6 \Pi_H c^{\scriptscriptstyle (0)}_3$ is constant on the compact $Y_3$ and can therefore be ignored in the equation of motion. Furthermore, using that on a K\"ahler manifold the Laplacian satisfies
$
\Delta_{\mathrm{d}}=2 \Delta_{\partial}=2\Delta_{\bar{\partial}}
$
one shows that
\begin{equation}
i \ast^{\scriptscriptstyle (0)}_{\scriptscriptstyle 6} \partial \bar{\partial} \xi = -\frac{1}{2}\Delta^{\scriptscriptstyle (0)} \ast^{\scriptscriptstyle (0)}_{\scriptscriptstyle 6} \left( J^{\scriptscriptstyle (0)} \wedge \xi \right),
\end{equation}
where $J^{\scriptscriptstyle (0)}$ is the K\"ahler form of $Y_3$ and $\Delta^{\scriptscriptstyle (0)}=\nabla^{\scriptscriptstyle  (0)} _k \nabla^{{\scriptscriptstyle (0)}k}$ is the Laplace-Beltrami operator. Having determined the non-trivial part of the correction to the internal Ricci tensor the equation determining $g^{\tbo}$ reads
\begin{equation}
R^{\tbo}_{m n}=384 (2\pi)^3\, \left( \nabla^{\scriptscriptstyle (0)}_m \nabla^{\scriptscriptstyle (0)}_n \nabla^{\scriptscriptstyle (0)}_k \nabla^{{\scriptscriptstyle (0)} k}+J^{{\scriptscriptstyle (0)} r}_m  J^{{\scriptscriptstyle (0)} s}_n \nabla^{\scriptscriptstyle (0)}_r \nabla^{\scriptscriptstyle (0)}_s \nabla^{\scriptscriptstyle (0)}_k \nabla^{{\scriptscriptstyle (0)} k} \right)\ast^{\scriptscriptstyle (0)}_{\scriptscriptstyle 6} \left( J^{\scriptscriptstyle (0)} \wedge \xi \right)
\end{equation}
whose solution can be checked to be
\begin{equation}\label{metsolution}
g^{\tbo}_{m n}=-768  (2\pi)^3\,\left( J_m^{{\scriptscriptstyle (0)} k}  J_n^{{\scriptscriptstyle (0)} l} \nabla_k \nabla_l +\nabla_m \nabla_n \right)\ast^{\scriptscriptstyle (0)}_{\scriptscriptstyle 6} \left( J^{\scriptscriptstyle (0)} \wedge \xi \right).
\end{equation}
We again observe in (\ref{metsolution}), that the correction to the metric is twice the derivative of a scalar function. We once more want to stress, that one of the key ingredients for the derivation of consistent two-derivative effective actions from dimensional reduction is the interplay between higher-derivative corrections and the fully backreacted background solution, as we will see in section \ref{Mthreduction}.

\subsection{Comments on M-theory - Type IIA duality on Calabi-Yau backgrounds at eight-derivative level}\label{MthIIAduality}

In the following we argue that we need to include another term in the eleven-dimensional action 
in order to obtain after compactification a five-dimensional effective action that is consistent with $\cN=2$ supergravity. 
Such a term is necessary due to the fact that the considered background solutions include a non-trivial 
correction to the ten-dimensional dilaton (\ref{dilatoncorr}). The additional part of the action then compensates for the absence of the dilaton in eleven dimensions. For the discussion in this section we will closely follow \cite{Lu:2003ze} and extend it by including the M-theory three-form. The additional piece of the eight-derivative action at order $\wh{\ax}$ we need is
\be\label{11dcorr}
\Delta S_{\text{M}}=256 \int_{\cM_{11}}\wh{Z}\, \wh{G}_4 \wedge \wh{\ast}\, \wh{G}_4\, ,
\ee
where $\wh{Z}$ is the generalization of the six-dimensional Euler density to eleven dimensions
\be\label{Z11}
\wh{Z}=\frac{1}{12}\big( {\wh{R}}_{M_1 M_2}^{~~~~~~~M_3 M_4} {\wh{R}}_{M_3 M_4}^{~~~~~~~M_5 M_6} {\wh{R}}_{M_5 M_6}^{~~~~~~~M_1 M_2}-2 {\wh{R}}_{M_1~~~M_3}^{~~~M_2~~~M_4}  {\wh{R}}_{M_2~~~M_4}^{~~~M_5~~~M_6} {\wh{R}}_{M_5~~~M_6}^{~~~M_1~~~M_2}\big)\,.
\ee
We will now argue that the corrected dilaton in ten dimensions can not be identified with the overall Weyl factor (\ref{MthPhi}) but requires the inclusion of  (\ref{11dcorr}). This indicates that the M-theory-Type IIA duality at eight derivatives has to be modified and thus deviates from the simple $S^1$ reduction of the classical case. As we showed in section \ref{Mthsolution} the corrected background solution in M-theory requires a Weyl factor (\ref{MthPhi}) whereas the Type IIA equations of motion can be solved without the latter. We therefore move to a frame in eleven dimensions which leads to the same Einstein equations for the backgrounds we are considering. This is achieved by the redefinition
\be\label{11dredef}
\wh{g}_{M N} \to \e^{-\frac{512}{3}\wh{\ax}\wh{Z}}\wh{g}_{M N}\, ,
\ee
such that the classical action picks up two additional terms from this field redefinition
\be
2 \kappa_{11}^2S_{\text{M}}^{\text{class}} \to 2 \kappa_{11}^2S_{\text{M}}^{\text{class}}-\int_{\cM_{11}}768\, \wh{\ax} \wh{Z}\, \wh{R}\, \wh{\ast}\, 1-128 \wh{\ax}\, \wh{Z}\, \wh{G}_4 \wedge \wh{\ast}\, \wh{G}_4\, .\label{Mthframe2}
\ee
It is now easy to check, that that the equations of motion for an ansatz involving five-dimensional Minkowski spacetime and a compact complex dimension three internal space in this frame can be solved by a direct product 
\begin{align}
\la \upd \wh{s}^2 \ra &=\eta_{\mu \nu} \upd x^\mu \upd x^\nu+2 \big(g^\tbz_{i \bar{\jmath}}+\wh{\ax} g^\tbo_{i \bar{\jmath}} \big)\upd zî \upd \bar{z}^{\bar{\jmath}}\, ,\label{Mthmet2}\\
\la \wh{G}_4\ra&=0\, ,
\end{align} 
where $g^{\tbo}_{i \bar{\jmath}}$ is simply (\ref{metsolution}) in complex indices. The additional term in (\ref{Mthframe2}) induces terms that become backreaction effects in the frame in which the Weyl factor has to be included. Therefore, the expectation is that with the choice $\cM_{10}=\bbM_{1,3} \times Y_3$ the circle reduction of (\ref{Mthframe2}) is equivalent to the classical Type IIA action up to a possible field redefinition of the dilaton. This redefinition of the dilaton is expected, since the background solution of Type IIA includes a non-constant dilaton (\ref{dilatoncorr}). Concretely, we will consider an eleven-dimensional ansatz of the form
\begin{align}
\upd \wh{s}^2 &=\e^{-\frac{2}{3} \wt{\phi}}\wt{g}_{A B} \upd x^A\upd x^B+\e^{\frac{4}{3}\wt{\phi}}\upd y^2\, ,\\
\wh{G}_4&= \wt{H}_3 \wedge \upd y\, ,
\end{align} 
with the circle coordinate $y \sim y+1$ and the ten-dimensional metric $\wt{g}_{A B}$ given by (\ref{Mthmet2}) with three-dimensional Minkowski spacetime. We will furthermore keep terms linear in the dilaton, since these are not expected to occur at higher-derivative level in the string frame and should therefore be captured by a field redefinition. The reduced action (\ref{Mthframe2}) including linear dilaton terms and the correction $\Delta S_{\text{M}}$ then reads
\begin{align}
2 \kappa_{11}^2 S^{\scriptscriptstyle{(10)}}&=\int_{\cM_{10}}\e^{-2 \wt{\phi}}\Big( \wt{R}\, \wt{\ast}\, 1+4 \upd \wt{\phi} \wedge \wt{\ast}\, \upd \wt{\phi}-\frac{1}{2}\wt{H}_3 \wedge \wt{\ast}\, \wt{H}_3 \Big)\\
&-\int_{\cM_{10}}768 \wh{\ax}\, \wt{Z}\, \wt{R} \, \wt{\ast}\, 1+3072 \wh{\ax}\, \wt{Z}\, \wt{\Box} \wt{\phi} \, \wt{\ast}\, 1-128 \wh{\ax}\, \wt{Z}\, \wt{H}_3 \wedge \wt{\ast}\, \wt{H}_3+\Delta S_{\text{M}}\, .
\end{align}
We then notice that we can recover the classical Type IIA action if we perform a redefinition of the dilaton according to 
\begin{equation}
\phi \to \phi-384 \wh{\ax}\, \wt{Z}
\end{equation}
and if we identify $\Delta S_{\text{M}}$ with the expression (\ref{11dcorr}). Consequently, the additional part of the eleven-dimensional action (\ref{11dcorr}) captures a part of the effect which the corrected dilaton has on the low energy effective action. We stress that this analysis relies on the background geometry stated in this section explicitly. It is therefore expected, that the general analysis requires more complicated structures. However, the analysis of \cite{Lu:2003ze} for the case, where the internal space is a $G_2$ manifold, shows the same properties with the quantity $\wh{Z}$ playing a major role. From now on we will consider the action (\ref{11dcorr}) as a part of the full eleven-dimensional action (\ref{Mthaction}).

\section{M-theory on Calabi-Yau threefolds}\label{Mthreduction}
We study the dimensional reduction of M-theory on the background solution found in section \ref{Mthsolution} preserving $\cN=2$ supersymmetry in five dimensions. We perturb the background solution and derive the two-derivative effective action as well as four-derivative operators quadratic in the lower dimensional fields.
\subsection{$\cN=2$ supergravity in five dimensions}\label{5dsugra}

For later reference the basic ingredients of five-dimensional $\cN=2$ ungauged supergravity are collected. In the dimensional reduction of M-theory we focus entirely on the massless sector. For this reason, the relevant massless multiplets   are given in Table \ref{5dMultiplets}. 
\begin{table}[h]
\begin{center}
{
\renewcommand{\arraystretch}{1.6}
\begin{tabular}[h]{|c||c||c|}
\hline
{\bf{multiplet}} & {\bf{bosonic field content}}& {\bf{\# of multiplets}}\\\hline
gravity multiplet & metric $g_{\mu \nu}$, graviphoton $A^0_{\mu}$ & 1\\\hline
tensor multiplet & tensor $B_{\mu \nu}$, real scalar $\phi$ & $n_T$ \\\hline
vector multiplet & vector $A^a_{\mu}$, real scalar $\Phi^a$ & $n_V$\\\hline
hypermultiplet & four real scalars $q^{u=1,\dotsb, 4}$& $n_H$\\\hline
\end{tabular}
}
\end{center}
\label{5dMultiplets}
\caption{multiplets of five-dimensional $\cN=2$ supergravity and their field content}
\end{table}
Note that the tensor multiplet can be dualized into a vector multiplet since in five dimensions a two form $B$ is dual to a vector. Let us now turn to the geometry of the scalar field space in the various multiplets. Since the main focus will lie on the vector- and gravity multiplet, we will only briefly discuss the hypermultiplet sector. The scalar field space $\cM_{\text{scalar}}$ is locally given as the direct product \cite{Ceresole:2000jd}
\begin{equation}
\cM_{\text{scalar}}=\cM_{\text{real sp.}} \times \cM_{\text{quat. K\"ahler}},
\end{equation} 
where $\M_{\text{quat. K\"ahler}}$ is a quaternionic K\"ahler manifold parametrized by the hypermultiplet scalars. The vector multiplet scalar geometry is encoded in a real very special manifold $\cM_{\text{real sp.}}$ with metric encoded by a cubic potential. This sector is highly restrictive and allows for precise tests of higher-derivative couplings based on supersymmetry. The vector multiplet scalar geometry is describe by the $(n_V+1)$ very special coordinates $L^a$, where $a=0,...,n_V$ exceeds the number counting the actual vector multiplet scalars by one. However in the end, the scalars $L^\lam$ parametrize only $n_V$ degrees of freedom. This can be understood in a geometric way as follows. The scalar sector of the vector multiplet can be interpreted as a $n_V$ dimensional submanifold embedded in an ambient $(n_V+1)-$dimensional manifold with coordinates $L^\lam$. The hypersurface spanned by the vector multiplet scalars is defined by a cubic polynomial, which in general takes the form
\begin{equation}
\cN(L)=\frac{1}{3!} C_{\lam \sig c}L^\lam L^\sig L^c,
\end{equation}
where $C_{\lam \sig c}$ is a constant and symmetric tensor. The hypersurface constraint that has to be satisfied by the very special coordinates $L^\lam$ is then simply given by
\begin{equation}\label{constraint}
\cN(L)=\frac{1}{3!} C_{\lam \sig c}L^\lam L^\sig L^c=1\,.
\end{equation}
The canonical $\cN=2$ supergravity action in the bosonic sector can then be written as \cite{Bergshoeff:2004kh,Bonetti:2011mw}
\begin{align}\label{can5daction}
S^{\tbfive}_{\text{can.}}=\int_{\cM_{5}} &\frac{1}{2} R \star 1- \frac{1}{2} G_{\lam \sig}\, \upd L^\lam \wedge \star\, \upd L^\sig- h_{u v}\,\upd q^u \wedge \star\, \upd q^v\\
&-\frac{1}{2} G_{\lam \sig} F^\lam \wedge \star \, F^\sig -\frac{1}{6}C_{\lam \sig c} A^\lam \wedge F^\sig \wedge F^{c}.\nonumber
\end{align}
The notation indicates that the vector in the gravity multiplet $A^0$ is included in a collective notation such that the index $\lam=0 ,..., n_V$. The hypermultiplet metric $h_{u v}$ does not play a role in the following and will therefore not be further discussed. 
The restrictive nature of $\cN=2$ supergravity in five dimensions  follows from the constraint (\ref{constraint})
and the fact that the metric for the vector multiplet scalars is determined from the cubic  polynomial $\cN$ as
\begin{equation}\label{vectormetric}
G_{\lam \sig}=\left. -\frac{1}{2} \pd_{L^\lam} \pd_{L^\sig} \log \cN \right|_{\cN=1}=\left.-\frac{1}{2}\cN_{\lam \sig}+\frac{1}{2} \cN_{\lam} \cN_{\sig}\right|_{\cN=1},
\end{equation}  
where the notation $\cN_{\lam}\equiv \pd_{L^\lam} \cN$ was introduced. Thus, the geometry of the vector multiplet space is fully determined by the cubic potential $\cN$.

\subsection{The two-derivative effective action}
The first step in our analysis will be the derivation of the two-derivative effective action for the gravity- and vector multiplet fields as well as a hypermultiplet scalar, which in the classical case is the volume modulus of $Y_3$. 
To perform the dimensional reduction of M-theory we perturb the background solution found in section \ref{Mthsolution}. 
A crucial observation is the fact that the correction to the Calabi-Yau metric $g^{\tbo}_{i \bar{\jmath}}$ drops out of the final expression, since it can be written as twice the derivative of a scalar function and therefore only contributes as a total derivative. So effectively the dimensional reduction of (\ref{Mthaction}) is performed on the metric background
{\setlength{\jot}{7pt}
\begin{align}
\mathrm{d}\wh{s}^2&=\text{e}^{\wh{\alpha} \Phi^{\tbo}} \Big( g\indices{_{\mu \nu}} \mathrm{d}x^{\mu}\mathrm{d}x^{\nu}+\big(g^{\scriptscriptstyle (0)}_{i \bar{\jmath}}-i \delta v^{a}\omega\indices{_{a\, i \bar{\jmath}}}\big)\mathrm{d}z^{i}\mathrm{d}\bar{z}^{\bar{\jmath}}\Big)\, ,\label{FullPertBackg}\\
\Phi^{\tbo}&=-\frac{512}{3}(2\pi)^3\, \ast_{\scriptscriptstyle 6} c_3=\langle \Phi^{\tbo} \rangle+\langle \partial_{a} \Phi^{\scriptscriptstyle  (2)} \rangle \delta v^{a}+\frac{1}{2} \langle \partial_{a} \partial_{b} \Phi^{\tbo} \rangle \delta v^{a} \delta v^{b}+\mathcal{O}(\delta v^3)\, ,\nonumber
\end{align}
}where the deformations of the \Kahler class of $Y_3$ parametrized by $\delta v^a$ are expanded in a real basis $\omega_a \in H^{1,1}_\pdb(Y_3),\, a=1, ...,h^{1,1}$ as
\be
\dx g_{i \bar{\jmath}}=-i \dx v^a \ox\inds{_{a i \bar{\jmath}}}\, ,\label{fluctuations}
\ee
and we have introduced the notation $\pd_a\equiv \pd_{\delta v^a}$. In the M-theory three-form zero-mode expansion we only keep terms contributing to the vector and gravity multiplet in five dimensions. More precisely, we only take into account modes giving rise to vectors $A^a$ in five dimensions. The expansion is thus
\beq
\wh{C}_3=A^a \wedge \omega_a\label{C3expansion}\, , \qquad 
\wh{G}_4= \upd \wh{C}_3=F^a \wedge \omega_a\, ,
\eeq
i.e.~along the $H^{1,1}_\pdb(Y_3)$ cohomology. In principle, the massless modes in the effective theory do not have to coincide with the ones from the classical reduction. The reason for this is the fact that the linearized equations of motion, which are solved by the massless modes, can receive non-trivial corrections. Along the lines of \cite{Grimm:2014efa}, using the fact that the massless deformations of the corrected background in the compactification ansatz should preserve the \Kahler condition as well as the Bianchi identity for the four-form field strength in the absence of M5-branes, it is possible to show on general grounds that the possible corrections to the massless fields at most contribute as total derivatives to the effective action and therefore decouple. Thus, we will ignore these corrections in the following and treat the perturbations as the ones of the classical M-theory reduction on a Calabi-Yau threefold. We will proceed by recording the results of the contributions of the classical and the eight-derivative action to the kinetic terms separately. Finally we will consider all contributions at quadratic order without any five-dimensional derivative, i.e.~terms contributing to a scalar potential.

\noindent{\bf Classical action.}

\noindent First let us perform the dimensional reduction of the classical Einstein-Hilbert term on the perturbed and $\ax'$-corrected background (\ref{FullPertBackg}). Focusing on terms carrying two derivatives in five dimensions, we obtain up to second order in the fluctuations the expression
\begin{align}
\int_{\cM_{11}} \wh{R}\, \wh{\ast}\, 1 \Big|_{\text{kin.}}&=\int_{\cM_{5}}\big(\Vm -768 \, (2 \pi)^3\, \wh{\ax}\, \chi(Y_3) \big)\, R\, \star 1\label{vkinclass}\\
&+\int_{\cM_{5}}\mathrm{d}\delta v^{a} \wedge \star\, \mathrm{d} \delta v^{b}\int_{Y_3}\Big( \frac{1}{2}\omega\indices{_{a\, i \bar{\jmath}}}\, \omega\indices{_{b}^{\bar{\jmath} i}}-\omega\indices{_{a\, i}^{i}}\, \omega\indices{_{b\, j}^{j}} \Big)\ast^{\tbz}_6 1\nonumber\\
&- \int_{\cM_5}768\, \wh{\alpha}\, \Big(\frac{1}{2}\Rcal_{a b}+\Tcal_{a b} \Big)\, \mathrm{d}\delta v^{a} \wedge \star\, \mathrm{d} \delta v^{b}\, ,\nonumber
\end{align}
where we made use of the shorthand notation
\begin{align}
\Vm &=\int_{Y_3} \Big[1-i \delta v^\lam \om\inds{_{\lam\, i}^i}+\frac{1}{2}\Big( \om\inds{_{\lam\, i \bar{\jmath}}}\, \om\inds{_{\sig}^{\bar{\jmath} i}}-  \om\inds{_{\lam\, i}^i}  \om\inds{_{\sig\, j}^j}\Big)\delta v^\lam \delta v^\sig \Big]\ast_6^\tbz 1, \label{omega0}\\
\mathscr{R}_{\lam \sig}&=(2\pi)^3\int_{Y_3} \om\inds{_{\lam\, i \bar{\jmath}}}\, \om\inds{_{\sig}^{\bar{\jmath} i}}\, c_3^\tbz,\label{CalR}\\
\mathscr{T}_{\lam \sig}&=(2 \pi)^3\int_{Y_3}\om\inds{_{\lam\, i}^i}  \om\inds{_{\sig\, j}^j}\, c_3^\tbz\, ,\label{CalT}
\end{align}
which we will use extensively throughout this work. From the classical action we furthermore pick up a correction to the kinetic terms of the vectors and a Chern-Simons term in five dimensions
\begin{align}
\int_{\cM_{11}}-\frac{1}{2} \wh{G} \wedge \wh{\ast}\, \wh{G}_4-\frac{1}{6} \wh{C}_3 \wedge \wh{G}_4 \wedge \wh{G}_4\Big|_{\text{kin.+C.S.}}&=\int_{\cM_{5}}\frac{1}{2}F^a \wedge \star\, F^b \int_{Y_3} \om\inds{_{a\, i \bar{\jmath}}}\, \om\inds{_{b}^{\jb i}} \ast^\tbz_6 1\nonumber\\
&-\int_{\cM_{5}}128\, \wh{\ax}\, \Rcal_{a b}\, F^a \wedge \star \, F^b\nonumber\\
&+\int_{\M_5}\frac{1}{6}\cK_{a b c}\, A^a \wedge F^b \wedge F^c\, ,\label{Fkinclass1}
\end{align} 
where we introduced the triple intersection numbers $\cK_{a b c}=\int_{Y_3} \om_a \wedge \om_b \wedge \om_c$ on $Y_3$, which appear as the coefficients of the Chern-Simons term. We furthermore record the reduction of (\ref{11dcorr}) yielding
\begin{equation}\label{Fkincorr}
\Delta S_{\text{M}}\big|_{\text{kin}}=-\int_{\cM_{5}}256\, \wh{\ax}\, \Rcal_{ab}\, F^a \wedge \star\, F^b\, .
\end{equation}

\noindent{\bf Eight-derivative action.}

\noindent We obtain further contributions to the kinetic terms of the five-dimensional theory by reducing the eight-derivative terms in the action (\ref{Mthaction}) on the lowest order Calabi-Yau background. The $\cR^4$-terms  (\ref{MthR4}) lead to a correction to the kinetic term of the \Kahler class deformations and a correction to the Ricci scalar
\begin{equation}\label{vkin2}
2 \kappa_{11}^2 S_{\wh{R}^4}\big|_{\text{kin.}}=\int_{\cM_5} 768\, (2 \pi)^3\, \chi(Y_3)\, R \star \, 1+384\, \Rcal_{ab}\, \upd \delta v^a \wedge \star \, \upd \delta v^b\, ,
\end{equation}
and from (\ref{GHCTerms1}) and (\ref{GHCTerms2}) we obtain the corrections to the kinetic terms of the vectors
\begin{equation}\label{Fkin2}
2\kappa_{11}^2 \big(S_{\wh{G}^2 \wh{R}^3}+S_{(\wh{\nabla}\wh{G})^2 R^2} \big)\big|_{\text{kin.}}=\int_{\cM_{5}}384 \, \Rcal_{ab}\, F^a \wedge \star \ F^b.
\end{equation}
Note that in order to obtain the result (\ref{Fkin2}) we had to fix $a_1=a_2$ in $\wh{s}_{18}$ . This is necessary to arrive at an expression, which is solely built of internal space Riemann tensors and harmonic (1,1)-forms without explicit derivatives after applying internal space total derivative identities. Then the final result can be shown to be independent of the unfixed coefficients $a_n$ by applying Schouten identities. 
Before putting the results obtained in this section together to obtain the five-dimensional two-derivative effective action we briefly comment on the scalar potential.

\noindent{\bf Scalar potential.}

\noindent Let us consider the higher curvature terms proportional to $\wh{R}^{4}$ once again. The dimensional reduction of these terms, focusing on the zero external derivative contributions, gives
\begin{align}
 2\kappa_{11}^2\, S_{\wh{R}^4} \big|_{\text{sc.p.}}&=-768 (2\pi)^3\, \int_{\cM_5} \delta v^\lam \delta v^\sig \star 1 \int_{Y_3} \nabla_k \nabla^k (\ast^\tbz_6 c_3^\tbz)\, \om\inds{_{\lam \, i \bar{\jmath}}}\, \om\inds{_{\sig}^{\bar{\jmath} i}} \ast_6^\tbz 1\, . \label{ScPot1}
\end{align}
This indeed looks like a mass term arising for the fluctuations $\delta v^\lam$. Another potential source for a scalar potential 
is the classical Einstein-Hilbert action. In fact, we will see that it induces a mass term if one performs the dimensional reduction 
on the $\wh{\ax}$-corrected background solution (\ref{FullPertBackg}). Indeed, the reduced Einstein-Hilbert term reads
\begin{equation}
\int_{\cM_{11}}\wh{R} \, \wh{\ast} \, 1 \Big|_{\text{sc.p.}}=768 (2\pi)^3\, \hat \alpha^2 \int_{\cM_5} \delta v^\lam \delta v^\sig \star 1 \int_{Y_3} \nabla_k \nabla^k (\ast^\tbz_6 c_3^\tbz)\, \om\inds{_{\lam \, i \bar{\jmath}}}\, \om\inds{_{\sig}^{\bar{\jmath} i}} \ast_6^\tbz 1\, , \label{ScPot2}
\end{equation}
which is exactly the contribution needed to cancel the one coming from the higher curvature terms (\ref{ScPot1}). From the reduction results (\ref{ScPot1}) and (\ref{ScPot2}) it is also possible to see that their contribution entirely stems from the non-harmonicity of the third Chern class given by $i \pd \pdb \xi$. Moreover, this cancellation shows that taking into account the backreaction and expanding the perturbations around a consistent background solution is crucial for the five-dimensional effective action. Note that this cancellation is already expected from a previous analysis on a fourfold in \cite{Grimm:2015mua}.

We are now in a position to collect the various pieces (\ref{vkinclass}), (\ref{Fkinclass1}), (\ref{Fkincorr}), (\ref{vkin2}), (\ref{Fkin2}) and merge them into the five-dimensional two-derivative effective action
\begin{align}
2\kappa_{11}^2 S^\tbfive&=\int_{\mathcal{M}_{5}} \Vm\, R\, \star 1+\int_{\mathcal{M}_{5}} \upd \dx v^\lam \wedge \star\, \upd \dx v^\sig \int_{Y_3} \Big( \frac{1}{2} \om\indices{_{\lam i \bar{\jmath}}} \om\indices{_{\sig}^{\bar{\jmath} i}}-\om\indices{_{\lam i}^i} \om\indices{_{\sig j}^j} \Big) \ast^\tbz_6 1\label{5d2derbefore1}\\
&- \int_{\mathcal{M}_5}768\, \wh{\ax}\, \Tcal_{\lam \sig}\, \upd \dx v^\lam \wedge \star\, \upd \dx v^\sig-\frac{1}{2} \int_{\mathcal{M}_5} F^\lam \wedge \star F^\sig \int_{Y_3}  \om_\lam \wedge \star\, \om_\sig\nonumber\\
&-\int_{\cM_5}\frac{1}{6} \cK_{\lam \sig c}\, A^\lam \wedge F^\sig \wedge F^c\,.\nonumber
\end{align}
Note that all terms involving the coupling $\Rcal_{ab}$ in (\ref{5d2derbefore1}) canceled. These cancellations are in fact crucial for compatibility with $\cN=2$ supergravity in five dimensions. This can be seen by noting that the coupling $\Rcal_{ab}$ is not proportional to the 
Euler characteristic $\chi(Y_3)$, since the non-harmonic part of the third Chern class prevents us from performing an integral split. The surviving coupling $\Tcal_{ab}$, however, satisfies
\begin{equation}\label{Tsplitting}
\Tcal_{ab}=-(2\pi)^3\,\frac{\chi(Y_3)}{\cV^\tbz}\, \cK^\tbz_a\, \cK^\tbz_{b}\, ,
\end{equation}  
since the traces of the harmonic (1,1)-forms are constant. The quantity $\cV^\tbz$ in (\ref{Tsplitting}) denotes the volume of the zeroth order Calabi-Yau manifold and $\cK_a^\tbz$, $\cK^\tbz_{a b}$ are contractions of the intersection numbers with the \Kahler moduli evaluated in the background, whose precise form can be found in appendix \ref{defs}. In the following we will denote quantities evaluated in the background with a zero superscript. We now perform a Weyl rescaling of (\ref{5d2derbefore1}) according to $g_{\mu \nu}\to \Vm^{-2/3}g_{\mu \nu}$ and the uplift from infinitesimal \Kahler class deformations to finite fields $v^a$ leading to the action in Einstein frame
{\setlength{\jot}{7pt}\begin{align}\label{5dEinst}
2\kappa_{11}^2 S^\tbfive= \int_{\cM_5} \Big[ R \star 1&+\frac{1}{2\cV}\big( \cK_{\lam \sig} -\frac{5}{3\cV} \cK_\lam \cK_\sig \big)\upd v^\lam \wedge \star\, \upd v^\sig\\
&+\frac{1}{2{\cV}^{\frac{1}{3}}} \big(\cK_{\lam \sig}-\frac{1}{2\cV} \cK_{\lam}\cK_{\sig} \big)F^\lam \wedge \star\, F^\sig -\frac{1}{6}\cK_{\lam \sig c}A^\lam \wedge F^\sig \wedge F^c \nonumber\\
&+768 (2\pi)^3\, \wh{\ax}\, \frac{\chi(Y_3)}{{\cV}^3} \cK_\lam \cK_\sig \upd  v^\lam \wedge \star\, \upd v^\sig\Big]\, .\nonumber
\end{align}}In (\ref{5dEinst}) we used the definition of the Calabi-Yau  volume $\V=\frac{1}{3!}\K_{abc}v^a v^b v^c$. 
Note that $\Vm$ is the volume $\cV$ expanded to second order in the fluctuations. 

We will now make contact with $\cN=2$ supergravity outlined in section \ref{5dsugra}.  It is already clear from (\ref{can5daction}) that we make the identification $C_{abc}=\cK_{abc}$, since there is no correction to the Chern-Simons term in five dimensions. This leads us 
to deduce the cubic constraint
\begin{equation}\label{Mthconstr}
\cN(L)=\frac{1}{3!}\cK_{abc}L^a L^b L^c = 1 \, ,
\end{equation}
which the physical scalars in the vector multiplet $L^a$ have to obey. This constraint is trivially solved by $L^a=\cV^{-\frac{1}{3}} v^a$,  which is identical to the definition in the uncorrected situation. Due to the relation (\ref{vectormetric}) this data is enough to 
completely fix the geometry on the vector multiplets and shows that there are no $\hat \alpha$-corrections present in this sector. The correction $\sim \chi(Y_3)$ in (\ref{5dEinst}) must therefore reside in the hypermultiplet sector. Using the explicit form of the physical scalars $L^a$ one can show that (\ref{5dEinst}) is equivalent to 
\begin{align}
\kappa_{11}^2 S^\tbfive&=\int_{\cM_{5}}\frac{1}{2}R \star 1-\frac{1}{2} G_{ab}(L)\, \upd L^a \wedge \star \, \upd L^b-\frac{1}{2} G_{ab}(L)\, F^a \wedge \star \, F^b- \frac{1}{4}\upd \log \cV\wedge \star \, \upd \log \cV\nonumber\\
&+\int_{\cM_{5}}384\, \wh{\ax}\,\frac{\chi(Y_3)}{\cV}\,\upd \log \cV \wedge \star \, \upd \log \cV-\frac{1}{6}\cK_{abc} A^a \wedge F^b \wedge F^c\, ,
\end{align}
where we used the metric $G_{ab}$ derived from the cubic potential (\ref{Mthconstr}) given by
\begin{equation}
G_{ab}=-\frac{1}{2} \pd_a \pd_b \log \cN(L) \Big|_{\cN=1}=-\frac{1}{2}\cK_{abc} L^c+\frac{1}{8}\cK_{acd}\cK_{bef}L^c L^d L^e L^f\, .
\end{equation}
Classically one identifies one of the hypermultiplet scalars $D$ with $D=-\frac{1}{2}\log \cV$ \cite{Bonetti:2011mw}, whereas when taking quantum corrections into account we find the corrected hypermultiplet scalar
\begin{equation}\label{VolRenomaliz}
D=-\frac{1}{2} \log \big( \V+768 \, \wh{\ax}\, \chi(Y_3) \big)\,,
\end{equation}
such that the final action is 
\begin{align}
\kappa_{11}^2 S^\tbfive&=\int_{\cM_{5}}\frac{1}{2}R \star 1-\frac{1}{2} G_{ab}(L)\, \upd L^a \wedge \star \, \upd L^b-\frac{1}{2} \mathscr{G}_{ab}(L)\, F^a \wedge \star \, F^b- \upd D \wedge \star \, \upd D\nonumber\\
&-\int_{\cM_{5}}\frac{1}{6}\cK_{abc} A^a \wedge F^b \wedge F^c\,.
\end{align}
We have thus shown that our dimensional reduction of M-theory at two-derivative level is compatible with $\cN=2$ supergravity. The metric of the vector multiplets coincides with the one of the classical reduction, such that the only net effect at two derivatives is a corrected field identification of one hypermultiplet scalar. This can in turn be interpreted as a renormalization  of the volume of $Y_3$ at order $\wh{\ax}$, see (\ref{VolRenomaliz}).

\subsection{Four-derivative terms of the \Kahler moduli}\label{Kahler4derivatives}\label{MthKahlerfour}

We now aim to include four-derivative terms in the effective action including at most two fluctuations $\dx v^a$. Obviously this truncation misses 
e.g.~the four-derivative interaction of the form $(\pd v^a)^4$,  since this would require an analysis to orders of at least $\dx v^4$, which is however technically very involved. In this section we therefore consider the background perturbations (\ref{fluctuations}) up to second order. We now introduce the coupling tensor $Z_{i \bar{\jmath} k \bar{l}}$ encoding many of the four-derivative terms for the \Kahler deformations and vectors in five dimensions. It is a non-topological, co-closed (2,2)-form $Z$ whose components are given by
\begin{equation}\label{ZetaTens}
Z_{i \bar{\jmath} k \bar{l}}=\varepsilon_{i \bar{\jmath} i_1 \bar{\jmath}_1 i_2 \bar{j}_2} \,  \varepsilon_{k \bar{l} k_1 \bar{l}_1 k_2 \bar{l}_2} \, {R}\indices{^{\bar{\jmath}_1 i_1 \bar{l}_1 k_1}}\, {R}\indices{^{\bar{\jmath}_2 i_2 \bar{l}_2 k_2}}
\end{equation}
satisfying the relations
\begin{align}
{Z}\inds{_{i \bar{\jmath} k}^k}&=-2i (2\pi)^2\, \,{(\ast_6\, c_2)}\inds{_{i \bar{\jmath}}}\,,\label{zetaid1}\\
Z\inds{_{l}^{\, l}_{\,k}^{\, k}}&= 2(2\pi)^2\, \ast_6\, \left(c_2 \wedge  J
\right)\,,\label{zetaid2}\\
Z\inds{_{i \bar{\jmath} k}^k}\om\inds{_{\lam}^{\bar{\jmath} i}}&=2i (2\pi)^2\, \ast_6 \left(c_2 \wedge \om_\lam \right)\label{zetaid3}.
\end{align}
This object was already recognized to play a role in the context of $\cN=2$ four-derivative couplings arising from string compactifications in \cite{Katmadas:2013mma,Weissenbacher:2016gey}. We determine the four-derivative couplings from the $\wh{R}^4$ terms by 
straightforward reduction to be 
{\setlength{\jot}{7pt}
\begin{align}
2 \kappa_{11}^2 S_{\wh{R}^4}\big|_{\text{four der.}}&= \int_{\mathcal{M}_5}192\, (2\pi)^2 \, \, \big[ {R}^2 \star 1  -4 {R}\indices{_{\mu \nu}}  {R}\indices{^{\mu \nu}} \star 1-16\, \text{Tr }\mathcal{R} \wedge \star \, \mathcal{R} \big]  \int_{Y_3} c_2^{\tbz} \wedge J\nonumber \\
& - \int_{\mathcal{M}_5}96\, \cZ^\tbz_{\lam \sig}\, \big[  R\, \mathrm{d}\del v^{a} \wedge \star\, \mathrm{d}\del v^{b} - 4 {R}\indices{^{\mu \nu}}\partial_{\mu} \del v^{a} \partial_{\nu} \del v^{b} \star 1 \nonumber\\
&\qquad \qquad\qquad\qquad\qquad\qquad\qquad\qquad + 2 (\Box \del v^{a})\, (\Box\del v^{b})\star 1 \big],\label{Kahler4derivatives}
\end{align}
}where we performed external spacetime integrations by parts and defined $\Box \equiv \nab_\mu \nab^\mu$. Additionally, we introduced the five-dimensional curvature two-form ${\mathcal{R}}\indices{^\mu_\nu}$ satisfying\\ ${R}\indices{_{\mu \nu \rho \sigma}} {R}\indices{^{\mu \nu \rho \sigma}} \star 1 = -8\, \text{Tr } \mathcal{R} \wedge \star\, \mathcal{R}$ and the shorthand notation
\begin{align}\label{zetaobj}
\cZ_{\lam \sig}&=\int_{Y_3} Z_{i \bar{\jmath} k \bar{l}}\; \omega\indices{_{a}^{\bar j i}}\omega\indices{_{b}^{\bar l k}} \ast_{\scriptscriptstyle 6} 1\,,\\
\cZ_{a}&=(2 \pi)^2 \int_{Y_3}c_2 \wedge \om_a\, ,\nonumber\\
\cZ&=(2\pi)^2 \int_{Y_3}c_2 \wedge J\, ,\nonumber
\end{align} 
which we will  use frequently in the following.\footnote{Note that we always denote evaluation on the background with a zero superscript.} Note that the \Kahler form in (\ref{Kahler4derivatives}) coupling to the second Chern class of $Y_3$ is with respect to the fluctuated metric $J=J^\tbz+\dx v^a \ox_a$ and the second order fluctuations of the terms built of two Riemann tensors in five dimensions cancel in a non-trivial way.

In the following we will discuss the Riemann squared terms in (\ref{Kahler4derivatives}) when moving to the five-dimensional Einstein frame. Note that the second line in (\ref{Kahler4derivatives}) is already second order in the fluctuations $\dx v^a$ such that the Weyl rescaling simply leads to an overall factor proportional to the zeroth order Calabi-Yau volume $\cV^\tbz$. The terms involving two five-dimensional Riemann tensors, however, come with at most linear terms in $\dx v^a$ and get therefore a less trivial modification from the Weyl rescaling considering terms up to order 
$(\dx v)^2$. The explicit form of the action after the Weyl rescaling in terms of the fluctuations is very involved and we will not display it here. Since we are in the end interested in the corresponding action of the finite fields we will give an action, which precisely reproduces the Weyl rescaled action when expanding it in infinitesimal fluctuations. We therefore introduce the scalar
\begin{equation}
u=\log \cV~~~~\text{with}~~~~\cV=\frac{1}{3!}\cK_{abc}v^a v^bv^c\,.
\end{equation}
The Riemann squared action in Einstein frame then reads
\begin{align}
2 \kappa_{11}^2 S_{R^2}^{\tiny{\text{Einst.}}}&=\int_{\cM_5}192\, \ah\, \Cbar \e^{-u/3}\Big[R^2 \star 1-4\, R\inds{_{\mu \nu}}R\inds{^{\mu \nu}}\star 1-16\, \text{Tr } \cR \wedge \star \, \cR \Big]\\
&+\int_{\cM_{5}}\frac{1}{3} \ah\, \Cbar\,\e^{-u/3}\Big[ 1536\, R\, (\Box u) \star 1-1536\, R\inds{^{\mu \nu}}\, \nabla_{\mu} \nabla_{\nu}u \star 1-512\, R\, \upd u \wedge \star \, \upd u\nonumber\\
&~~~~~~~~~~~~~~~~~~~~~~~~+256\, R^{\mu \nu}\, \pd_{\mu}u\, \pd_\nu u\star 1+1024\, (\Box u)^2 \star 1 \Big]\, ,\nonumber
\end{align}
where we now have the moduli dependent coupling
\be
\Cbar\equiv \Cbar(v^a)=(2\pi)^2\int_{Y_3}c_2 \wedge J\, ,~~~~\text{with}~~~~J=v^a \ox_a\, .
\ee
The uplift of the terms in the second line in (\ref{Kahler4derivatives}) is however not obvious, due to the non-topological nature of the coupling $\cZ^\tbz_{ab}$. The naive guess would be to to promote $\cZ^\tbz_{ab} \to \cZ_{ab}\equiv\cZ_{ab}(v^a)$ to its moduli dependent counterpart and to lift $\dx v^a \to v^a$. A higher order analysis in the fluctuations $\dx v^a$ might provide further evidence for this claim.

\subsection{Four-derivative terms of the five-dimensional vectors}\label{vector4derivatives}

Before we continue with the results of the four-derivative terms involving five-dimensional vectors, or rather two powers of their field strength to be more precise, let is note that it is well known that there is also a five-dimensional gauge-gravitational Chern-Simons term present \cite{Antoniadis:1997eg, Ferrara:1996hh, Bonetti:2011mw}. This contribution to the five-dimensional effective action arises upon dimensional reduction of (\ref{MthCX8}) and is in our conventions
\be
S_{\wh{C} \wh{X}_8}=768 \,\int_{\cM_5}\cZ^\tbz_a \,A^a \wedge \text{Tr }\cR \wedge \cR .
\ee
It was worked out in \cite{Hanaki:2006pj} that the coefficients of this gauge-gravitational Chern-Simons term and the coefficient of the $R_{\mu \nu \rho \sigma}R^{\mu \nu \rho \sigma}$ term in (\ref{Kahler4derivatives}) are related by supersymmetry and matches our computation. This known fact serves as a crosscheck at this point.

We proceed by dimensionally reducing the action
\be\label{Gaction}
S_{\wh{G}_4}=S_{\wh{G}^2 \wh{R}^2}+S_{(\wh{\nabla} \wh{G})^2 \wh{R}^2}
\ee
keeping terms containing four derivatives in $\cM_5$ and up to two five-dimensional field strengths $F^a=\upd A^a$. We will keep the coefficients $a_n$ completely generic, however keeping in mind that the discussion of the two-derivative action forced us to impose $a_1=a_2$. 

 The part of the five-dimensional effective action containing the Ricci scalar and two field strengths of the vectors is
\begin{equation}\label{RSF2}
\left. 2 \kappa_{11}^2 S_{\wh{G}_4} \right|_{R\, F^\lam \wedge \star F^\sig}=-96 \int_{\cM_5} \cZ^\tbz_{\lam \sig} R \, F^\lam \wedge \star \,F^\sig \, ,
\end{equation}
and the terms involving the five-dimensional Ricci tensor reduce to
\begin{equation}\label{RtensF2}
\left. 2 \kappa_{11}^2 S_{\wh{G}_4} \right|_{{R}\inds{^{\mu \nu}}\,{F^\lam}{F^\sig}}=192 \int_{\cM_5} \cZ^\tbz_{\lam \sig}\, {R}\inds{^{\mu \nu}}\,{F^\lam}\inds{_{\rho \mu}}{F}\inds{^{b\, \rho}_{\nu}} \star 1\, .
\end{equation}
The treatment of the terms containing one five-dimensional Riemann tensor fully contracted on two field strength tensors $F^a$ is more complicated. In addition to internal space total derivative identities we made use of the first Bianchi identity for the five-dimensional Riemann tensor and introduced a new object $\cX^\tbz_a=\cX^\tbz_{a i \bar{\jmath}}\, \upd z^i \wedge \upd \bar{z}^{\bar{\jmath}}$ whose components are given by
\be \label{XTens}
\cX^\tbz_{a i \bar{\jmath}}=R^\tbz_{i \bar{\jmath} k \bar{l}}\, \,\om\inds{_{a}^{\bar{l} k}}\,.
\ee
It is interesting to note that with this new building block the reduction can be performed without having to fix any of the parameters $a_n$. We will therefore give the general result, however, keeping in mind that $a_1=a_2$. We then obtain the following additional coupling of the vectors to the five-dimensional Riemann tensor
\begin{align}
 2 \kappa_{11}^2 S_{\wh{G}_4} \big|_{{R}\inds{^{\mu \nu \rho \sigma}}{F^\lam}{F^\sig}}&=\int_{\M_5} {R}\inds{^{\mu \nu \rho \sigma}}F\inds{^{\lam}_{\mu \nu}}F\inds{^{\sig}_{\rho \sigma}} \star 1\label{RiemannFF}\\
&\times \int_{Y_3} \Big[f^\tba_1\,\X^\tbz_{\lam \, i \bar{\jmath}} \X_{\sig}^{\tbz \, \bar{\jmath} i}
+f^\tba_2\,Z_{i  \bar{\jmath}  k}^{\tbz\, k}\, \om\inds{_{\lam}^{\bar{\jmath} n}} \om\inds{_{\sig \, n}^{i}}+f^\tba_3\, Z^{\tbz \, k~l}_{~\;\;k ~l}\, \om\inds{_{\lam\, i \bar{\jmath}}}\,\om\inds{_{\sig}^{\bar{\jmath} i}}\nonumber\\
&~~~~~+f^\tba_4\, Z_{i  \bar{\jmath}  k}^{\tbz\, k}\, \om\inds{_{\lam}^{\bar{\jmath} i}} \om\inds{_{\sig\, l}^l}+f^\tba_5\, Z^{\tbz \, k~l}_{~\;\;k ~l}\, \om\inds{_{\lam\, i}^i}\, \om\inds{_{\sig\, j}^j}+f^\tba_6\, Z^\tbz_{i \bar{\jmath} k \bar{l}}\, \om\inds{_{\lam}^{\bar{\jmath} i}} \om\inds{_{\sig}^{\bar{l} k}}\Big] \ast^\tbz_6 1.\nonumber
\end{align}
The coefficients $f_i^\tba,\, i=1,...,6$ depend on the unfixed $a_n$ from the definition of the tensor $\wh{s}_{18}$ and are given by the linear relations
{\setlength{\jot}{7pt}
\begin{align}
f_1^\tba &=-96a_1+24a_2-36 a_3 -24 a_4+4 a_5 - 2a_6\, ,\\
f_2^\tba &=-192a_1+72 a_2-66 a_3-48a_4+6a_5-4a_6\, ,\nonumber\\
f_3^\tba &=48+96a_1-24 a_2+36 a_3+24 a_4-4 a_5+2 a_6\, ,\nonumber\\
f_4^\tba &= 96 a_1-24a_2+36a_3+24a_4-4a_5+2a_6\ ,\nonumber\\
f_5^\tba &= -48 a_1+12a_2-18a_3-12a_4+2a_5-a_6\, , \nonumber\\
f_6^\tba &= -48+48a_1-36a_2+12a_3+12a_4+a_6\, .\nonumber
\end{align}
}Observe that we have $f^\tba_1=-f^\tba_4=2 f^\tba_5$. The last contribution we are lacking is the structure with the schematic form $\nabla F^a \nabla F^b$. The reduction reveals, that there are two different structures present, which are however related to each other by exploiting the Bianchi identity of the vectors $A^a$ given by $\upd F^a=0$. In components this means $3 \nabla_{[\rho} F^a_{\mu \nu]}=0$ giving us the identity
\be
3 \nabla^\mu {F^\lam}\inds{^{\nu \rho}}\nabla_{[\mu} {F^\sig}\inds{_{\nu \rho]}}=0~~~~\Rightarrow~~~~ \nabla_\nu {F^\lam}\inds{_{\mu \rho}}\nabla^\rho {F^\sig}\inds{^{\mu \nu}}=\frac{1}{2}\nabla_\mu {F^\lam}\inds{_{\nu \rho}}\nabla^\mu {F^\sig}\inds{^{\nu \rho}}\, ,
\ee
which allows us to to eliminate one of the two structures. The resulting piece in the five-dimensional Lagrangian is then
\begin{align}
 2 \kappa_{11}^2 S_{G_4} \big|_{\nab F^\lam \nab F^\sig}&=\int_{\M_5} \nab_{\mu}F\inds{^{\lam}_{\rho \sigma}}\nab^{\mu}F\inds{^{\sig \,\rho \sigma}} \star 1 \label{CDFCDF}\\
&\times \int_{Y_3} \Big[f_1^\tbb\,\X^\tbz_{\lam \, i \bar{\jmath}}\, \X_{\sig}^{\tbz \, \bar{\jmath} i}+f_2^\tbb\,Z_{i  \bar{\jmath}  k}^{\tbz\, k}\, \om\inds{_{\lam}^{\bar{\jmath} l}} \om\inds{_{\sig \, l}^{i}}+f_3^\tbb\,Z^{\tbz \, k~l}_{~\;\;k ~l}\, \om\inds{_{\lam\, i \bar{\jmath}}}\,\om\inds{_{\sig}^{\bar{j} i}}\nonumber\\
&~~~~~+f_4^\tbb\,  Z_{i  \bar{\jmath}  k}^{\tbz\, k}\, \om\inds{_{\lam}^{\bar{\jmath} i}} \om\inds{_{\sig\, l}^l}+f_5^\tbb \,Z^{\tbz \, k~l}_{~\;\;k ~l}\, \om\inds{_{\lam\, i}^i}\, \om\inds{_{\sig\, j}^j}+f_6^\tbb\, Z^\tbz_{i \bar{\jmath} k \bar{l}}\, \om\inds{_{\lam}^{\bar{\jmath} i}} \om\inds{_{\sig}^{\bar{l} k}}\Big] \ast^\tbzero_6 1\, \nonumber
\end{align}
and its corresponding coefficients in terms of the $a_n$ are
{\setlength{\jot}{7pt}\begin{align}
f^\tbb_1 &=-48a_1-24a_3-12a_4+4a_5\, ,\\
f^\tbb_2 &=-72a_1+24a_2-24a_3-12a_4\, ,\nonumber\\
f^\tbb_3 &=24a_1+12a_3+6a_4\, ,\nonumber\\
f^\tbb_4 &=24a_1+12a_3\, ,\nonumber\\
f^\tbb_5 &=12a_1+6a_3\, ,\nonumber\\
f^\tbb_6 &=-12+24a_1-24a_2+6a_3+6a_4\, ,\nonumber
\end{align}
}where we again find a linear relation among the coefficients $f_4^\tbb=2 f_5^\tbb$. We can now put the results together and use the identities (\ref{zetaid1})-(\ref{zetaid3}) to obtain the four-derivative action of the vectors quadratic in $F^a$ given by 
{\setlength{\jot}{9pt}\begin{align}\label{vecHDres}
2 \kappa_{11}^2 S^\tbfive_{F}&=\int_{\M_5}\zeta_{\lam \sig}\,\big[192\, {R}\inds{^{\mu \nu}}\, {F^\lam}\inds{_{\rho \mu}}{F^\sig}\inds{^{\rho}_{\nu}} \star 1 -96\, R \, F^\lam \wedge \star F^\sig \big]\\
&+\int_{\M_5}\!\!\!\! {R}\inds{^{\mu \nu \rho \sigma}}F\inds{^{\lam}_{\mu \nu}}F\inds{^{\sig}_{\rho \sigma}} \star 1\int_{Y_3} (2\pi)^2 \Big[-\tilde{f}_1^\tba\,\X^\tbz_a \wedge \ast_6^\tbzero \X^\tbz_b
-2i f_2^\tba\,{c_2^\tbzero}\inds{_{i \bar{\jmath}}}\, \om\inds{_{\lam}^{\bar{\jmath} k}} \om\inds{_{\sig \, k}^{i}}\ast_6^\tbzero 1\nonumber\\
&~~~~~~~~~~~~~~~~~~~~~~~~~~~~~~~~~~~~~~~+2 f_3^\tba\,\om\inds{_{\sig\, i \bar{\jmath}}}\, \om\inds{_{\sig}^{i \bar{\jmath}}}\, c_2^\tbzero \wedge J^\tbzero +2i f_4^\tba \,  \om\inds{_{\sig\, l}^l}\, c_2^\tbzero \wedge \om_\lam \nonumber\\[0.15cm]
&~~~~~~~~~~~~~~~~~~~~~~~~~~~~~~~~~~~~~~~+2f_5^\tba\, \om\inds{_{\sig\, i}^i}\, \om\inds{_{\sig\, j}^j}\, c_2^\tbzero \wedge J^\tbzero+\tilde{f}_6^\tba\,  Z^\tbz_{i \bar{\jmath} k \bar{l}}\, \om\inds{_{\lam}^{\bar{\jmath} i}} \om\inds{_{\sig}^{\bar{l} k}} \ast_6^\tbzero 1\Big] \nonumber\\
&+\int_{\M_5}\!\!\!\! \nab_{\mu}F\inds{^{\lam}_{\rho \sigma}}\nab^{\mu}F\inds{^{\sig \,\rho \sigma}} \star 1\int_{Y_3}(2\pi)^2 \Big[-\tilde{f}_1^\tbb\,\X^\tbz_a \wedge \ast_6^\tbzero \X^\tbz_b
-2i f_2^\tbb \, {c_2^\tbzero}\inds{_{i \bar{\jmath}}}\, \om\inds{_{\lam}^{\bar{\jmath} k}} \om\inds{_{\sig \, k}^{i}}\ast_6^\tbzero 1\nonumber\\
&~~~~~~~~~~~~~~~~~~~~~~~~~~~~~~~~~~~~~~~+2f_3^\tbb\, \om\inds{_{\lam\, i \bar{\jmath}}}\,\om\inds{_{\sig}^{\bar{\jmath} i}}\, c_2^\tbzero \wedge J^\tbzero+2i f_4^\tbb\, \om\inds{_{\sig\, l}^l}\, c_2^\tbzero \wedge \om_\lam\nonumber\\[0.15cm]
&~~~~~~~~~~~~~~~~~~~~~~~~~~~~~~~~~~~~~~~+2 f_5^\tbb \, \om\inds{_{\sig\, i}^i}\, \om\inds{_{\sig\, j}^j}\, c_2^\tbzero \wedge J^\tbzero +\tilde{f}_6^\tbb\, Z^\tbz_{i \bar{\jmath} k \bar{l}}\, \om\inds{_{\lam}^{\bar{\jmath} i}} \om\inds{_{\sig}^{\bar{l} k}}\ast_6^\tbzero 1\Big].\nonumber
\end{align}
}We furthermore introduced $\tilde{f}_1^{\scriptscriptstyle{(\ax, \bx)}}=(2 \pi)^{-2}{f}_1^{\scriptscriptstyle{(\ax, \bx)}}$. A general four-derivative $\cN=2$ supergravity theory combined with taking into account the possibility to perform higher-derivative field redefinitions could in principle, if available, be used to constrain or even completely fix the coefficients $a_n$.

\section{Type IIA supergravity on Calabi-Yau threefolds}\label{IIAsection}

In this section we derive the four-dimensional two-derivative effective action of Type IIA supergravity including both tree-level and one-loop eight-derivative terms in ten dimensions. We stress that higher-derivative terms of the ten-dimensional dilaton and the RR-fields are not known and therefore not taken into account. We therefore focus on the $\cR^4$ terms in ten dimensions and and the conjectured $\wt{H}_3$ completion of the one-loop Type IIA terms from \cite{Liu:2013dna}. Since these conjectured terms are based on introducing a connection with torsion, upon which the superinvariants $\cJ_{0,1}$ are evaluated, we assume for this section that the tree-level terms are completed in the same fashion. We already saw in the context of the M-theory $Y_3$ reduction that the one-loop terms are compatible with $\cN=2$ supegravity in five dimensions. 
One consistency check of our computations will be to match the corrections to the prepotential from which the metric of the complexified \Kahler moduli $t^a=b^a+i v^a$ is derived. The expected prepotential takes the form
\be \label{pre-pot}
f(t)=f_{\text{class}}(t)-i \frac{ \ze(3)}{2 (2 \pi)^3}\, \chi(Y_3)=\frac{1}{3!}\, \cK_{abc}t^a t^b t^c-i \frac{ \ze(3)}{2 (2 \pi)^3}\, \chi(Y_3)\,+... \, ,
\ee
where the ellipses denote terms $\sim \, k_{ab}t^at^b+c_a t^a$, which do not contribute to the \Kahler potential, as well as non-perturbative  contributions stemming from worldsheet instantons. The form of the prepotential \eqref{pre-pot} is well known \cite{ Antoniadis:1997eg, Candelas:1990rm, Grisaru:1986dk}, we however show a first derivation which is solely based on a dimensional reduction of Type IIA supergravity.
\subsection{The two-derivative effective action }
The purpose of this section is to dimensionally reduce the ten-dimensional Type IIA supergravity action including the eight-derivative corrections introduced in section \ref{IIAderivatives} to four dimensions. We are considering the modified $Y_3$ solution (\ref{IIAansatz}) including deformations of the \Kahler class parametrized by the fluctuations $\dx v^a$ and in addition scalars $b^a$ from the zero-mode expansion of the NS-NS two-form
\be
\wt{B}_2=b^a \ox_a
\ee
in harmonic $(1,1)$-forms. Our focus will lie on the kinetic terms of these scalar modes, since they combine into the complexified \Kahler moduli $t^a=b^a+i v^a$. The geometry on the moduli space parametrized by $t^a$ is specified by a prepotential $f(t^a)$. We will split again the reduction into two separate parts: the reduction of the classical Type IIA action on the corrected background (\ref{IIAansatz}) and the reduction of the eight-derivative terms in ten dimensions on the lowest order Calabi-Yau background.  

\noindent{\bf Classical action.}

\noindent The classical action gives rise to the following contribution to the kinetic terms in four dimensions
\begin{align}\label{IIAclassred}
2 \kappa_{10}^2 S_{\text{IIA}}^{\text{class}}\big|_{\text{kin.}}&=\int_{\cM_4} \big[\Vm-1536\, (2\pi)^3\, \ax\, \chi(Y_3)\, (\lz+\lo) \big]\, R  \star 1\\
&+\int_{\M_4}\e^{-2 \phi_0}\, \upd \delta v^\lam \wedge \star \upd \delta v^\sig \int_{Y_3} \Big(\frac{1}{2} \om\inds{_{\lam\, i \bar{\jmath}}}\, \om\inds{_{\sig}^{\bar{\jmath} i}}-\om\inds{_{\lam\, i}^i}  \om\inds{_{\sig\, j}^j} \Big)\ast_6^\tbzero 1\nonumber\\
&- \int_{\M_4}768\,\alpha \,  (\lz+\lo)\,\big(\mathscr{T}_{\lam \sig}+\frac{1}{2}\Rcal_{ab}\big)\, \upd \delta v^\lam \wedge \star\, \upd \delta v^\sig \, \nonumber\\
&+\int_{\M_4}\frac{1}{2}  \e^{-2 \phi_0} \upd b^\lam \wedge \star\, \upd b^\sig \int_{Y_3} \om\inds{_{\lam\, i \bar{\jmath}}}\, \om\inds{_{\sig}^{\bar{\jmath} i}} \ast_6^\tbzero 1-384\, \alpha\, (\lz+\lo) \,\mathscr{R}_{\lam \sig}\, \upd b^\lam \wedge \star \, \upd b^\sig\, ,\nonumber
\end{align}
where we defined the constants
\be\label{Zcal}
\lz=\ze(3)\e^{-2 \phi_0}\, , ~~~~~~\lo=\frac{\pi^2}{3}\, ,
\ee
and made use of the definition (\ref{omega0}).

\noindent{\bf Eight-derivative action.}

\noindent The reduction of the eight-derivative terms in ten dimensions yields the action
\begin{align}\label{IIAhdred}
2 \kappa_{10}^2 \big( S^{\text{tree}}_{\wt{R}^4}+S^{\text{loop}}_{\wt{R}^4}+S_{\wt{H}^2} \big)\big|_{\text{kin.}} &=\int_{\cM_{4}}-768\, (2 \pi)^2\,(\lz-\lo)\, \chi(Y_3)\, R \star 1\\
&+\int_{\M_4}384\, (\lz+\lo)\, \Rcal_{ab}\, \upd b^a \wedge \star \, \upd b^b\nonumber\\
&+\int_{\cM_{4}}384\,(\lz+\lo)\, \Rcal_{ab}\, \upd \dx v^a \wedge \star \, \upd \dx v^b\nonumber\, ,
\end{align}
which again has the property that, combining it with the reduction from the classical action (\ref{IIAclassred}), all couplings involving $\Rcal_{ab}$ cancel. Summing up the two contributions (\ref{IIAclassred}) and (\ref{IIAhdred}) of the four-dimensional action leads to
\begin{align}\label{IIAeffstring}
2 \kappa_{10}^2\, S^{\tbf}&=\int_{\M_4} \Va R \star 1+\frac{1}{2}\e^{-2 \phi_0} \Big(\cK^\tbzero_{\lam \sig}-\frac{1}{{\V^\tbzero}}\cK^\tbzero_{\lam} \cK^\tbzero_{\sig} \Big)\upd b^\lam \wedge \star \upd b^\sig\\
&+\int_{\M_4}\frac{1}{2} \e^{-2 \phi_0} \Big(\cK^\tbzero_{\lam \sig} + \frac{1}{\V^\tbzero} \cK^\tbzero_{\lam} \cK^\tbzero_{\sig} \Big) \upd \del v^\lam \wedge \star \upd \del v^\sig\nonumber\\
&+\int_{\M_4}768\,(2\pi)^3\,(\lz+\lo)\,\frac{\chi(Y_3)}{{\V^\tbzero}^2}\,  \cK^\tbzero_{\lam} \cK^\tbzero_{\sig}\upd \del v^\lam \wedge \star \upd \del v^\sig\, .\nonumber
\end{align}
We furthermore introduced the notation for the prefactor of the four-dimensional Ricci scalar
\be
\Va=\Vm \e^{-2 \phi_0}-1536\, (2\pi)^3\, \ax\,\lz\, \chi(Y_3)\,\, , 
\ee
which can be removed by a Weyl rescaling of the metric $g_{\mu \nu} \to \Va^{-1}\, g_{\mu \nu}$. Performing this rescaling as well as the uplift from infinitesimal to finite fields $v^a$ leads to the effective action in Einstein frame
\begin{align}\label{IIAeffaction}
 S^{\tbf}&=S^\tbfour_{\scriptscriptstyle\ax'^0}+{\alpha'}^3 \,S^\tbfour_{\scriptscriptstyle \ax'^3}\, ,
\end{align}
where we have restored the explicit $\ax'$-dependence by using (\ref{IIAalpha}). We split the action (\ref{IIAeffaction})
 into a classical and quantum (tree-level and one-loop) corrected part which are explicitly given by
\begin{align}
2 \kappa_{10}^2 S^\tbfour_{\scriptscriptstyle\ax'^0}&=\int_{\cM_{4}} R \star 1 +\frac{1}{2 {\V}}\Big(\K_{\lam \sig}-\frac{1}{\V}\K_{\lam}\K_{\sig} \Big)\, \upd b^\lam \wedge \star\, \upd b^\sig\\
&+\int_{\M_4}\frac{1}{ {\V}}\Big(\frac{1}{2}\K_{\lam \sig}-\frac{1}{\V}\K_{\lam}\K_{\sig} \Big)\, \upd  v^\lam \wedge \star \, \upd v^\sig \, ,\nonumber
\end{align}
and 
\begin{align}
2 \kappa_{10}^2 S^\tbfour_{\scriptscriptstyle \ax'^3}&=\int_{\cM_4}\frac{\chi(Y_3)\, \zeta(3)\, \pi^3}{{\V}^2} \Big(\K_{\lam \sig}-\frac{1}{\V} \K_{\lam}\K_{\sig} \Big)\, \upd b^\lam \wedge \star\, \upd b^\sig\label{IIAeffactcorr}\\
&+\int_{\cM_4}\frac{\chi(Y_3)\, \zeta(3)\, \pi^3}{{\V}^2} \Big(\K_{\lam \sig}-\frac{4}{\V} \K_{\lam}\K_{\sig} \Big)\, \upd  v^\lam \wedge \star\, \upd  v^\sig\nonumber\\
&+\int_{\cM_{4}}\frac{\chi(Y_3)\, \pi^5}{3{\V}^3}\, \e^{2 \phi_0}\, \K_{\lam} \K_{\sig}\, \upd  v^\lam \wedge \star\, \upd  v^\sig\nonumber.
\end{align}
Let us now comment on the origin of the various corrections. It is evident in (\ref{IIAeffstring}) that the corrections to the kinetic term of the scalars $b^a$ cancel and can thus only come from the Weyl rescaling. This in turn implies that the correction is at tree-level in $g_s$, since the contributions to the Ricci scalar for the one-loop terms precisely cancel. Originally, this can be traced back to the fact that the $\cR^4$ terms 
in Type IIA suffer from a relative sign flip when comparing tree-level to one-loop. 
The \Kahler moduli receive both tree-level and one-loop corrections. The one-loop corrections are marked in (\ref{IIAeffactcorr}) with a factor proportional to the dilaton vev. 

Now let us discuss the possible corrections of the vector multiplets containing the complexified \Kahler moduli $t^a$. The vector
multiplet metric can not depend on a scalar  in a hypermultiplet, in particular the dilaton, which identifies the term $\sim \e^{2 \phi_0}$ in (\ref{IIAeffactcorr}) as a correction to a hypermultiplet. It is furthermore well known, that the prepotential in Type IIA contains a tree-level (in $g_s$) correction according to
\be\label{prepcorr}
f(t)=\frac{1}{3!}\cK_{abc}\, t^a t^b t^c-i \, a \, \chi(Y_3)\, , ~~~~a \in \mathbb{R}\, .
\ee
The constant $a$ is already completely fixed from the terms $\sim \cK_{ab}$ in (\ref{IIAeffactcorr}), while terms $\sim \cK_a \cK_b\, \upd v^a \wedge \star\, \upd v^b$ can in principle be absorbed in the definition of a hypermultiplet scalar. One can then compute the metric following from the prepotential (\ref{prepcorr}) and compare the coefficient to the one of the $\sim \cK_{ab}$ terms in (\ref{IIAeffactcorr}), which leads to 
\be
a=4\, \ax'^3 \, \pi^3\, \zx(3)\, .
\ee
Choosing units in which $\ell_s=2\pi \sqrt{\ax'}=1$ one obtains the desired result for the prepotential correction
\be\label{corrprep}
f(t)=\frac{1}{3!}\, \cK_{abc}t^a t^b t^c-i \frac{ \ze(3)}{2 (2 \pi)^3}\, \chi(Y_3)\, .
\ee
Note however, that a correction $\sim \cK_a\, \cK_b\, \upd b^a \wedge \star\, \upd b^b$ cannot be absorbed in the definition of any 
hypermultiplet scalar $\phi_4$, e.g.~of the form $\phi_4=\wt{\phi}+\cK_a b^a+...$, without spoiling the shift symmetry of $b^a$. 
In order to match also the $b^a$ sector with the metric derived from (\ref{corrprep}), we therefore need the additional contribution
\be
\Delta S^\tbf_{b}=-\int_{\cM_4}\frac{\ze(3)}{8 (2\pi)^3}\chi(Y_3)\, \cV^{-3}\, \cK_a \cK_b\, \upd b^a \wedge \star \, \upd b^b\label{deltaS}
\ee
from the reduction of the tree-level eight-derivative terms including $\wt{H}_3$.  This missing structure strongly indicates that the logic of obtaining the $\wt{B}_2$-field completion of the eight-derivative terms employed in \cite{Liu:2013dna} has to be modified for the tree-level terms. In particular, we find that the modified replacement \eqref{newrep} containing the new structure $\wt{t_8} \wt{t}_8 \wt{H}_3^2 \wt{R}^3$ gives precisely the lacking contribution \eqref{deltaS}.

\subsection{Four-derivative terms of the \Kahler moduli}

We proceed by determining the four-derivative terms of Type IIA supergravity in a similar way to section \ref{MthKahlerfour}. 
This analysis will not include a complete treatment of the dilaton, since a complete eight-derivative action of the dilaton in ten dimensions is 
not known. Furthermore, we only reduce the action
\be
S^{\text{IIA}}_{\wt{R^4}}=S^{\text{tree}}_{\wt{R}^4}+S^{\text{loop}}_{\wt{R}^4}\, ,
\ee
on the classical solution, since the classical action on the $\alpha'$-corrected solution 
does not lead to further four-derivative couplings in four dimensions. However, let us add two comments on 
the importance of the classical action for the derivation of the four-derivative couplings. After the reduction the four-derivative 
terms at order $\ax'^3$ are not in Einstein frame. Performing a Weyl rescaling of the Einstein-Hilbert term given by $\sim \e^{-2 \phi} \cV R$ one obtains the Einstein frame four-derivative couplings. Notice that any order $\ax'^3$ contribution to the Ricci scalar does not alter the form of the four-derivative terms at order $\ax'^3$. 

\newpage
\noindent{\bf Tree-level terms.}

\noindent The tree-level terms give rise to the four-derivative couplings
\begin{align}
2 \kappa_{10}^2\,\lz^{-1}\, S^{\text{tree}}_{\wt{R}^4}\big|_{\text{four der.}}&= \int_{\cM_{4}} 768\,\cZ^\tbz_a\, \big[R\, (\Box \del v^a) -2 R^{\mu \nu}\, \nab_{\mu} \nab_{\nu} \del v^a \big] \star 1\nonumber\\
&+ \int_{\M_4}768\, (2\pi)^2 \,\big[R_{\mu \nu}R^{\mu \nu}-4 R^2 \big]\star 1 \int_{Y_3}c_2^\tbz\wedge J \label{IIAfour}\\
&+\int_{\M_4}96\, \cZ^\tbz_{a b} \big[R \, \upd \del v^a \wedge \star \, \upd \del v^b-4\, R^{\mu \nu}\, \pd_{\mu} \del v^a \pd_\nu \del v^b \star 1 \big]\nonumber\\
&+ \int_{\M_4}192\,\cZ^\tbz_{a b} \big[(\Box \del v^a) (\Box \del v^b)\star 1 -2 \nab_{\mu}\nab_{\nu}\del v^a \nab^{\mu}\nab^{\nu}\del v^b\big]\star 1\, ,\nonumber
\end{align}
where $J=J^\tbz+\del v^a \om_a$. Notice that the terms in the first line in (\ref{IIAfour}) coupling to divisor integrals of the second Chern class vanish up to derivative terms of the dilaton by virtue of the contracted second Bianchi identity. Performing the Weyl rescaling $g_{\mu \nu} \to\e^{2 \phi} \V^{-1} g_{\mu \nu}$ one obtains the effective action in Einstein frame. Due to the appearance of the $R^2$ terms in four dimensions the Weyl rescaling is rather involved. We therefore refrain from spelling out the effective action in terms of fluctuations but rather give an action in terms of the fields after the continuation to finite moduli $v^a$, which precisely reproduces the action obtained in terms of the fluctuations $\dx v^a$. By defining $\mathcal{G}_{\mu \nu} =R_{\mu \nu} - \frac{1}{4} R g_{\mu \nu} $,  in close analogy to the Einstein tensor\footnote{ The Einstein tensor  is given by $\mathcal{G}_{\mu \nu} =R_{\mu \nu} - \frac{1}{2} R g_{\mu \nu} $.}  one finds
 \begin{align}
 2 \kappa_{10}^2\,\lz^{-1}\, S^{\text{tree}}_{\wt{R}^4}\big|_{\text{four der.}}&=\int_{\M_4}768\, \ax \,\Big[   \, \mathcal{G}_{\mu\nu} \mathcal{G}^{\mu \nu} \mathcal{Z}  +  \tfrac{2}{\cV} \cK_a  \mathcal{Z} \, \mathcal{G}_{\mu\nu}\,  \nabla^\nu  \nabla^\mu  v^a\,\Big]\star1\nonumber\\
 &-\int_{\M_4}768\, \ax \,  \Big( \tfrac{1}{2}  \mathcal{Z}_{ab}  - \tfrac{2}{\cV} \cK_{ab}    \mathcal{Z}  +\tfrac{1}{\cV^2} \cK_a \cK_b   \mathcal{Z} \Big) \mathcal{G}_{\mu\nu}\, \nabla^\mu  v^a\nabla^\nu  v^b\nonumber\\
 &+\int_{\M_4}192 \, \ax \,\,\Big(  \mathcal{Z}_{ab}  - \tfrac{1}{ \cV^2}\cK_a \cK_b \mathcal{Z} \Big)  (\Box v^a) \, (\Box v^b) \star 1\nonumber\\
 &-\int_{\M_4}384\, \ax  \,\Big(  \mathcal{Z}_{ab}  - \tfrac{2}{ \cV^2}\ \cK_a \cK_b \mathcal{Z} \Big) \nabla_\mu\nabla_\nu  v^a \;\;  \nabla^\mu \nabla^\nu  v^b \star 1\nonumber\\
 &+ \int_{\cM_{4}} 768\, \ax\,\mathcal{Z}_a \big[R\, (\Box  v^a) -2 R^{\mu \nu}\, \nab_{\mu} \nab_{\nu}  v^a \big] \star 1\, .\label{IIAtreefourfinal}
 \end{align}
The last line in (\ref{IIAtreefourfinal}) involving the coupling to the divisor integrals of the second Chern class again vanishes up to dilaton terms upon integrating by parts. We once more stress, that the trivial continuation $\cZ^\tbz_{ab} \to\cZ_{ab}$ is not necessarily the complete answer and the role the tensor $Z^\tbz_{i \bar{\jmath} k \bar{l}}$ plays both from a mathematical and a physical point of view is not clear.

\newpage
\noindent{\bf One-loop terms.}

\noindent The one-loop terms in ten dimensions give rise to the four-derivative terms
\begin{align}
2 \kappa_{10}^2 \, \lo^{-1}\, S^{\text{loop}}_{\wt{R}^4}\big|_{\text{four der.}}&= \int_{\cM_{4}} 768\,\cZ^\tbz_a \big[ 2 R^{\mu \nu}\, \nab_{\mu} \nab_{\nu} \del v^a -R\, (\Box \del v^a)\big] \star 1\nonumber\\
&+\int_{\M_4}192\, (2 \pi)^2\,\big[2 R_{\mu \nu \rho \sigma}R^{\mu \nu \rho \sigma}-4 R_{\mu \nu}R^{\mu \nu}+ R^2 \big]\star 1 \int_{Y_3}c_2^\tbz\wedge J\nonumber\\
&-\int_{\M_4}96\, \cZ^\tbz_{a b} \big[R \, \upd \del v^a \wedge \star \, \upd \del v^b-4\, R^{\mu \nu}\, \pd_{\mu} \del v^a \pd_\nu \del v^b \star 1\nonumber\\
&~~~~~~~~~~~~~~~~~~~~~~~~~~~~~~~~~~~~~~~~+2 (\Box \del v^a) (\Box \del v^b) \big]\, .\label{IIAfourDmod2}
\end{align}
In the following we will again consider the Weyl rescaling of the squared Riemann terms in (\ref{IIAfourDmod2}). To do this we redefine the metric as $g_{\mu \nu} \to \e^{2 \phi} \V^{-1} g_{\mu \nu}$ leading to a canonically normalized Einstein-Hilbert term. Performing the rescaling and the uplift to finite fields one obtains
\begin{align}
2 \kappa_{10}^2 \, \lo^{-1}\, S^{\text{loop}}_{\wt{R}^4}\big|_{\text{four der.}}&=\int_{\M_4}\!\!384 \, \Cbar\, \Big[R_{\mu \nu \rho \sigma}R^{\mu \nu \rho \sigma}+\tfrac{1}{\V}R\, \cK_a\,(\Box v^a)-2\, \cG_{\mu \nu}\cG^{\mu \nu} \Big]\star 1\nonumber\\
&+\int_{\M_4}\!\!384 \Big[\Cbar_{ab}\,\cG_{\mu \nu}+\tfrac{1}{\V}\Cbar \, R \, \K_{ab}\, g_{\mu \nu}-\tfrac{3}{2 \V^2}\Cbar \, R\, \K_a \, \K_b\, g_{\mu \nu} \Big] \nab^\mu v^a \nab^\nu v^b\star 1\nonumber\\
&-\int_{\M_4}\!\!192 \Big[\Cbar_{ab}-\tfrac{3}{\V^2} \Cbar \, \K_a \, \K_b \Big] (\Box v^a)\, (\Box v^b) \star 1\, .
\end{align}

\subsection{Type IIB supergravity on Calabi-Yau threefolds at four derivatives}

In this section we summarize the four-derivative couplings  involving K\"{a}hler moduli derived in \cite{Weissenbacher:2016gey} via the same logic as in the previous section but starting from type IIB supergravity in ten dimensions.  The intention is to present a complete discussion in this work. The two-derivative discussion of type IIB supergravity at order $\alpha'^3$ was recently revisited in \cite{Bonetti:2016dqh} for generic $h^{1,1}$, and reproduces the well known Euler-characteristic correction to the K\"{a}hler potential \cite{Becker:2002nn} 
upon truncation to $\cN=1$.  
This discussion involves the parametrisation of the higher-derivative dilaton action and is thus beyond the treatment in this work.
 However, let us note that the complete dilaton-axion dependence of the $\cR^4$-terms in type IIB is known to be
\beq\label{R42}
S^{\text{IIB}}_{\wt{R}^4} =  \fr{1}{4 \kappa_{10}^2}\int  E_{\frac{3}{2}}(\tau,\bar \tau)  
\Big( \wt{t}_8  \wt{t}_8 + \tfrac{1}{8}  \epsilon_{10}  \epsilon_{10} \Big) \wt{R}^4 \wt * 1 \;\; ,
\eeq
where $E_{\frac{3}{2}}(\tau,\bar \tau)$ is the $SL(2,\mathbb Z)$-invariant  Eisenstein series   given by 
\beq\label{Eisen32}
E_{\frac{3}{2}}(\tau,\bar \tau) =
 \sum_{(m,n) \neq(0,0)} \frac{\tau_2^{3/2}}{|m + n \, \tau|^3} \ ,
\eeq
with  $\tau =  \tilde C_0 + i e^{ -   \tilde \phi} := \tau_1 + i \tau_2$ the dilaton-axion.
When performing the  large $\tau_2$ limit, corresponding to the small
string coupling limit \eqref{Eisen32} becomes 
\beq \label{f0expansion}
E_{\frac{3}{2}}(\tau,\bar \tau) = 2 \zeta(3) \, \tau_2^{3/2} + \tfrac{2\pi^2}{3} \tau_2^{-1/2}
+ \cO(e^{-2\pi \tau_2}) \ .
\eeq 
In the following discussion, we will use this approximation  in \eqref{R42}  and only consider the leading order $g_s$ contribution, given by
\beq\label{R4}
S^{\text{IIB}}_{\wt R^4} =  \fr{1}{2 \kappa_{10}^2}  \int e^{-\frac{3}{2} \tilde \phi} 
\big( \wt{t}_8  \wt{t}_8 + \tfrac{1}{8}  \epsilon_{10}  \epsilon_{10} \big)  \wt{R}^4  \wt{\ast} 1  \;\; .
\eeq
The four-derivative corrections arising from the ten-dimensional $\cR^4$-terms result in
\begin{align}
2 \kappa_{10}^2 \lz^{-1} S^{\text{IIB}}_{\wt{R}^4}\big|_{\text{four der.}}&=192 \int_{\M_4} \!\!\! (2\pi)^2 \e^{-\frac{3}{2}  \phi}  \big[  4  \, R_{\mu \nu} R^{\mu \nu} - R^2  \big]\star 1 \int_{Y_3} c_2^\tbzero\wedge J\label{R4Red}   \\
 &+192\int_{\M_4}\!\!\!\e^{-\frac{3}{2} \phi} \cZ^\tbz_{ab}\Big[   \big( - 2 R_{\mu \nu }   + \tfrac{1}{2} R g_{\mu \nu} \big)\nabla^\mu \delta v^a\nabla^\nu \delta v^b  \Big]\star 1  \nonumber \\
 &+192\int_{\M_4}\!\!\!\e^{-\frac{3}{2} \phi} \cZ^\tbz_{ab}\,\Big[(\Box \delta v^a)\; (\Box \delta v^b)-2\nabla_\mu \nabla_\nu \delta v^a\; \nabla^\mu \nabla^\nu \delta v^b\Big] \star 1\, ,\nonumber
\end{align}
where we once again have $J=J^\tbz+\dx v^a \om_a$
and we have dropped the tilde on $\tilde \phi$ to indicate that it is now a 
four-dimensional field.  The final result is derived by combing the Weyl rescaling of the reduction of tyhe classical Einstein-Hilbert action, with the uplift of the reduction result   \eqref{R4Red}. The action one obtains is then
 \begin{align}
 2 \kappa_{10}^2 \lz^{-1}S_{\text{kin}}^\tbfour&= \int_{\M_4}  R     + \tfrac{1}{ \cV}   \Big( \tfrac{1}{2}\cK_{ab} - \tfrac{1}{\cV}\cK_a\cK_b\Big) \upd v^a \wedge \star \, \upd v^b \nonumber\\
 &+\int_{\M_4}768\, \ax \, \e^{-\frac{3}{2} \phi}\Big[   \, \mathcal{G}_{\mu\nu} \mathcal{G}^{\mu \nu} \mathcal{Z}  +  \tfrac{2}{\cV} \cK_a  \mathcal{Z} \, \mathcal{G}_{\mu\nu}\,  \nabla^\nu  \nabla^\mu  v^a\,\Big]\star1\nonumber\\
 &-\int_{\M_4}768\, \ax \, \e^{-\frac{3}{2} \phi} \Big( \tfrac{1}{2}  \mathcal{Z}_{ab}  - \tfrac{2}{\cV} \cK_{ab}    \mathcal{Z}  +\tfrac{1}{\cV^2} \cK_a \cK_b   \mathcal{Z} \Big) \mathcal{G}_{\mu\nu}\, \nabla^\mu  v^a\nabla^\nu  v^b\nonumber\\
 &+\int_{\M_4}192 \, \ax \,\e^{-\frac{3}{2} \phi}\,\Big(  \mathcal{Z}_{ab}  - \tfrac{1}{ \cV^2}\cK_a \cK_b \mathcal{Z} \Big)  (\Box v^a) \, (\Box v^b) \star 1\nonumber\\
 &-\int_{\M_4}384\, \ax \,\e^{-\frac{3}{2} \phi} \,\Big(  \mathcal{Z}_{ab}  - \tfrac{2}{ \cV^2}\ \cK_a \cK_b \mathcal{Z} \Big) \nabla_\mu\nabla_\nu  v^a \;\;  \nabla^\mu \nabla^\nu  v^b \star 1\, .\label{IIBfourfinal}
 \end{align}
As expected, since the tree-level $\cR^4$-terms of both Type IIA and Type IIB  have the same structure, the results of the dimensional reduction coincide, see (\ref{IIAtreefourfinal}) and (\ref{IIBfourfinal}). In addition, compared to the one-loop $\cR^4$-terms in Type IIB, the corresponding one-loop $\cR^4$-terms in Type IIA suffer from a relative sign flip between the two basic structures given by $\wt{t}_8\wt{t}_8 \wt{R}^4$ and $\epsilon_{10} \epsilon_{10}\wt{R}^4$. The two main differences between one-loop contribution to the four dimensional action at four derivatives are: the appearance of a $R_{\mu \nu \rho \sigma}^2$ term in Type IIA, whereas this term is not present in the case of Type IIB and the lack of a term with the structure $\nab_{\mu} \nab_\nu v^a \, \nab^{\mu} \nab^\nu v^b$ in the case of Type IIA compared to Type IIB. Both cases can be traced back to the different sign structures of the one-loop terms in ten dimensions leading to non-trivial cancellations.

\section{Threefold reduction up to four derivatives - the one modulus case}

In the following we simplify the discussion and assume that the background geometries under consideration have only one modulus. The advantage of the simplified discussion is, that all computations can be done exactly yet leading to non-trivial results. We consider the $\cR^4$-terms in M-theory and both Type II theories, where for the latter we consider tree-level and one-loop terms.

\subsection{M-theory one modulus reduction}\label{Mthonemod}

In this section we comment on the M-theory compactification on a Calabi-Yau threefold including higher-derivative terms for the case of a single (volume) modulus. We only consider the gravitational terms in eleven dimensions in the following discussion. The compactification ansatz for the metric is then given by
\begin{align}
\upd \wh{s}_{11}^2&=\e^{\wh{\ax} \Phi^\tbo} \left(g_{\mu \nu} \upd x^\mu \upd x^\nu+2 \e^{u/3} g^\tbz_{i \bar{\jmath}} \upd z^i \upd \bar{z}^{\bar{\jmath}} \right),\label{Bbackgr}\\
\Phi^\tbo&= -\frac{512}{3}(2\pi)^3\, \ast^\tbz_6 \, c^\tbz_3\, ,
\end{align}
where in (\ref{Bbackgr}) the Weyl factor $\Phi^\tbo$ is computed using the metric $g_{i \bar{\jmath}}=\e^{u/3} g^\tbz_{i \bar{\jmath}}$. The metric $g_{i \bar{\jmath}}$ is furthermore normalized to unit volume, such that the relation
\begin{equation} \label{V-def}
\int_{Y_3} \ast_6 1=\int_{Y_3} \upd^6 y \sqrt{g}=\e^{u} \int_{Y_3} \upd^6 y \sqrt{g^\tbz}=\e^{u}
\end{equation}
holds. Note that the scalar field $u(x)$ is related to the volume modulus by $u(x)=\log \cV(x)$. As a first step we dimensionally reduce the action on the background (\ref{Bbackgr}) including the $\cR^4$-terms in eleven dimensions up to four external spacetime derivatives. One obtains in five dimensions the following action
{\setlength{\jot}{7pt}\begin{align}\label{Baction1}
2 \kappa_{11}^2 S^\tbfive&=\int_{\cM_5} \e^u R \star 1+\frac{5}{6}\e^{u} \upd u \wedge \star\, \upd u+768 (2\pi)^3\, \wh{\ax} \chi(Y_3)\upd u \wedge \star\, \upd u\\
&+\int_{\cM_5}\wh{\ax}\,\Ccal\,\e^{u/3}\left[ 384\, {R}\inds{_{\mu \nu \rho \sigma}}{R}\inds{^{\mu \nu \rho \sigma}}-768\, {R}\inds{_{\mu \nu }}{R}\inds{^{\mu \nu }} +192 {R}^2 \right]\star 1\nonumber\\
&+\int_{\cM_5}\wh{\ax}\,\Ccal\,\e^{u/3} \bigg[ \frac{256}{3}{R}\inds{^{\mu \nu}}\,\pd_{\mu} u\, \pd_\nu u \star 1+512 {R}\inds{^{\mu \nu}}\nabla_{\mu}\nabla_{\nu}u \star 1\nonumber\\
&~~~~~~~~~~~~~~~~~~~~~ -256 R\, (\Box u )\star 1 -64 R\, \upd u \wedge \star\, \upd u \bigg]\nonumber\\
&+\int_{\cM_5}\wh{\ax}\,\Ccal\, \e^{u/3} \bigg[\frac{128}{3} (\Box u)^2 \star 1+\frac{128}{9} (\Box u)\,\upd u \wedge \star\, \upd u+\frac{80}{9}(\pd u)^4 \star 1 \bigg]\nonumber.
\end{align}}
Note that the quantity $\Ccal$ is computed using the metric $g^\tbz_{i \bar{\jmath}}$ which is normalized to unit volume. We furthermore made use of the schematic notation $(\pd u)^4\equiv \pd_{\mu}u \,\pd^{\mu}u\, \pd_{\nu}u\, \pd^{\nu}u$. For a canonical normalization of the five-dimensional Einstein-Hilbert term we perform a Weyl rescaling 
\begin{align}\label{BWeyl}
g_{\mu \nu} &\to \e^\sigma g_{\mu \nu},\\
\sigma &= -\frac{2}{3}u\nonumber,
\end{align}
which results in the action
\begin{align}\label{Bactioneinst}
2 \kappa_{11}^2 S^\tbfive&=\int_{\cM_5} R \star 1 -\frac{1}{2}\upd u \wedge \star\, \upd u+768 (2\pi)^3\, \wh{\alpha} \chi(Y_3)\e^{-u}\upd u \wedge \star\, \upd u\\
&+\int_{\cM_5}\wh{\alpha} \, \Ccal\,\left[ 384\, {R}\inds{_{\mu \nu \rho \sigma}}{R}\inds{^{\mu \nu \rho \sigma}}-768\, {R}\inds{_{\mu \nu }}{R}\inds{^{\mu \nu }} +192 {R}^2 \right]\star 1\nonumber\\
&+\int_{\cM_5}\wh{\alpha}\, \Ccal\,\left[384 \left(\Box u \right)^2 \star 1-\frac{1280}{3}\left( \Box u \right)\,\upd u \wedge \star\, \upd u+\frac{368}{3} \left( \pd u \right)^4 \star 1\right]\nonumber\\
&+\int_{\cM_5}\wh{\alpha}\, \Ccal\,\left[256 R\, \left( \Box u \right) \star 1-\frac{448}{3}R\, \upd u \wedge \star\, \upd u \right]\nonumber
\end{align}
in Einstein frame, where we have furthermore performed integrations by parts. It is clear, that higher-derivative actions of the form (\ref{Bactioneinst}) suffer from ambiguities due to the possibility of performing higher-derivative field redefinitions and integrations by parts. Nevertheless, we will propose new field variables such that the action (\ref{Bactioneinst}) takes a rather simple form. If one redefines the five-dimensional metric $g_{\mu \nu}$ and the scalars $u$ as
\begin{align}
g_{\mu \nu}&\rightarrow g_{\mu \nu}+\am_1 \, \ax \,\Ccal\, (\Box u)\, g_{\mu \nu}+\am_2\, \ah \, \Ccal \, (\pd u)^2\, g_{\mu \nu}+\am_3 \, \ah \, \Ccal\, R \, g_{\mu \nu} \\[0.2cm]
&~~~~~~\,\,~~+\am_4 \, \ah\, \Ccal \, R_{\mu \nu}+\am_5\, \ah \, \Ccal\, \pd_{\mu} u \, \pd_\nu u+\am_6 \, \ax \, \Ccal \, \nab_\mu \nab_\nu u \, ,\nonumber\\[0.3cm]
u&\rightarrow u+\am_7\, \ah \, \Ccal \, (\Box u)+\am_8 \, \ah \, \Ccal\, (\pd u)^2+\am_9 \, \ah \, \Ccal \, R\, ,
\end{align}
where the nine coefficients $\am_i$ take the values
\be
\bgroup
\def\arraystretch{2.0}
\begin{tabular}{l l l} 
$\am_1=-\frac{512}{3}~~~~~~~~~~$&$\am_2=\frac{320}{9}~~~~~~~~~~$&$\am_3=-128$\\
$\am_4=768~~~~~~~~~~$&$\am_5=384~~~~~~~~~~$&$\am_6=0$\\
$\am_7=-384~~~~~~~~~~$&$\am_8=\frac{896}{3}~~~~~~~~~~$&$\am_9=0\, ,$
\end{tabular}
\egroup
\ee
the five-dimensional action boils down to
\begin{align}\label{Bactioneinst2}
2 \kappa_{11}^2 S^\tbfive&=\int_{\cM_5} R \star 1 -\frac{1}{2}\upd u \wedge \star\, \upd u+768 (2\pi)^3\, \wh{\alpha} \chi(Y_3)\e^{-u}\upd u \wedge \star\, \upd u\\\vspace{0.5cm}
&+\int_{\cM_5}384\,\wh{\alpha} \, \Ccal\,\left[  {R}\inds{_{\mu \nu \rho \sigma}}{R}\inds{^{\mu \nu \rho \sigma}}-4\, {R}\inds{_{\mu \nu }}{R}\inds{^{\mu \nu }} + {R}^2 \right]\star 1\nonumber\\
&+\int_{\cM_5}192\, \wh{\ax}\,\Ccal \, (\pd u)^4 \star 1.\nonumber
\end{align}
The only four-derivative couplings in (\ref{Bactioneinst2}) surviving this field redefinition are the squared Riemann terms in the Gauss-Bonnet combination and the $(\pd u)^4$ interaction which leads to a particularly simple form of the effective action. Effectively, one obtains a massless scalar field $u$ coupled to Gauss-Bonnet (super-)gravity and a $(\pd u)^4$ like interaction term. The four-derivative terms couple to the theory by means of the divisor integral
\be 
\Ccal=\int_{Y_3}c^\tbz_2 \wedge J^\tbz=\int_{D}c^\tbz_2\, ,
\ee
where $D \equiv [J]_{\text{PD}}$ is the divisor Poincare dual to $J$. At this stage the coupling $\Ccal$ is independent of the modulus $u=\log \cV$, since it is computed using the metric $g_{i \bar{\jmath}}$. The modulus dependent coupling $\Cbar$ 
\begin{equation}\label{moduluscoupling}
\Cbar=\int_{Y_3} c_2 \wedge J=\e^{u/3}\int_{Y_3}c^\tbz_2 \wedge J^\tbz=\e^{u/3} \,\Ccal\, ,
\end{equation}
where the quantities $c_2$ and $J$ are now associated to the metric $g_{i \bar{\jmath}}=\e^{u/3}g^\tbz_{i \bar{\jmath}}\,$, can now be implemented in (\ref{Bactioneinst2}) in a straightforward way using (\ref{moduluscoupling}).

\subsection{Type IIA one modulus reduction}\label{IIAonemodulus}

Now we turn to the discussion of the one-modulus case in Type IIA. We therefore consider in a similar fashion as in the M-theory context shown above the ansatz for the metric
\begin{align}
\upd \wt{s}_{10}^2&=g_{\mu \nu} \upd x^\mu \upd x^\nu+2 \e^{u/3} g^\tbz_{i \bar{\jmath}} \upd z^i \upd \bar{z}^{\bar{\jmath}}\, ,\label{BbackgrIIA}\\
\wt{\phi}&= \phi_0+ \ax\, \la {\phi}^\tbo \ra\, ,
\end{align}
where we again compute $\la \phi^\tbo \ra$ given in (\ref{IIAansatz}) using the metric $g_{i \bar{\jmath}}=\e^{u/3} g^\tbz_{i \bar{\jmath}}$. Performing the dimensional reduction of the Type IIA action including tree-level and one-loop corrections results in a four-dimensional effective action of the form
\be\label{onemodIIAaction}
S^\tbf=S^\tbf_{\text{2 der.}}+\ax \,S^\tbf_{R^2}+\ax \,S^\tbf_{R u}+ \ax \, S^\tbf_{u}\, .
\ee
The various contributions to (\ref{onemodIIAaction}) are the two-derivative action $S^\tbf_{\text{2 der.}}$ , the piece containing the quadratic Riemann tensor terms $S^\tbf_{R^2}$, the part of the action containing the mixed Riemann tensor and $u$ terms $S^\tbf_{R u}$ and finally the contribution where all four derivatives act on the $u$ scalars, denoted by $S^\tbf_u$. The various pieces are
\begin{align}
2 \kappa_{10}^2 S^\tbf_{\text{2 der.}}&=\int_{\cM_{4}}\big[\e^{-2 \phi_0} \e^{u}-1536\, \ax\,(2\pi)^3\, \chi(Y_3)\,\lz\big] R\star 1+\frac{5}{6}\, \e^{-2 \phi_0}\, \e^{u}\, \upd u \wedge \star \, \upd u\nonumber\\
&~~~~~~~~~~+768\, \ax \,(\lz+\lo)\, \chi(Y_3)\, (2\pi)^3\, \upd u \wedge \star \, \upd u\big]\,,\\[0.2cm]
2\kappa_{10}^2 S^\tbf_{R^2}&=\int_{\M_4}\Cbar\, \big[384\, \lo\, R_{\mu \nu \rho \sigma} R^{\mu \nu \rho \sigma}+(\lz-\lo) \, R_{\mu \nu}R^{\mu \nu}-192\, (\lz-\lo) \,R^2 \big]\star 1\, ,\\[0.2cm]
2 \kappa_{10}^2 S^\tbf_{R u}&=\int_{\M_4} \Cbar\, \Big[-\frac{256}{3}\, (\lz-\lo) \, R^{\mu \nu}\pd_{\mu}u\, \pd_{\nu}u \star 1-512\, (\lz-\lo)\, R^{\mu \nu}\, \nabla_{\mu}\nabla_{\nu} u\nonumber\\
&~~~~~~~~~~~~~+256\, (\lz-\lo)\, R\, (\Box u) \star 1+64 \, (\lz-\lo)\, R\, \upd u \wedge \star \, \upd u\Big]\, ,\\[0.2cm]
2 \kappa_{10}^2 S^\tbf_{u}&=\int_{\M_4}\Cbar\, \Big[-\frac{128}{3}\, (\lz-\lo)\, (\Box u)^2 \star 1-\frac{128}{9}\, (\lz-\lo)\, (\Box u)\, \upd u \wedge \star \, \upd u\nonumber \\
&~~~~~~~~~~~~ +\frac{80}{9}\, (\lz+\lo) \, (\pd u)^4 \star 1+\frac{256}{9}\,\lz\, \pd^{\mu}u \, \nabla_{\mu}\nabla_{\nu} u\, \pd^{\nu}u \star 1 \nonumber\\
&~~~~~~~~~~~~+\frac{256}{3}\, \lz\, \nabla_{\mu} \nabla_{\nu}u\,  \nabla^{\mu} \nabla^{\nu}u \star 1 \Big]\, ,
\end{align}
where we made use of the shorthand notation (\ref{Zcal}) and (\ref{moduluscoupling}).  A Weyl rescaling according to
\be
g_{\mu \nu} \to \big[\e^{-2 \phi_0} \e^{u}-1536\, \ax\,(2\pi)^3\, \chi(Y_3)\lz \big]^{-1}\, g_{\mu \nu}
\ee
transforms the action (\ref{onemodIIAaction}) into the four dimesnional Einstein frame action
\be\label{onemodIIAaction2}
S^\tbf_{\text{IIA}}=S^\tbf_{\text{class}}+ \ax\, \lz\, S^\tbf_{\text{tree}}+\ax \, \lo\, S^\tbf_{\text{loop}}\, ,
\ee
where we split the action (\ref{onemodIIAaction2}) in a classical piece $S^\tbf_{\text{class}}$, a tree-level (in $g_s$) part $S^\tbf_{\text{tree}}$ and a one-loop correction $S^\tbf_{\text{loop}}$. Using external space total derivative identities on can show that the relations
\begin{align}
\int_{\M_4}\upd^4 x \sqrt{-g}\,\e^{u/3}\,\nab^\mu u\, \nab_{\mu}\nab_\nu u \, \nab^\nu u&=-\frac{1}{2}\int_{\M_4}\upd^4 x \sqrt{-g}\, \e^{u/3} \Big[\frac{1}{3} (\pd u)^4+(\Box u)\, (\pd u)^2 \Big]
\end{align}
as well as
\begin{align}
\int_{\M_4}\upd^4 x \sqrt{-g}\,\e^{u/3}\, \nab_{\mu}\nab_\nu u\, \nab^{\mu} \nab^\nu u&=\int_{\M_4}\upd^4 x \sqrt{-g}\, \e^{u/3}\Big[(\Box u)^2+\frac{1}{2}\, (\Box u)\, (\pd u)^2 \\
&~~~~~~~~~~~~~~~~~~~~~~~~~~~~~~+\frac{1}{18}\, (\pd u)^4-R^{\mu \nu}\, \pd_{\mu}u\, \pd_\nu u\Big]\nonumber
\end{align}
hold. The constituents of (\ref{onemodIIAaction2}) can then be shown to take the form
\begin{align}
2 \kappa_{10}^2 S^\tbf_{\text{class}}&=\int_{\M_4} R \star 1-\frac{2}{3} \upd u \wedge \star \, \upd u\, ,\\[0.2cm]
2\kappa_{10}^2 S^{\tbf}_{\text{tree}}&=\int_{\M_4}\Cbar\, \big[768\, R_{\mu \nu} R^{\mu \nu}-192\, R^2 \big]\star 1-2560\,(2\pi)^3\, \chi(Y_3)\, \e^{2 \phi_0}\, \e^{-u}\, \upd u \wedge \star \, \upd u\nonumber\\
&+\int_{\M_4}\Cbar\, \Big[\frac{6592}{9}\, (\Box u)^2 \star 1-128\, (\Box u)\, \upd u \wedge \star\, \upd u+\frac{2048}{81}\, (\pd u)^4 \star 1 \Big]\nonumber\\
&+\int_{\M_4} \Cbar\, \Big[\frac{2048}{9}\, R^{\mu \nu}\, \pd_\mu u \, \pd_\nu u \star 1+1024\, R^{\mu \nu} \nab_{\mu}\nab_\nu u \star 1\nonumber\\
&~~~~~~~~~~~~~~-128\, R\,(\Box u) \star 1-128\, R\, \upd u \wedge \star\, \upd u \Big]\, ,\\[0.2cm]
2 \kappa_{10}^2 S^\tbf_{\text{loop}}&=\int_{\cM_4}\Cbar \, \Big[384\, R_{\mu \nu \rho \sigma}R^{\mu \nu \rho \sigma}-768\, R_{\mu \nu}R^{\mu \nu}+192\, R^2 \Big]\star 1\nonumber\\
&+\int_{\M_4}768\, (2\pi)^3\, \e^{2 \phi_0}\, \chi(Y_3)\, \e^{-u}\, \upd u \wedge \star \, \upd u\nonumber\\
&+\int_{\M_4} \Cbar \, \Big[448 (\Box u)^2 \star 1-\frac{6016}{9}\, (\Box u)\, \upd u \wedge \star \, \upd u+\frac{4928}{27}\, (\pd u^4) \star 1\Big]\nonumber\\
&+\int_{\M_4}\Cbar \, \Big[-256\, R^{\mu \nu}\, \pd_\mu u \pd_\nu u \star 1+512 \, R^{\mu \nu}\nab_\mu \nab_\nu u \star 1\nonumber\\
&~~~~~~~~~~~~~+128 \, R\, (\Box u) \star 1-256\, R\, \upd u \wedge \star \, \upd u \Big]\, .
\end{align}
Similar to the M-theory discussion in section \ref{Mthonemod} we can again consider higher-derivative field redefinitions. The general ansatz for these redefinitions reads
\begin{align}
g_{\mu \nu}&\rightarrow g_{\mu \nu}+\aa_1 \, \ax \,\Ccal\,\e^{u/3}\, (\Box u)\, g_{\mu \nu}+\aa_2\, \ah \, \Ccal \,\e^{u/3}\, (\pd u)^2\, g_{\mu \nu}+\am_3 \, \ah \, \Ccal\,\e^{u/3}\, R \, g_{\mu \nu} \label{IIredef1}\\[0.2cm]
&~~~~~~\,\,~~+\aa_4 \, \ah\, \Ccal \,\e^{u/3}\, R_{\mu \nu}+\aa_5\, \ah \, \Ccal\,\e^{u/3}\, \pd_{\mu} u \, \pd_\nu u+\aa_6 \, \ax \, \Ccal \,\e^{u/3}\, \nab_\mu \nab_\nu u \, ,\nonumber\\[0.3cm]
u&\rightarrow u+\aa_7\, \ah \, \Ccal \,\e^{u/3}\, (\Box u)+\aa_8 \, \ah \, \Ccal\,\e^{u/3}\, (\pd u)^2+\aa_9 \, \ah \, \Ccal \,\e^{u/3}\, R\, .\label{IIredef2}
\end{align}
The coefficients $\aa_i$ can then have tree-level and one-loop contributions proportional to $\ze(3)\e^{-2 \phi_0}$ and $\pi^2/3$ respectively which we again choose in a way, such that the action takes a particularly simple form. The factors multiplying these tree-level and one-loop coefficients in the various parameters $\aa_i\equiv \ax_i \, \lz+\bx_i\, \lo$ are listed in Table \ref{coefftab}. 

\begin{table}[h]
\centering
\bgroup
\def\arraystretch{1.5}
\begin{tabular}{|c||c|c|c|c|c|c|c|c|c|}
\hline
{}  &  $i=1$  &  $i=2$  &  $i=3$  &  $i=4$  &  $i=5$  &  $i=6$  &  $i=7$  &  $i=8$  &  $i=9$\\\hline\hline
$\ax_i$ &  $-384$  &  $\frac{2048}{9}$  &  $-192$  &  $768$  &  $\frac{512}{9}$  &  $1024$  &  $-\frac{3184}{3}$  &  $-96$  &  $0$  \\\hline
$\bx_i$ &  $-384$  &  $\frac{1280}{3}$  &  $-192$  &  $768$  &  $-\frac{256}{3}$  &  $512$  &  $-592$  &  $\frac{928}{3}$  &  $0$  \\\hline
\end{tabular}
\egroup
\caption{Our choice of the coefficients in the field redefinitions.}
\label{coefftab}
\end{table}

\noindent The various pieces in the four-dimensional action we obtain after these redefinitions are then 
\begin{align}
2 \kappa_{10}^2 S^\tbf_{\text{class}}&=\int_{\M_4} R \star 1-\frac{2}{3} \upd u \wedge \star \, \upd u\, ,\\
2 \kappa_{10}^2 S^\tbf_{\text{tree}}&=-\int_{\M_4}2560\,(2 \pi)^3\, \chi(Y_3)\, \e^{2 \phi_0}\, \e^{-u}\, \upd u \wedge \star \, \upd u\,+\frac{5632}{81}\, \Cbar\, (\pd u)^4 \star 1 ,\\
2 \kappa_{10}^2 S^\tbf_{\text{loop}}&=\int_{\M_4}384 \, \Cbar\, \Big[R_{\mu \nu \rho \sigma}R^{\mu \nu \rho \sigma}-4 \, R_{\mu \nu}R^{\mu \nu}+R^2 \Big]\star 1\\
&+\int_{\M_4}768\, (2\pi)^3\, \e^{2 \phi_0}\, \chi(Y_3)\, \e^{-u}\, \upd u \wedge \star\, \upd u-\frac{3008}{27}\, \Cbar\, (\pd u)^4 \star 1\, .\nonumber
\end{align}

\subsection{Type IIB one modulus reduction}\label{IIB1mod}

We finally aim to include the eight-derivative $\cR^4$-terms for the case of one modulus in Type IIB. We will take into account the classical Einstein-Hilbert action, which is included in Type IIB supergravity, as well as the tree-level and one-loop $\cR^4$-corrections in ten dimensions. Since the higher-derivative terms of the dilaton and the NS-NS two-form are not known in Type IIB, we do not include them as dynamical fields. The action we are considering thus takes the form
\be
S^{\text{IIB}}=S_{\text{class}}^{\text{IIB}}+\ax \, S_{\wt{R}^4}^{\text{IIB}}\, ,
\ee
with the classical action
\be \label{IIBclass}
S_{\text{class}}^{\text{IIB}}=\int_{\M_{10}}\wt{R}\,\wt{\ast}\, 1-\frac{1}{2 \tau_2^2}\upd \tau \wedge \wt{\ast}\, \upd \bar{\tau}+...\, .
\ee
and the leading order $\cR^4$-action defined in (\ref{R42}). The ellipses in (\ref{IIBclass}) stand for the NS-NS two-form and the R-R fields which we do not display here. The ansatz for the ten-dimensional fields is the background solution found in \cite{Bonetti:2016dqh}
\begin{align}
\upd \wt{s}^2&=\e^{\ax \Phi^\tbo}\Big(g_{\mu \nu}\upd x^\mu \upd x^\nu+2\e^{u/3}g^\tbz_{i \bar{\jmath}} \upd z^i \upd \bar{z}^{\bar{\jmath}} \Big)\label{IIBonemod}\\
\Phi^\tbo&=-192\,(2\pi)^3\, \mathscr{Y}\, (\ast^\tbz_{6}\, c^\tbz_3)\nonumber\\
\wt{\phi}&=\phi_0+\cO(\ax)\, ,\nonumber
\end{align}
where we have defined 
\be
\mathscr{Y}\equiv \ze(3)\,\e^{-3 \phi_0/2}+\frac{\pi^2}{3}\,\e^{\phi_0/2}\, .
\ee
As in the M-theory and Type IIA reduction, we introduced the modulus $u$, such that the internal Calabi-Yau metric $g^\tbz_{i \bar{\jmath}}$ is normalized to unit volume and the overall Weyl factor $\Phi^\tbo$ is again computed using $g_{i \bar{\jmath}}=\e^{u/3}g^\tbz_{i \bar{\jmath}}$. There is furthermore a correction to the dilaton $\wt{\phi}^\tbo \sim \ast_6\, c_3$, which however only contributes at order $\ax^2$ to the action and can therefore be ignored.

The results of the dimensional reduction are very similar to the tree-level terms of Type IIA presented in section \ref{IIAonemodulus}. We therefore refrain from giving the results before the Weyl rescaling and simply state the result in Einstein frame. The four dimension effective action is then
\begin{align}
2 \kappa_{10}^2 S^\tbf_{\text{IIB}}&=\int_{\M_4} R \star 1-\frac{2}{3} \upd u \wedge \star \, \upd u-2560\,(2\pi)^3\,\ax\, \chi(Y_3)\, \mathscr{Y}\, \e^{-u}\, \upd u \wedge \star \, \upd u\label{IIBEinst}\\
&+\int_{\M_4}\ax\,\Cbar\,\Ycal\, \big[768\, R_{\mu \nu} R^{\mu \nu}-192\, R^2 \big]\star 1\nonumber\\
&+\int_{\M_4} \ax \,\Cbar\,\Ycal\, \Big[\frac{2048}{9}\, R^{\mu \nu}\, \pd_\mu u \, \pd_\nu u \star 1+1024\, R^{\mu \nu} \nab_{\mu}\nab_\nu u \star 1\nonumber\\
&~~~~~~~~~~~~~~~~~~~-128\, R\,(\Box u) \star 1-128\, R\, \upd u \wedge \star\, \upd u \Big]\nonumber\\
&+\int_{\M_4}\ax\,\Cbar\,\Ycal\, \Big[\frac{6592}{9}\, (\Box u)^2 \star 1-128\, (\Box u)\, \upd u \wedge \star\, \upd u+\frac{2048}{81}\, (\pd u)^4 \star 1 \Big]\, .\nonumber
\end{align}
We can furthermore use the field redefinitions (\ref{IIredef1}) and (\ref{IIredef2}) and the coefficients in the first column of Table \ref{coefftab} to simplify (\ref{IIBEinst}). Note that we also have to replace $\ze(3)\, \e^{-2 \phi_0} \to \Ycal$. This finally leads to the action
\begin{align}\label{db4}
2 \kappa_{10}^2 S^\tbf_{\text{IIB}}&=\int_{\M_4}R \star 1-\frac{2}{3}\upd u \wedge \star\, \upd u-2560 \, (2 \pi)^3\, \ax\, \chi(Y_3)\, \Ycal\, \e^{-u}\, \upd u \wedge \star \, \upd u\nonumber\\
&-\int_{\M_4}\ax \, \frac{5632}{81}\, \Cbar\, \Ycal\, (\pd u)^4 \star 1\, .
\end{align}
Note that compared to (\ref{onemodIIAaction}) the couplings in the action (\ref{IIBEinst}) involve different powers of the dilaton vev $\phi_0$. This is due to the fact, that we started on the one hand in the ten-dimensional string frame for Type IIA, whereas on the other hand our starting point for Type IIB was already the Einstein frame in ten dimensions.  We can obtain the same powers of the dilaton as in (\ref{onemodIIAaction}) if we perform the shift $u \to u-\frac{3}{2} \phi_0$.

\subsection{Comments on the $\cN=1$ orientifold truncation}

In this final subsection we will discuss the orientifold truncation of the $\cN=2$ compactification performed in the previous 
subsection. The resulting theory is expected to be a $\cN=1$ supergravity theory and we will focus on the subsector of the 
theory involving the K\"ahler structure deformations. In addition to this restricted focus we will not try to complete the compactification 
to a fully consistent $\cN=1$ setup. In fact, the presence of orientifold planes would require a more complete treatment 
involving D-branes, which themselves eventually contribute higher-derivative terms to the four-dimensional theory. 
With these caveats in mind we can nevertheless try to push our $\cN=2$ results and comment on the recent proposal of \cite{Ciupke:2015msa}
to stabilize moduli. This procedure of direct truncation from $\cN=2$ to $\cN=1$ has been used before in the determination of 
corrections to the $\cN=1$ K\"ahler potential in \cite{Becker:2002nn,Bonetti:2016dqh}.

In the $\cN=1$ settings one faces similar difficulties as for in the discussion of the $\cN=2$ Calabi-Yau threefold compactifications 
of the previous sections. In particular, a general $\cN=1$ four-derivative on-shell action to which one can compare the effective action after performing a reduction is currently not available. There are, however, partial results extracted by expanding certain 
four-derivative terms in superspace \cite{Koehn:2012ar}. Therefore, it is tempting to compare the $\cN=1$ truncated one modulus reduction 
of section \ref{IIB1mod} to the action of \cite{Koehn:2012ar} as suggested in \cite{Ciupke:2015msa}. 
We will therefore briefly review the required results. The main idea is to take into account a four-derivative $\cN=1$ Lagrangian and infer from the four-derivative coupling a corresponding scalar potential. The relevant four-derivative Lagrangian is \cite{Koehn:2012ar,Ciupke:2015msa}
\begin{align} 
\frac{\cL_{\text{bos.}}}{\sqrt{-g}}&=\tfrac{1}{2}R- G_{i \jb}(A, \bar{A})\, \pd_{\mu}A^i \pd^\mu \bar{A}^\jb-2 \e^K\, T^{\bar l ~\,k}_{~i~\,\bar \jmath}\, D_k W\,  \overline{D}_{\bar{l}} \overline{W}\, \pd_{\mu}A^i \pd^\mu \bar{A}^\jb\label{action1}\\
&+T_{i j \bar k \bar l}\, \, \pd_{\mu}A^i \pd^\mu A^j\, \pd_{\nu}\bar{A}^{\bar{k}} \pd^\nu \bar{A}^{\bar{l}}-V(A, \bar{A})\, .\nonumber
\end{align}
In \eqref{action1} $A^i$ are complex scalars in chiral multiplets, $W$ is the holomorphic superpotential, $T_{i j \bar k \bar l}$ is the coupling tensor for the four-derivative interaction and $D_j$ is the \Kahler covariant derivative. Additionally, the \Kahler metric $G_{i \jb}$ is given in terms of a \Kahler potential $G_{i \jb}= \pd_i \pd_\jb K(A, \bar A)$. The scalar potential in \eqref{action1} consists of two terms 
$V(A, \bar A)=V_{(0)}+V_{(1)}$, where
\begin{align}
V_{(0)}&=\e^{K} \Big(G^{i \jb}D_i W \, \overline{D}_\jb \overline{W}- 3|W|^2 \Big)\, ,\nonumber\\
V_{(1)}&=-\e^{2 K} T^{\ib \jb k l}\overline{D}_\ib \overline{W}\,\overline{D}_\jb \overline{W}\, {D}_k {W}\,{D}_l {W}\, .\label{Vcorr}
\end{align}
In a simplified setup with only a single \Kahler modulus sitting in a chiral multiplet after the $\cN=1$ truncation we
have to compare  \eqref{action1} with the reduction result \eqref{db4}.

In order to perform the suggested comparison we first have to determine the correct $\cN=1$ coordinates. 
It is well-known \cite{Giddings:2001yu,Becker:2002nn,Grimm:2004uq} that at leading order one has to 
introduce complex fields $A \equiv \rho+i \, \e^{\frac{2}{3} u}$, where $\rho$ is the appropriately re-scaled scalar arising from the 
R-R four-form and we recall $u=\log \cV$ is the logarithm of the Einstein-frame volume of $Y_3$ as seen in \eqref{V-def}.
Taking into account the  order $\ax'^3$ corrections at the two-derivative level they potentially correct the $\cN=1$ 
coordinates. However, it was argued in \cite{Becker:2002nn,Bonetti:2016dqh} that the above $A$ and the complex dilaton-axion $\tau = C_0+ie^{-\phi}$ are still the correct 
$\cN=1$ coordinates. In the following we will freeze $\tau$ and only consider the dynamics of the field $A$. 

In order to match the action \eqref{action1} with the reduction result \eqref{db4} we now have to assume, that the complexified coordinates $A$, especially the scalars $\rho$, arrange themselves in way, such that only the contribution $(\pd A)^2 (\pd \bar{A})^2$ enters the four dimensional action. 
Comparing the action in the correct field variables with \eqref{action1} leads to \footnote{In this expression we have set $2\pi \sqrt{\alpha'}=1$.}
\begin{equation}\label{coeff}
T_{A A \bar{A} \bar{A}} =-\frac{11}{384}\frac{(2\pi)^{-4}}{\cV^{8/3}} \zeta(3) \big(\text{Im} \tau \big)^{3/2}\int_{Y_3}c_2 \wedge J\, .
\end{equation}
This fixes, at least under the stated assumptions, the numerical factor discussed in \cite{Ciupke:2015msa}.
Let us stress two points. First, we have used a non-trivial coordinate redefinition to obtain \eqref{db4}
and it would be desirable to study its significance in this $\cN=1$ truncated scenario.
Second, it would be desirable to compute the terms involving $\rho$ in order 
to justify the crucial assumption about the dependence on the complex moduli $A$. 
In this computation one would have to use the same non-trivial coordinate redefinition, which
would provide a non-trivial check. 

Let us close this section by some further comments. Preliminary results suggest that 
a generalization of the derivation of $T_{i j \bar k \bar l}$ and other couplings for the kinetic terms of $v^i$ for more than one 
modulus are not only computationally involved, but also induce new non-topological couplings. Whether or 
not these can be absorbed into the definition of the $\cN=1$ coordinates remains to be explored 
in the future. Also the derivation of the actual
scalar potential seems currently difficult, since many of the required higher-derivative terms in ten 
dimensions are not known. While \eqref{coeff} seems to suggest that a potential $V_{(1)}$ given 
in \eqref{action1} is induced if a non-trivial superpotential is included, we are not able to give any further 
evidence for that. It has been shown, for example, in \cite{Bielleman:2016grv} that it 
can happen in string reductions that the potential vanishes if one finds other structures for the kinetic terms
not appearing in \eqref{action1}. It would be desirable to explore this further.


\newpage
\appendix
\section{Definitions and conventions}\label{defs}

The metric signature of the ten and eleven-dimensional spacetime  is $(-,+,\dots,+)$.
Our conventions for the totally 
antisymmetric tensor in Lorentzian signature
 in an orthonormal frame are $\epsilon_{012...9 (10)} = \epsilon_{0123(4)}=+1$. 
The epsilon tensor in $d$ dimensions then satisfies
\begin{align}
\epsilon^{R_1\cdots R_p N_{1 }\ldots N_{d-p}}\epsilon_{R_1 \ldots R_p M_{1} \ldots M_{d-p}} &= (-1)^s (d-p)! p! 
\delta^{N_{1}}{}_{[M_{1}} \ldots \delta^{N_{d-p}}{}_{M_{d-p}]} \,, 
\end{align}
where  $s=0$ if the metric has Euclidean signature and $s=1$ for a Lorentzian metric.

We adopt the following conventions for the Christoffel symbols and Riemann tensor 
\begin{align}
\G^R{}_{M N} & = \fr12 g^{RS} ( \pa_{M} g_{N S} + \pa_N g_{M S} - \pa_S g_{M N}  ) \, , &
R_{M N} & = R^R{}_{M R N} \, , \nonumber\\
R^{M}{}_{N R S} &= \pa_R \G^M{}_{S N}  - \pa_{S} \G^M{}_{R N} + \G^M{}_{R  T} \G^T{}_{S N} - \G^M{}_{ST} \G^T{}_{R N} \,, &
R & = R_{M N} g^{M N} \, , 
\end{align}
with equivalent definitions on the internal and external spaces. Written in components, the first and second  Bianchi identity are
\begin{align}\label{Bainchiid}
{R^O}_{PMN} + {R^O}_{MNP}+{R^O}_{NPM} & =  0 \nonumber\\
\nabla_L R^O{}_{PMN} + \nabla_M R^O{}_{PNL} + \nabla_N R^O{}_{PLM} & =  0 \;\;\; .
\end{align}
Differential $p$-forms are expanded in a basis of differential one-forms as
\begin{equation}
\Lambda = \frac{1}{p!} \Lambda_{M_1\dots M_p} \upd x^{M_1}\wedge \dots \we \upd x^{M_p} \;\; .
\end{equation}
The wedge product between a $p$-form $\Lambda^{(p)}$ and a $q$-form $\Lambda^{(q)}$ is given by
\begin{equation}
(\Lambda^{(p)} \we \Lambda^{(q)})_{M_1 \dots M_{p+q}} = \frac{(p+q)!}{p!q!} \Lambda^{(p)}_{[M_1 \dots M_p} \Lambda^{(q)}_{M_1 \dots M_q]} \;\; .
\end{equation}
Furthermore, the exterior derivative on a $p$-form  $\Lambda$ reads 
\begin{equation}
 ( d \Lambda)_{N M_1\dots M_p} = (p+1) \pa_{[N}\Lambda_{ M_1\dots M_p]} \;\;,
\end{equation}
while the Hodge star of   $p$-form  $\Lambda$ in $d$ real coordinates is given by
\begin{equation}
(\ast_d \Lambda)_{N_1 \dots N_{d-p}} = \frac{1}{p!} \Lambda^{M_1 \dots M_p}\epsilon_{M_1 \dots M_p N_1\dots N_{d-p}} \;\; .
\end{equation}
Moreover, the identity
\begin{equation}\label{idwestar}
 \Lambda^{(1)} \we \ast \Lambda^{(2)} = \frac{1}{p!}\Lambda^{(1)}_{M_1\dots M_p} \Lambda^{(2)}{}^{M_1\dots M_p} \ast 1 \;\;
\end{equation}
  holds  for two arbitrary $p$-forms $\Lambda^{(1)}$ and $\Lambda^{(2)}$. 

Let us specify in more detail our conventions regarding complex coordinates
in the internal space.
For a 
complex Hermitian manifold $\M$ with complex dimension $n$
the complex coordinates $z^1 , \dots, z^n$ and 
the underlying real coordinates $\xi^1, \dots , \xi^{2n}$ are related by
\begin{equation}
( z^1,...,z^n ) = \left(  \frac{1}{\sqrt{2}}(\xi^1 + i \xi^2), \dots ,  \frac{1}{\sqrt{2}}(\xi^{2n-1} + i \xi^{2n}) \right) \,.
\end{equation}
Using these conventions one finds
\begin{equation}
\sqrt{g}  \upd\xi^1 \wedge ... \wedge \upd\xi^{2n} = \sqrt{g} (-1)^{\frac{(n-1)n}{2}} i^n  \upd z^1\wedge...\wedge \upd z^n 
\wedge \upd\bar z^1 \wedge...\wedge \upd\bar z^n = \frac{1}{n!} J^n \,,
\end{equation}
with $g$ the determinant of the metric in real coordinates and  $\sqrt{\det g_{m n}} = \det g_{i \bar j} $. The K\"{a}hler form is given by
\begin{equation}
\label{eq:Kform}
J = i g_{i\bar{\jmath} } \upd z^i \wedge \upd\bar z^{\bar{\jmath} } \, .
\end{equation}
Let $\omega_{p,q}$ be a $(p,q)$-form, then its Hodge dual is the $(n-q,n-p)$ form
\begin{align} \label{eq:pgform}
\ast \omega_{p,q} & = \frac{ (-1)^{\frac{n(n-1) }{2}  } \, i^n \, (-1)^{pn}}  {p!q!(n-p)!(n-q)!}   
\omega_{m_1 \dots m_p \bar{n} _1 \dots \bar{n} _q} 
\epsilon^{m_1 \dots m_p}_{\phantom{m_1 \dots m_p} \bar r_1 \dots \bar r_{n-p}} \nonumber \\
& \quad \times \epsilon^{\bar{n} _1 \dots \bar{n} _q}_{\phantom{\bar \beta_1 \dots \bar{n} _q}  s_1 \dots  s_{n-q}} 
\upd z^{ s_1}\wedge \dots \wedge \upd z^{ s_{n-q}} \wedge \upd \bar z^{\bar r_1} \wedge \dots \wedge \upd \bar z^{\bar  r^{n-p}}.
\end{align}

Finally, let us record our conventions regarding Chern forms.
To begin with, 
 we define the curvature two-form for Hermitian manifolds to be
\begin{equation}\label{curvtwo}
 {\cR^i}_j  =  R\inds{^{i}_{j k \bar{l}}} \,\upd z^ k \wedge \upd\bar{z}^{\bar{l}}
\end{equation}
and we set
 \begin{align} \label{defR3}
 \Tr{\cR}\;\;&  =& {{R^ m }_ m }_{ r \bar{s}}\,\upd z^ r \wedge \upd\bar{z}^{\bar{s}} \;,\nonumber \\
 \Tr{\cR^2} &= & {{R^{ m }}_{n }}_{ r \bar{s}} {{R^{n }}_{ m }}_{ r_1 \bar{s}_1}\,\upd z^{ r}
 \wedge \upd \bar{z}^{\bar{s}}\wedge \upd z^{ r_1} \wedge \upd \bar{z}^{\bar{s}_1} \;,\nonumber  \\
 \Tr{\cR^3} &=& {{R^{ m }}_{n }}_{ r \bar{s}}  R^{n }{}_{n _1  r_1 \bar{s}_1}
 {{R^{n _1}}_{ m }}_{ r_2 \bar{s}_2}\,\upd z^{ r} \wedge \upd \bar{z}^{\bar{s}}\wedge \upd z^{ r_1} \wedge \upd \bar{z}^{\bar{s}_1}\wedge \upd z^{ r_2} \wedge \upd \bar{z}^{\bar{s}_2} \; .
 \end{align}
 The Chern forms can then  be expressed in terms of the curvature two-form as
\begin{align} \label{Chernclasses}
 c_0 &= 1 \nonumber \;, \\
 c_1 &= \frac{i}{2\pi} \Tr{ \mathcal{R}} \nonumber\;, \\
 c_2 &= \frac{1}{(2\pi)^2} \frac{1}{2}\left( \Tr{\cR^2} -  (\Tr{\cR})^2 \right)\;, \\
 c_3 &=   \frac{1}{3}c_1\wedge c_2 + \frac{1}{(2\pi)^2} \frac{1}{3} c_1 \wedge \Tr \cR^2 - 
 \frac{1}{(2\pi)^3}\frac{i}{3} \Tr \cR^3 
   \;,\nonumber \\
 c_4 &= \frac{1}{24} \Big( c_1^4 - \frac{6}{(2\pi)^2} c_1^2 \wedge \Tr\cR^2 
 - \frac{8 i}{(2\pi)^3}  c_1 \wedge \Tr\cR^3\Big) 
 + \frac{1}{(2\pi)^4} \frac{1}{8}\big( (\Tr\cR^2)^2 - 2 \Tr\cR^4 \big)   \,. \nonumber 
\end{align}
The Chern forms of an $n$-dimensional Calabi-Yau manifold $Y_n$ reduce to
\begin{equation}\label{chern34}
c_3 (Y_{n \geq 3}) =  - \frac{1}{(2\pi)^3} \frac{i}{3}  \Tr{\cR^3} \;\; \text{and} \;\;
 c_4 (Y_{n \geq 4}) =\frac{1}{(2\pi)^4} \frac{1}{8}\big((\Tr\cR^2)^2 - 2 \Tr\cR^4\big)\;.
\end{equation}
We  furthermore introduce the intersection numbers $\K_{abc}$ as
\begin{equation}
\K_{abc}=D_a \cdot D_b \cdot D_c=\int_{Y_3} \om_a \wedge \om_b \wedge \om_c\, ,
\end{equation}
where $D_a=[\om_a]_{\text{PD}}$ are divisors Poincare dual to the harmonic (1,1)-forms $\om_a$. Contractions of the intersection numbers with \Kahler moduli $v^a$, such as the four-cycle volumes $\K_a$ are defined as
\begin{align}
\K_a&=\frac{1}{2}\K_{abc}v^b v^c\\
\K_{ab}&=\K_{abc}v^c\nonumber\,.
\end{align}
The corresponding quantities on the backgroud Calabi-Yau are denoted by $\K^\tbz_{ab}, \, \K^\tbz_a$ are defined in an analogous way. 

\section{Eight-derivative terms in ten and eleven dimensions}\label{higherappendix}

\subsection{Eight-derivative terms in ten dimensions}
Here we collect the explicit forms of the eight-derivative terms relevant for the discussion in the main part. The relevant structures containing four Riemann tensors are the familiar $\tilde{t}_8 \wt{t}_8 {\wt{R}}^4$ and $\epsilon_{10}\epsilon_{10}\wt{R}^4$ combinations. Their explicit form in terms of ten-dimensional indices is 
\begin{align}
\tilde{t}_8 \wt{t}_8 \wt{R}^4&={\tilde{t}}\indices{^{A_1 \dotsb A_8}}\,{\tilde{t}}\indices{_{B_1 \dotsb B_8}}{\wt{R}\indices{^{B_1 B_2}_{A_1 A_2}}}\dotsb {\wt{R}\indices{^{B_7 B_8}_{A_7 A_8}}}\\
\epsilon_{10}\epsilon_{10}\wt{R}^4&={\epsilon_{10}}\inds{^{\!\!\!C_1 C_2 A_1 \dotsb A_8}}{\epsilon_{10}}\inds{_{\,C_1 C_2 B_1 \dotsb B_8}}{\wt{R}\indices{^{B_1 B_2}_{A_1 A_2}}}\dotsb {\wt{R}\indices{^{B_7 B_8}_{A_7 A_8}}}\, ,
\end{align}
where $\epsilon_{10}$ is the ten-dimensional curved spacetime Levi-Civita tensor and the tensor $\wt{t}_8$ has the explicit representation in terms of the metric tensor
\begin{align}
\wt{t}_8^{A_1 \dotsb A_8}=&\frac{1}{16}\Big[ -2 \big( \wt{g}^{A_1 A_3}\wt{g}^{A_2 A_4}\wt{g}^{A_5 A_7}\wt{g}^{A_6 A_8}+\wt{g}^{A_1 A_5}\wt{g}^{A_2 A_6} \wt{g}^{A_3 A_7} \wt{g}^{A_4 A_8}+\wt{g}^{A_1 A_7}\wt{g}^{A_2 A_8}\wt{g}^{A_3 A_5} \wt{g}^{A_4 A_6}\big)\nonumber\\
&+8\big(\wt{g}^{A_2 A_3}\wt{g}^{A_4 A_5} \wt{g}^{A_6 A_7}\wt{g}^{A_8 A_1}+\wt{g}^{A_2 A_5}\wt{g}^{A_6 A_3} \wt{g}^{A_4 A_7} \wt{g}^{A_8 A_1} +\wt{g}^{A_2 A_5}\wt{g}^{A_6 A_7}\wt{g}^{A_8 A_3}\wt{g}^{A_4 A_1}\big)\nonumber\\
&-(A_1 \leftrightarrow A_2)-(A_3 \leftrightarrow A_4)-(A_5 \leftrightarrow A_6)-(A_7 \leftrightarrow A_8) \Big]\,.\label{10dt8}
\end{align}
The $\wt{B}_2$ completion is then obtained by introducing the connection with torsion as outlined in section \ref{IIAderivatives}
\be
{\Omega_{\pm}}\inds{_{A}^{\ax \bx}}=\Omega\inds{_{A}^{\ax \bx}} \pm \frac{1}{2} \wt{H}_3{}_A{}^{\ax \bx}\, ,
\ee
where $\Omega_{A}{}^{\ax \bx}$ are the components of the $\mathfrak{so}(1,9)$ - valued connection one-form corresponding to the Levi-Civita connection. In this notation $\ax, \bx$ are flat tangent space indices of $\cM_{10}$. The structure $\epsilon_{10}\epsilon_{10}\wt{H}_{3}^2 \wt{R}^3$ which enters the replacement in section \ref{IIAderivatives} has the component form
\begin{align}
\epsilon_{10}\epsilon_{10}\wt{H}_{3}^2 \wt{R}^3(\Omega_{\pm})=\epsilon_{10\, C_1 A_0 \dotsb A_8}\, \epsilon_{10}^{C1 B_0 \dotsb B_8} &{\wt{H}_3}{}^{A_1 A_2}{}_{B_0}{\wt{H}_3}{}_{B_1 B_2}{}^{A_0}\label{repl1}\\
&\times {\wt{R}(\Omega_+)}\inds{^{A_3 A_4}_{B_3 B_4}}\dotsb {\wt{R}(\Omega_+)}\inds{^{A_7 A_8}_{B_7 B_8}}\, .\nonumber
\end{align}

\subsection{Eight-derivative terms in eleven dimensions}

The classical eleven-dimensional supergravity action gets corrected by different contributions which should be discussed in the following. The most prominent higher curvature term in M-theory is the sector containing four Riemann tensors. These involve two different structures namely
\begin{equation}\label{R4terms}
S_{R^4}=\frac{1}{2 \kappa^2_{11}} \int_{\mathcal{M}_{11}} \Big(\wh{t}_8\wh{t}_8-\frac{1}{24} \epsilon_{\scriptscriptstyle 11} \epsilon_{\scriptscriptstyle 11}\Big){\wh{R}}^4 \wh{\ast} 1.
\end{equation}
In (\ref{R4terms}) the two quantities $\wh{t}_8\, \wh{t}_8 {\wh{R}}^4$ and $\epsilon_{11}\, \epsilon_{11}{\wh{R}}^4$ have the index representation
\begin{align}
\wh{t}_8\,\wh{t}_8 {\wh{R}}^4&={t_8}\inds{^{M_1 \dotsb M_8}}{t_8}\inds{_{N_1 \dotsb N_8}}{\wh{R}}\inds{^{N_1 N_2}_{M_1 M_2}} \dotsb{\wh{R}}\inds{^{N_7 N_8}_{M_7 M_8}}\label{Mtht8}\\
\epsilon_{11}\epsilon_{11}{\wh{R}}^4&={\epsilon_{11}}\inds{^{R_1 R_2 R_3 M_1 \dotsb M_8}}{\epsilon_{11}}\inds{_{R_1 R_2 R_3 N_1 \dotsb N_8}}{\wh{R}}\inds{^{N_1 N_2}_{M_1 M_2}} \dotsb{\wh{R}}\inds{^{N_7 N_8}_{M_7 M_8}}\label{Mthe11}.
\end{align}
The tensor $\wh{t}_8$ is defined in a completely analogous way as in (\ref{10dt8}).  These $\cR^4$ - terms are furthermore supplemented by another term quartic in the Riemann tensor. This term however also comprises a three form $\wh{C}_3$. This piece of the higher-derivative action then has the form
\begin{equation}
S_{C_3 X_8}= -\frac{3^2 2^{13}}{2 \kappa_{11}^2} \int_{\mathcal{M}_{11}}\wh{C}_3 \wedge \wh{X}_8
\end{equation}
where eight form $\wh{X}_8$ is defined as
\begin{equation}\label{C3X8}
\wh{X}_8=\frac{1}{192}\left[ \Tr {\wh{\mathcal{R}}}^4-\frac{1}{4}\left(  \Tr {\wh{\mathcal{R}}}^2 \right)^2   \right]
\end{equation}
which is in terms of the (real) curvature two form
\begin{equation}
\wh{\mathcal{R}}\inds{^{M}_{N}}=\frac{1}{2} {\wh{R}}\inds{^{M}_{N N_1 N_2}} \upd x^{N_1} \wedge \upd x^{N_2}.
\end{equation}
In addition to these quartic Riemann tensor terms it was conjectured in \cite{Liu:2013dna} that the complete $\wh{G}_4$ dependence at $\mathcal{O}(\wh{G}_4^2)$ is captured by introducing 
\begin{align}
\wh{t}_8\, \wh{t}_8\, {\wh{G}_4}^2\, {\wh{R}}^3=&{\wh{t}_8}^{M_1 \dotsb M_8}\,{\wh{t}_8}{}_{N_1 \dotsb N_8}{\wh{G}_4}{}^{N_1}{}_{M_1 R_1 R_2}{\wh{G}_4}{}^{N_2}{}_{M_2}{}^{R_1 R_2}{\wh{R}}\inds{^{N_3 N_4}_{M_3 M_4}}\dotsb{\wh{R}}\inds{^{N_7 N_8}_{M_7 M_8}}\label{t8t8G2R3}\\
\epsilon_{11}\epsilon_{11}\wh{G}^2_4\, {R}^3=&{\epsilon_{11}}\inds{^{R M_1 \dotsb M_{10}}}{\epsilon_{11}}\inds{_{R N_1 \dotsb N_{10}}}{\wh{G}_4}{}^{N_1 N_2}{}_{M_1 M_2}{\wh{G}_4}{}^{N_3 N_4}{}_{M_3 M_4}{\wh{R}}\inds{^{N_5 N_6}_{M_5 M_6}}\dotsb{\wh{R}}\inds{^{N_9 N_{10}}_{M_9 M_{10}}}\, .\label{e11e11G2R3}
\end{align}
The last eleven-dimensional eight-derivative contribution involves the tensor $\wh{s}_{18}$ parametrized by six unknown coefficients $a_n \in \mathbb{R}$. We then have the additional coupling of the form
\begin{align}
\wh{s}_{18} (\wh{\nabla} \wh{G}_{4})^2 {\wh{R}}^2&={\wh{s}_{18}}{}^{N_1 \dotsb N_{18}} {\wh{R}}\inds{_{N_1 \dotsb N_4}} {\wh{R}}\inds{_{N_5 \dotsb N_8}} \wh{\nabla}_{N_9} {\wh{G}_4}{}_{N_{10} \dotsb N_{13}} \wh{\nabla}_{N_{14}} {\wh{G}_4}{}_{N_{15} \dotsb N_{18}}\nonumber\\
&=\mathcal{A}+\sum_{n=1}^{6}a_n\, \mathcal{Z}_n\, ,
\end{align}
where the quantities $\mathcal{A}, \mathcal{Z}_n$ are defined by
\begin{align}
\mathcal{A}&=-24 B_5-48B_8-24B_{10}-6B_{12}-12 B_{13}+12B_{14}+8B_{16}-4 B_{20}+B_{22}+4 B_{23}+B_{24}\nonumber\\
\mathcal{Z}_1&=48 B_1+48B_2-48B_3+36B_4+96 B_6+48 B_7-48 B_8+96 B_{10}+12 B_{12}+24 B_{13}\nonumber\\
\mathcal{Z}_2&=-48 B_1 -48 B_2-24 B_4-24 B_5+48 B_6-48 B_8-24 B_9-72 B_{10}-24 B_{13}+24 B_{14}\nonumber\\
&-B_{22}+4 B_{23}\nonumber\\
\mathcal{Z}_3&=12 B_1+12 B_2-24 B_3+9 B_4+48 B_6+24 B_7-24 B_8+24 B_{10}+6 B_{12}+6 B_{13}+4 B_{15}\nonumber\\
&-4 B_{17}+3 B_{19}+2 B_{21}\nonumber\\
\mathcal{Z}_{4}&=12 B_1+12 B_2-12 B_3+9 B_4+24 B_6+12 B_7-12 B_8+24 B_{10}+3 B_{12}+6 B_{13}\nonumber\\
&+4 B_{15}-4 B_{17}+2 B_{20}\nonumber\\
\mathcal{Z}_5&=4B_3-8B_6-4B_7+4B_8-B_{12}-2B_{14}+4B_{18}\nonumber\\
\mathcal{Z}_6&=B_4+2B_{11}. \label{Zbasis}
\end{align}
The elements $B_i$ form a basis of the terms with the structure $(\wh{\nab} \wh{G}_4)^2 \wh{R}^2$ given in ( \ref{def-Bi}).
\vspace*{-1cm}
\begin{center}
\resizebox{1\textwidth}{!}{
\hspace*{-.1cm}\begin{minipage}[h]{\textwidth}
\begin{align}
B_1&=R\inds{_{N_1 N_2 N_3 N_4}} R\inds{_{N_5 N_6 N_7 N_8}} \nabla^{N_5} G\inds{^{N_1 N_7 N_8}_{N_9}} \nabla^{N_3} G\inds{^{N_2 N_4 N_6 N_9}}&~~~~~~
B_2&=R\inds{_{N_1 N_2 N_3 N_4}} R\inds{_{N_5 N_6 N_7 N_8}} \nabla^{N_5} G\inds{^{N_1 N_3 N_7}_{N_9}} \nabla^{N_8} G\inds{^{N_2 N_4 N_6 N_9}}\nonumber\\
B_3&=R\inds{_{N_1 N_2 N_3 N_4}} R\inds{_{N_5 N_6 N_7 N_8}} \nabla^{N_5} G\inds{^{N_1 N_3 N_7}_{N_9}} \nabla^{N_6} G\inds{^{N_2 N_4 N_8 N_9}}&~~~~~~
B_4&=R\inds{_{N_1 N_2 N_3 N_4}} R\inds{_{N_5 N_6 N_7 N_8}} \nabla_{N_9} G\inds{^{N_3 N_4 N_7 N_8}} \nabla^{N_6} G\inds{^{N_9 N_1 N_2 N_5}}\nonumber\\
B_5&=R\inds{_{N_1 N_2 N_3 N_4}} R\inds{_{N_5 N_6 N_7}^{N_4}} \nabla^{N_1} G\inds{^{N_2 N_3}_{N_8 N_9}} \nabla^{N_5} G\inds{^{N_6 N_7 N_8 N_9}}&~~~~~~
B_6&=R\inds{_{N_1 N_2 N_3 N_4}} R\inds{_{N_5 N_6 N_7}^{N_4}} \nabla^{N_1} G\inds{^{N_2 N_5}_{N_8 N_9}} \nabla^{N_3} G\inds{^{N_6 N_7 N_8 N_9}}\nonumber\\
B_7&=R\inds{_{N_1 N_2 N_3 N_4}} R\inds{_{N_5 N_6 N_7}^{N_4}} \nabla^{N_1} G\inds{^{N_2 N_5}_{N_8 N_9}} \nabla^{N_7} G\inds{^{N_3 N_6 N_8 N_9}}&~~~~~~
B_8&=R\inds{_{N_1 N_2 N_3 N_4}} R\inds{_{N_5 N_6 N_7}^{N_4}} \nabla^{N_1} G\inds{^{N_3 N_5}_{N_8 N_9}} \nabla^{N_2} G\inds{^{N_6 N_7 N_8 N_9}}\nonumber\\
B_9&=R\inds{_{N_1 N_2 N_3 N_4}} R\inds{_{N_5 N_6 N_7}^{N_4}} \nabla^{N_1} G\inds{^{N_3 N_5}_{N_8 N_9}} \nabla^{N_6} G\inds{^{N_2 N_7 N_8 N_9}}&~~~~~~
B_{10}&=R\inds{_{N_1 N_2 N_3 N_4}} R\inds{_{N_5 N_6 N_7}^{N_4}} \nabla_{N_9} G\inds{^{N_3 N_5 N_7}_{N_8}} \nabla^{N_9} G\inds{^{N_1 N_2 N_6 N_8}}\nonumber\\
B_{11}&=R\inds{_{N_1 N_2 N_3 N_4}} R\inds{_{N_5 N_6 N_7}^{N_4}} \nabla_{N_8} G\inds{^{N_1 N_2 N_6}_{N_9}} \nabla^{N_9} G\inds{^{N_3 N_5 N_7 N_8}}&~~~~~~
B_{12}&=R\inds{_{N_1 N_2 N_3 N_4}} R\inds{_{N_5 N_6 N_7}^{N_4}} \nabla^{N_3} G\inds{^{N_5 N_6}_{N_8 N_9}} \nabla^{N_7} G\inds{^{N_2 N_1 N_8 N_9}}\nonumber\\
B_{13}&=R\inds{_{N_1 N_2 N_3 N_4}}  R\inds{_{N_5}^{N_1}_{N_6}^{N_3}} \nabla_{N_9} G\inds{^{N_2 N_6}_{N_7 N_8}} \nabla^{N_9} G\inds{^{N_4 N_5 N_7 N_8}}&~~~~~~
B_{14}&=R\inds{_{N_1 N_2 N_3 N_4}}  R\inds{_{N_5}^{N_1}_{N_6}^{N_3}} \nabla_{N_9} G\inds{^{N_2 N_4}_{N_7 N_8}} \nabla^{N_5} G\inds{^{N_4 N_7 N_8 N_9}}\nonumber\\
B_{15}&=R\inds{_{N_1 N_2 N_3 N_4}}  R\inds{_{N_5}^{N_1}_{N_6}^{N_3}} \nabla^{N_2} G\inds{^{N_6}_{N_7 N_8 N_9}} \nabla^{N_5} G\inds{^{N_4 N_7 N_8 N_9}}&~~~~~~
B_{16}&=R\inds{_{N_1 N_2 N_3 N_4}}  R\inds{_{N_5}^{N_1}_{N_6}^{N_3}} \nabla^{N_2} G\inds{^{N_4}_{N_7 N_8 N_9}} \nabla^{N_5} G\inds{^{N_6 N_7 N_8 N_9}}\nonumber\\
B_{17}&=R\inds{_{N_1 N_2 N_3 N_4}}  R\inds{_{N_5}^{N_1}_{N_6}^{N_3}} \nabla^{N_2} G\inds{^{N_5}_{N_7 N_8 N_9}} \nabla^{N_4} G\inds{^{N_6 N_7 N_8 N_9}}&~~~~~~
B_{18}&=R\inds{_{N_1 N_2 N_3 N_4}}  R\inds{_{N_5}^{N_1}_{N_6}^{N_3}} \nabla_{N_9} G\inds{^{N_5 N_6}_{N_7 N_8}} \nabla^{N_4} G\inds{^{N_2 N_7 N_8 N_9}}\nonumber\\
B_{19}&=R\inds{_{N_1 N_2 N_3 N_4}}  R\inds{_{N_5 N_6}^{N_3 N_4}} \nabla_{N_9} G\inds{^{N_1 N_5}_{N_7 N_8}} \nabla^{N_9} G\inds{^{N_2 N_6 N_7 N_8}}&~~~~~~
B_{20}&=R\inds{_{N_1 N_2 N_3 N_4}}  R\inds{_{N_5 N_6}^{N_3 N_4}} \nabla^{N_1} G\inds{^{N_5}_{N_7 N_8 N_9}} \nabla^{N_2} G\inds{^{N_6 N_7 N_8 N_9}}\nonumber\\
B_{21}&=R\inds{_{N_1 N_2 N_3 N_4}}  R\inds{_{N_5 N_6}^{N_3 N_4}} \nabla^{N_1} G\inds{^{N_5}_{N_7 N_8 N_9}} \nabla^{N_6} G\inds{^{N_2 N_7 N_8 N_9}}&~~~~~~
B_{22}&=R\inds{_{N_1 N_2 N_3 N_4}}  R\inds{_{N_5}^{N_1 N_3 N_4}} \nabla^{N_2} G\inds{_{N_6 N_7 N_8 N_9}} \nabla^{N_5} G\inds{^{N_6 N_7 N_8 N_9}}\nonumber\\
B_{23}&=R\inds{_{N_1 N_2 N_3 N_4}}  R\inds{_{N_5}^{N_1 N_3 N_4}} \nabla_{N_9} G\inds{^{N_2}_{N_6 N_7 N_8}} \nabla^{N_9} G\inds{^{N_5 N_6 N_7 N_8}}&~~~~~~
B_{24}&=R\inds{_{N_1 N_2 N_3 N_4}}  R\inds{^{N_1 N_2 N_3 N_4}} \nabla_{N_5} G\inds{_{N_6 N_7 N_8}} \nabla^{N_6} G\inds{^{N_5  N_7 N_8 N_9}}\nonumber
\end{align}
\end{minipage}
}
\vspace{-.1cm}
\begin{equation}
 \ \label{def-Bi}
\end{equation}
\end{center}
\bibliographystyle{utcaps}
\newpage
\bibliography{references}

\providecommand{\href}[2]{#2}\begingroup\raggedright\begin{thebibliography}{10}

\bibitem{Blumenhagen:2013fgp}
R.~Blumenhagen, D.~Lüst, and S.~Theisen, {\em {Basic concepts of string
  theory}}.
\newblock Theoretical and Mathematical Physics. Springer, Heidelberg, Germany,
2013.
\newblock

\bibitem{Ibanez:2012zz}
L.~E. Ibanez and A.~M. Uranga, {\em {String theory and particle physics: An
  introduction to string phenomenology}}.
\newblock Cambridge University Press,
2012.
\newblock

\bibitem{Blumenhagen:2006ci}
R.~Blumenhagen, B.~Kors, D.~Lust, and S.~Stieberger, ``{Four-dimensional String
  Compactifications with D-Branes, Orientifolds and Fluxes},'' {\em Phys.
  Rept.} {\bf 445} (2007) 1--193,
\href{http://arXiv.org/abs/hep-th/0610327}{{\tt hep-th/0610327}}.

\bibitem{Gross:1986iv}
D.~J. Gross and E.~Witten, ``{Superstring Modifications of Einstein's
  Equations},'' {\em Nucl. Phys.} {\bf B277} (1986)
1.

\bibitem{Gross:1986mw}
D.~J. Gross and J.~H. Sloan, ``{The Quartic Effective Action for the Heterotic
  String},'' {\em Nucl. Phys.} {\bf B291} (1987)
41--89.

\bibitem{Duff:1995wd}
M.~J. Duff, J.~T. Liu, and R.~Minasian, ``{Eleven-dimensional origin of
  string-string duality: A One loop test},'' {\em Nucl. Phys.} {\bf B452}
  (1995) 261--282,
\href{http://arXiv.org/abs/hep-th/9506126}{{\tt hep-th/9506126}}.

\bibitem{Green:1997di}
M.~B. Green and P.~Vanhove, ``{D instantons, strings and M theory},'' {\em
  Phys. Lett.} {\bf B408} (1997) 122--134,
\href{http://arXiv.org/abs/hep-th/9704145}{{\tt hep-th/9704145}}.

\bibitem{Green:1997as}
M.~B. Green, M.~Gutperle, and P.~Vanhove, ``{One loop in eleven-dimensions},''
  {\em Phys. Lett.} {\bf B409} (1997) 177--184,
\href{http://arXiv.org/abs/hep-th/9706175}{{\tt hep-th/9706175}}.

\bibitem{Kiritsis:1997em}
E.~Kiritsis and B.~Pioline, ``{On R**4 threshold corrections in IIb string
  theory and (p, q) string instantons},'' {\em Nucl. Phys.} {\bf B508} (1997)
  509--534,
\href{http://arXiv.org/abs/hep-th/9707018}{{\tt hep-th/9707018}}.

\bibitem{Kehagias:1997cq}
A.~Kehagias and H.~Partouche, ``{On the exact quartic effective action for the
  type IIB superstring},'' {\em Phys. Lett.} {\bf B422} (1998) 109--116,
\href{http://arXiv.org/abs/hep-th/9710023}{{\tt hep-th/9710023}}.

\bibitem{Kehagias:1997jg}
A.~Kehagias and H.~Partouche, ``{D instanton corrections as (p,q) string
  effects and nonrenormalization theorems},'' {\em Int. J. Mod. Phys.} {\bf
  A13} (1998) 5075--5092,
\href{http://arXiv.org/abs/hep-th/9712164}{{\tt hep-th/9712164}}.

\bibitem{Policastro:2006vt}
G.~Policastro and D.~Tsimpis, ``{R**4, purified},'' {\em Class. Quant. Grav.}
  {\bf 23} (2006) 4753--4780,
\href{http://arXiv.org/abs/hep-th/0603165}{{\tt hep-th/0603165}}.

\bibitem{Policastro:2008hg}
G.~Policastro and D.~Tsimpis, ``{A Note on the quartic effective action of type
  IIB superstring},'' {\em Class. Quant. Grav.} {\bf 26} (2009) 125001,
\href{http://arXiv.org/abs/0812.3138}{{\tt 0812.3138}}.

\bibitem{Liu:2013dna}
J.~T. Liu and R.~Minasian, ``{Higher-derivative couplings in string theory:
  dualities and the B-field},'' {\em Nucl. Phys.} {\bf B874} (2013) 413--470,
\href{http://arXiv.org/abs/1304.3137}{{\tt 1304.3137}}.

\bibitem{Minasian:2015bxa}
R.~Minasian, T.~G. Pugh, and R.~Savelli, ``{F-theory at order $\alpha'^3$},''
  {\em JHEP} {\bf 10} (2015) 050,
\href{http://arXiv.org/abs/1506.06756}{{\tt 1506.06756}}.

\bibitem{Russo:1997mk}
J.~G. Russo and A.~A. Tseytlin, ``{One loop four graviton amplitude in
  eleven-dimensional supergravity},'' {\em Nucl. Phys.} {\bf B508} (1997)
  245--259,
\href{http://arXiv.org/abs/hep-th/9707134}{{\tt hep-th/9707134}}.

\bibitem{Tseytlin:2000sf}
A.~A. Tseytlin, ``{R**4 terms in 11 dimensions and conformal anomaly of (2,0)
  theory},'' {\em Nucl. Phys.} {\bf B584} (2000) 233--250,
\href{http://arXiv.org/abs/hep-th/0005072}{{\tt hep-th/0005072}}.

\bibitem{Peeters:2005tb}
K.~Peeters, J.~Plefka, and S.~Stern, ``{Higher-derivative gauge field terms in
  the M-theory action},'' {\em JHEP} {\bf 08} (2005) 095,
\href{http://arXiv.org/abs/hep-th/0507178}{{\tt hep-th/0507178}}.

\bibitem{Hyakutake:2005rb}
Y.~Hyakutake and S.~Ogushi, ``{R**4 corrections to eleven dimensional
  supergravity via supersymmetry},'' {\em Phys. Rev.} {\bf D74} (2006) 025022,
\href{http://arXiv.org/abs/hep-th/0508204}{{\tt hep-th/0508204}}.

\bibitem{Hyakutake:2006aq}
Y.~Hyakutake and S.~Ogushi, ``{Higher derivative corrections to eleven
  dimensional supergravity via local supersymmetry},'' {\em JHEP} {\bf 02}
  (2006) 068,
\href{http://arXiv.org/abs/hep-th/0601092}{{\tt hep-th/0601092}}.

\bibitem{Hyakutake:2007sm}
Y.~Hyakutake, ``{Toward the Determination of R**3 F**2 Terms in M-theory},''
  {\em Prog. Theor. Phys.} {\bf 118} (2007) 109,
\href{http://arXiv.org/abs/hep-th/0703154}{{\tt hep-th/0703154}}.

\bibitem{Martin-Garcia:2007bqa}
J.~M. Martin-Garcia, R.~Portugal, and L.~R.~U. Manssur, ``{The Invar Tensor
  Package},'' {\em Comput. Phys. Commun.} {\bf 177} (2007) 640--648,
\href{http://arXiv.org/abs/0704.1756}{{\tt 0704.1756}}.

\bibitem{Nutma:2013zea}
T.~Nutma, ``{xTras : A field-theory inspired xAct package for mathematica},''
  {\em Comput. Phys. Commun.} {\bf 185} (2014) 1719--1738,
\href{http://arXiv.org/abs/1308.3493}{{\tt 1308.3493}}.

\bibitem{Martin-Garcia}
J.~M. Martin-Garcia, ``{xPerm: fast index canonicalization for tensor computer
  algebra},'' {\em Comput. Phys. Commun.} {\bf 179} (2008) 597--603,
\href{http://arXiv.org/abs/0803.0862}{{\tt 0803.0862}}.

\bibitem{Paulos:2008tn}
M.~F. Paulos, ``{Higher derivative terms including the Ramond-Ramond
  five-form},'' {\em JHEP} {\bf 10} (2008) 047,
\href{http://arXiv.org/abs/0804.0763}{{\tt 0804.0763}}.

\bibitem{Candelas:1990rm}
P.~Candelas, X.~C. De~La~Ossa, P.~S. Green, and L.~Parkes, ``{A Pair of
  Calabi-Yau manifolds as an exactly soluble superconformal theory},'' {\em
  Nucl. Phys.} {\bf B359} (1991)
21--74.

\bibitem{Antoniadis:1993ze}
I.~Antoniadis, E.~Gava, K.~S. Narain, and T.~R. Taylor, ``{Topological
  amplitudes in string theory},'' {\em Nucl. Phys.} {\bf B413} (1994) 162--184,
\href{http://arXiv.org/abs/hep-th/9307158}{{\tt hep-th/9307158}}.

\bibitem{Antoniadis:1997eg}
I.~Antoniadis, S.~Ferrara, R.~Minasian, and K.~S. Narain, ``{R**4 couplings in
  M and type II theories on Calabi-Yau spaces},'' {\em Nucl. Phys.} {\bf B507}
  (1997) 571--588,
\href{http://arXiv.org/abs/hep-th/9707013}{{\tt hep-th/9707013}}.

\bibitem{Becker:2002nn}
K.~Becker, M.~Becker, M.~Haack, and J.~Louis, ``{Supersymmetry breaking and
  alpha-prime corrections to flux induced potentials},'' {\em JHEP} {\bf 06}
  (2002) 060,
\href{http://arXiv.org/abs/hep-th/0204254}{{\tt hep-th/0204254}}.

\bibitem{Bonetti:2016dqh}
F.~Bonetti and M.~Weissenbacher, ``{The Euler characteristic correction to the
  Kaehler potential - revisited},''
\href{http://arXiv.org/abs/1608.01300}{{\tt 1608.01300}}.

\bibitem{Hanaki:2006pj}
K.~Hanaki, K.~Ohashi, and Y.~Tachikawa, ``{Supersymmetric Completion of an R**2
  term in Five-dimensional Supergravity},'' {\em Prog. Theor. Phys.} {\bf 117}
  (2007) 533,
\href{http://arXiv.org/abs/hep-th/0611329}{{\tt hep-th/0611329}}.

\bibitem{Ozkan:2013nwa}
M.~Ozkan and Y.~Pang, ``{All off-shell $R^{2}$ invariants in five dimensional
  $\mathcal{N} =$ 2 supergravity},'' {\em JHEP} {\bf 08} (2013) 042,
\href{http://arXiv.org/abs/1306.1540}{{\tt 1306.1540}}.

\bibitem{Castro:2008ne}
A.~Castro, J.~L. Davis, P.~Kraus, and F.~Larsen, ``{String Theory Effects on
  Five-Dimensional Black Hole Physics},'' {\em Int. J. Mod. Phys.} {\bf A23}
  (2008) 613--691,
\href{http://arXiv.org/abs/0801.1863}{{\tt 0801.1863}}.

\bibitem{Ciupke:2016agp}
D.~Ciupke, ``{Scalar Potential from Higher Derivative $\mathcal{N} = 1$
  Superspace},''
\href{http://arXiv.org/abs/1605.00651}{{\tt 1605.00651}}.

\bibitem{Ciupke:2015msa}
D.~Ciupke, J.~Louis, and A.~Westphal, ``{Higher-Derivative Supergravity and
  Moduli Stabilization},'' {\em JHEP} {\bf 10} (2015) 094,
\href{http://arXiv.org/abs/1505.03092}{{\tt 1505.03092}}.

\bibitem{Grimm:2014xva}
T.~W. Grimm, T.~G. Pugh, and M.~Weissenbacher, ``{On M-theory fourfold vacua
  with higher curvature terms},'' {\em Phys. Lett.} {\bf B743} (2015) 284--289,
\href{http://arXiv.org/abs/1408.5136}{{\tt 1408.5136}}.

\bibitem{Grimm:2014efa}
T.~W. Grimm, T.~G. Pugh, and M.~Weissenbacher, ``{The effective action of
  warped M-theory reductions with higher derivative terms — part I},'' {\em
  JHEP} {\bf 01} (2016) 142,
\href{http://arXiv.org/abs/1412.5073}{{\tt 1412.5073}}.

\bibitem{Grimm:2015mua}
T.~W. Grimm, T.~G. Pugh, and M.~Weissenbacher, ``{The effective action of
  warped M-theory reductions with higher-derivative terms - Part II},'' {\em
  JHEP} {\bf 12} (2015) 117,
\href{http://arXiv.org/abs/1507.00343}{{\tt 1507.00343}}.

\bibitem{Cremmer:1978km}
E.~Cremmer, B.~Julia, and J.~Scherk, ``{Supergravity Theory in
  Eleven-Dimensions},'' {\em Phys. Lett.} {\bf B76} (1978)
409--412.

\bibitem{Cadavid:1995bk}
A.~C. Cadavid, A.~Ceresole, R.~D'Auria, and S.~Ferrara, ``{Eleven-dimensional
  supergravity compactified on Calabi-Yau threefolds},'' {\em Phys. Lett.} {\bf
  B357} (1995) 76--80,
\href{http://arXiv.org/abs/hep-th/9506144}{{\tt hep-th/9506144}}.

\bibitem{Becker:1996gj}
K.~Becker and M.~Becker, ``{M theory on eight manifolds},'' {\em Nucl. Phys.}
  {\bf B477} (1996) 155--167,
\href{http://arXiv.org/abs/hep-th/9605053}{{\tt hep-th/9605053}}.

\bibitem{Becker:2001pm}
K.~Becker and M.~Becker, ``{Supersymmetry breaking, M theory and fluxes},''
  {\em JHEP} {\bf 07} (2001) 038,
\href{http://arXiv.org/abs/hep-th/0107044}{{\tt hep-th/0107044}}.

\bibitem{Lu:2004ng}
H.~Lu, C.~N. Pope, K.~S. Stelle, and P.~K. Townsend, ``{String and M-theory
  deformations of manifolds with special holonomy},'' {\em JHEP} {\bf 07}
  (2005) 075,
\href{http://arXiv.org/abs/hep-th/0410176}{{\tt hep-th/0410176}}.

\bibitem{Lu:2003ze}
H.~Lu, C.~N. Pope, K.~S. Stelle, and P.~K. Townsend, ``{Supersymmetric
  deformations of G(2) manifolds from higher order corrections to string and M
  theory},'' {\em JHEP} {\bf 10} (2004) 019,
\href{http://arXiv.org/abs/hep-th/0312002}{{\tt hep-th/0312002}}.

\bibitem{Ceresole:2000jd}
A.~Ceresole and G.~Dall'Agata, ``{General matter coupled N=2, D = 5 gauged
  supergravity},'' {\em Nucl. Phys.} {\bf B585} (2000) 143--170,
\href{http://arXiv.org/abs/hep-th/0004111}{{\tt hep-th/0004111}}.

\bibitem{Bergshoeff:2004kh}
E.~Bergshoeff, S.~Cucu, T.~de~Wit, J.~Gheerardyn, S.~Vandoren, and
  A.~Van~Proeyen, ``{N = 2 supergravity in five-dimensions revisited},'' {\em
  Class. Quant. Grav.} {\bf 21} (2004) 3015--3042,
  \href{http://arXiv.org/abs/hep-th/0403045}{{\tt hep-th/0403045}}.
[Class. Quant. Grav.23,7149(2006)].

\bibitem{Bonetti:2011mw}
F.~Bonetti and T.~W. Grimm, ``{Six-dimensional (1,0) effective action of
  F-theory via M-theory on Calabi-Yau threefolds},'' {\em JHEP} {\bf 05} (2012)
  019,
\href{http://arXiv.org/abs/1112.1082}{{\tt 1112.1082}}.

\bibitem{Katmadas:2013mma}
S.~Katmadas and R.~Minasian, ``{$\mathcal{N} =$ 2 higher-derivative couplings
  from strings},'' {\em JHEP} {\bf 02} (2014) 093,
\href{http://arXiv.org/abs/1311.4797}{{\tt 1311.4797}}.

\bibitem{Weissenbacher:2016gey}
M.~Weissenbacher, ``{On four-derivative terms in IIB Calabi-Yau orientifold
  reductions},''
\href{http://arXiv.org/abs/1607.03913}{{\tt 1607.03913}}.

\bibitem{Ferrara:1996hh}
S.~Ferrara, R.~R. Khuri, and R.~Minasian, ``{M theory on a Calabi-Yau
  manifold},'' {\em Phys. Lett.} {\bf B375} (1996) 81--88,
\href{http://arXiv.org/abs/hep-th/9602102}{{\tt hep-th/9602102}}.

\bibitem{Grisaru:1986dk}
M.~T. Grisaru, A.~E.~M. van~de Ven, and D.~Zanon, ``{Two-Dimensional
  Supersymmetric Sigma Models on Ricci Flat Kahler Manifolds Are Not Finite},''
  {\em Nucl. Phys.} {\bf B277} (1986)
388--408.

\bibitem{Koehn:2012ar}
M.~Koehn, J.-L. Lehners, and B.~A. Ovrut, ``{Higher-Derivative Chiral
  Superfield Actions Coupled to N=1 Supergravity},'' {\em Phys. Rev.} {\bf D86}
  (2012) 085019,
\href{http://arXiv.org/abs/1207.3798}{{\tt 1207.3798}}.

\bibitem{Giddings:2001yu}
S.~B. Giddings, S.~Kachru, and J.~Polchinski, ``{Hierarchies from fluxes in
  string compactifications},'' {\em Phys. Rev.} {\bf D66} (2002) 106006,
\href{http://arXiv.org/abs/hep-th/0105097}{{\tt hep-th/0105097}}.

\bibitem{Grimm:2004uq}
T.~W. Grimm and J.~Louis, ``{The Effective action of N = 1 Calabi-Yau
  orientifolds},'' {\em Nucl. Phys.} {\bf B699} (2004) 387--426,
\href{http://arXiv.org/abs/hep-th/0403067}{{\tt hep-th/0403067}}.

\bibitem{Bielleman:2016grv}
S.~Bielleman, L.~E. Ibanez, F.~G. Pedro, I.~Valenzuela, and C.~Wieck, ``{The
  DBI Action, Higher-derivative Supergravity, and Flattening Inflaton
  Potentials},'' {\em JHEP} {\bf 05} (2016) 095,
\href{http://arXiv.org/abs/1602.00699}{{\tt 1602.00699}}.

\end{thebibliography}\endgroup

\end{document}